\documentclass[a4paper,11pt]{article}
\pdfoutput=1 

\usepackage{jcappub} 

\usepackage[T1]{fontenc} 

\usepackage{amsmath}
\usepackage{amssymb}
\usepackage{amsfonts}
\usepackage{bm}
\usepackage{verbatim}
\usepackage{tablefootnote}
\usepackage{enumerate}
\usepackage{afterpage}
\usepackage{xcolor}
\usepackage{natbib, hyperref}

\newcommand{\orcid}[1]{\href{https://orcid.org/#1}{\,\includegraphics[width=8px]{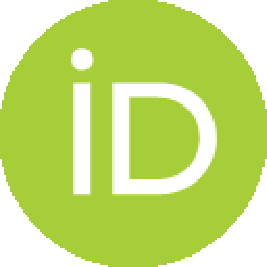}}}

\title{\boldmath Model-agnostic assessment of dark energy after DESI DR1 BAO}

\author[a,1]{Bikash R. Dinda\orcid{0000-0001-5432-667X}\note{Corresponding author.}}
\author[a,b,c]{Roy Maartens\orcid{0000-0001-9050-5894}}

\affiliation[a]{Department of Physics \& Astronomy, University of the Western Cape, Cape Town 7535, South Africa}

\affiliation[b]{Institute of Cosmology \& Gravitation, University of Portsmouth, Portsmouth PO1 3FX, United Kingdom}

\affiliation[c]{National Institute for Theoretical \& Computational Science, Cape Town 7535, South Africa}
\emailAdd{bikashrdinda@gmail.com}
\emailAdd{roy.maartens@gmail.com}

\abstract{Baryon acoustic oscillation measurements by the Dark Energy Spectroscopic Instrument (Data Release 1) have revealed exciting results that show evidence for dynamical dark energy at $\sim3\sigma$ when combined with cosmic microwave background and type Ia supernova observations. These measurements are based on the $w_0w_a$CDM model of dark energy. The evidence is less in other dark energy models such as the $w$CDM model. In order to avoid imposing a dark energy model, we reconstruct the distance measures and the equation of the state of dark energy independent of any dark energy model and driven only by observational data. Our results show that the model-agnostic (in terms of late-time models) evidence for dynamical dark energy from DESI is not significant. Our analysis also provides model-independent constraints on cosmological parameters such as the Hubble constant and the matter-energy density parameter at present. Although we used CMB distance priors (not full CMB data) from a $\Lambda$CDM early-time model, our results remain largely similar for other cosmological models, provided that these models do not differ significantly from the standard model.}

\begin{document}
\maketitle
\flushbottom

\section{Introduction}
\label{sec-intro}

Data Release 1 (DR1) from the new Dark Energy Spectroscopic Instrument (DESI) survey has produced exciting results from baryonic acoustic oscillation (BAO) measurements that seem to imply a degree of tension with the standard $\Lambda$CDM model of cosmology \citep{DESI:2024mwx}. 
The initial DESI results have been extensively examined and used in follow-up work (e.g.~\cite{Tada:2024znt,Yin:2024hba,Berghaus:2024kra,Shlivko:2024llw,Ramadan:2024kmn,Gialamas:2024lyw,Patel:2024odo,Wang:2024hks,Yang:2024kdo,Escamilla-Rivera:2024sae,Chudaykin:2024gol,Luongo:2024fww,Cortes:2024lgw,Colgain:2024xqj,Carloni:2024zpl,Wang:2024rjd,Giare:2024smz,Seto:2024cgo,Li:2024qso,Qu:2024lpx,Wang:2024dka,Park:2024jns,DESI:2024aqx,Dinda:2024kjf,DESI:2024kob,Mukherjee:2024ryz,Pogosian:2024ykm,Roy:2024kni,Jia:2024wix,Heckman:2024apk,Notari:2024rti,Lynch:2024hzh,Liu:2024gfy,Orchard:2024bve,Hernandez-Almada:2024ost,Pourojaghi:2024tmw,Mukhopadhayay:2024zam,Ye:2024ywg,Giare:2024gpk}).
The original DESI paper \citep{DESI:2024mwx} favours a phantom behaviour of dark energy ($w<-1$) over a significant redshift range, with a preference for crossing to the non-phantom region at lower redshift. This conclusion arises when the dark energy equation of state in a late-time, spatially flat Friedmann-Lema\^itre-Robertson-Walker (FLRW) model is parametrised as \citep{Chevallier:2000qy,Linder:2002et}
\begin{align}\label{wowa}
    w(a)=w_0+w_a(1-a)\,.
\end{align}
Here $a=(1+z)^{-1}$ is the scale factor, where $z$ is the cosmological redshift. This dark energy model generalises the standard $\Lambda$CDM model ($w_0=-1, w_a=0$), allowing for dynamical (evolving) dark energy at the cost of only 2 parameters.  
In addition, many possible dark energy models can be approximated by the $w_0w_a$CDM model.

However, there are various issues associated with using \eqref{wowa} to constrain dark energy evolution (see \cite{Wolf:2023uno,Gialamas:2024lyw,Patel:2024odo}). Although it facilitates the computation of a wide variety of dark energy evolutions, it is a phenomenological ansatz that is not based on a physical and self-consistent model of dark energy. In particular, there is no obstacle to the phantom regime $w<-1$, which is unphysical in general relativity (and in some modified gravity theories \cite{Ye:2024ywg}), and the speed of sound of dark energy is arbitrary. By contrast, quintessence models of dark energy \citep{Caldwell:1997ii,Zlatev:1998tr,Tsujikawa:2013fta,Dinda:2016ibo,Garcia-Garcia:2019cvr,Yin:2024hba,Tada:2024znt,Berghaus:2024kra,Shlivko:2024llw,Gialamas:2024lyw,Ramadan:2024kmn} are physically self-consistent: there is no phantom regime and the quintessence speed of sound is always $c_s=c$ (speed of light in vacuum), which ensures that dark energy does not cluster and that causality is respected \cite{Ellis:2007ic}. 

A disadvantage of assuming a physical quintessence model is that this rules out other non-standard models which can produce late-time acceleration. Examples are dark energy that is non-gravitationally coupled to dark matter \citep{Amendola:2002bs,Clemson:2011an,Pourtsidou:2013nha,Murgia:2016ccp,DiValentino:2019ffd,Seto:2024cgo,Giare:2024smz,Li:2024qso},  and modified gravity models which produce late-time acceleration from modifications to general relativity \citep{Tsujikawa:2010zza,Clifton:2011jh,Koyama:2015vza,Wang:2024hks,Yang:2024kdo,Escamilla-Rivera:2024sae,Chudaykin:2024gol}.

A model-agnostic approach is to avoid choosing a particular type of model so as to allow the data itself to constrain the evolution of $w$ \citep{Perenon:2022fgw,DESI:2024aqx,Dinda:2024xla}.
This approach can also incorporate non-standard models. In the case that modified gravity drives the late-time acceleration, $w$ becomes an equation of state for the effective dark energy -- i.e. the modified gravitational degree of freedom that mimics dark energy in the flat FLRW background. We limit our analysis to late-time accelerating models -- i.e., models in which the (effective) dark energy does not affect the cosmic microwave background (CMB) or earlier processes in the young universe.

Machine learning provides various model-agnostic methods, including Gaussian Process (GP) regression \citep{williams1995gaussian,GpRasWil,Shafieloo:2012ht,Seikel:2012uu,Haridasu:2018gqm,Mukherjee:2020ytg,Keeley:2020aym,Perenon:2021uom,Dinda:2022vmb,Sabogal:2024qxs}, 
which enable data-driven reconstructions of the trends in the evolution of $w(z)$. In this paper, we apply GP regression to the DESI DR1 BAO data, combined with other data, in order to derive the behaviour of $w(z)$ that is consistent with the data sets and their errors.

The paper is organized as follows. In Section~\ref{sec-basic}, we present the relevant spacetime and dynamical quantities in a flat late-time FLRW universe whose expansion is accelerating. We also express
$w$ in terms of BAO and supernova (SNIa) observables. Section~\ref{sec-data} summarises the relevant observational data, including  DESI DR1 and other BAO data, CMB distance data, and  SNIa data. Appendix~\ref{sec-CMB_dist_priors} gives details of  CMB distance priors. Our results are given in Section~\ref{sec-result}, using the multi-task GP methodology. Both the simple posterior (single-task) and the multi-task GP analysis are summarised in Appendix~\ref{sec-method}, with Appendix~\ref{sec-GPR_short_notations} listing the notation used. Some applications of single-task GP to PantheonPlus data are discussed in Appendix~\ref{sec-pantheon_plus_data_related}. We also address the effect on the results of (1)~different CMB distance priors, in Appendix~\ref{sec-different_cmb_dist_prior},  and (2)~different GP mean functions and hyperparameters, in  Appendix~\ref{sec-gpr_different_means}. Our conclusions are presented in Section~\ref{sec-conclusion}.

\section{Background equations}
\label{sec-basic}

Neglecting the contribution from radiation, in the flat FLRW metric, the Hubble parameter $H$ for the late-time evolution is 
\begin{equation}
\frac{H^2(z)}{H_0^2} = \Omega_{\rm m0}(1+z)^3+(1-\Omega_{\rm m0}) \exp{ \left[ 3 \int_{0}^{z} \frac{1+w(\tilde{z})}{1+\tilde{z}} d\tilde{z} \right] } ,
\label{eq:flat_hubble}
\end{equation}
where $\Omega_{\rm m0}$ is the present value of the matter energy density parameter.
The comoving radial distance and the Hubble radius are
\begin{align}
D_M(z) &= c \int_{0}^{z} \frac{d\tilde{z}}{H(\tilde{z})} ,
\label{eq:flat_DM}\\
D_H(z) &= \frac{c}{H(z)} ,
\label{eq:defn_DH}
\end{align}
so that
\begin{equation}
D'_M(z) = D_H(z) ,
\label{eq:DM_prime}
\end{equation}
where a prime denotes $d/dz$.

From \eqref{eq:flat_hubble}, we find $w$ in terms of $H$ and $H'$:
\begin{eqnarray}
w(z) &=& \frac{2(1+z)H(z)H'(z)-3H^2(z)}{3 \left[ H^2(z)-\Omega_{\rm m0}H_0^2(1+z)^3 \right] } \nonumber\\
&=& -1 + 
\frac{2(1+z)H(z)H'(z)-3\alpha(1+z)^3}{3\big[H^2(z)-\alpha(1+z)^3\big]} ,
\label{eq:w_DE}
\end{eqnarray}
where 
\begin{equation}
\alpha = \Omega_{\rm m0}H_0^2 = 10^4 \, \omega_{\rm m0} ~~~ ({\rm km/s/Mpc})^2.
\label{eq:alpha}
\end{equation}
Using \eqref{eq:defn_DH} in \eqref{eq:w_DE} leads to
\begin{equation}
w(z) = -1 - 
\frac{2(1+z)D'_H(z)+3\beta(1+z)^3\,D^3_H(z)}{3\big[D_H(z)-\beta(1+z)^3\,D^3_H(z)\big]} ,
\label{eq:w_DE_2}
\end{equation}
where 
\begin{equation}
\beta = \frac{\alpha}{c^2}
\approx 1.1126 \times 10^{-7} \omega_{\rm m0}~~~{\rm Mpc}^{-2},  
\label{eq:beta}
\end{equation}
on using
${H_0} \approx {ch}/(2998~{\rm Mpc})$.
From \eqref{eq:DM_prime} and \eqref{eq:w_DE_2}, we find an alternative expression 
\begin{equation}
w(z) = -1 -  \frac{2(1+z)D''_M(z)+3\beta(1+z)^3\,D^{\prime\, 3}_M(z)}{3\big[D'_M(z)-\beta(1+z)^3\,D^{\prime\, 3}_M(z)\big]} .
\label{eq:w_DE_3}
\end{equation}

The anisotropic BAO measurements have data corresponding to the two observables
\begin{eqnarray}
\tilde{D}_M(z) = \frac{D_M(z)}{r_d} ,\qquad
\tilde{D}_H(z)= \frac{D_H(z)}{r_d} ,
\label{eq:DM_tilde} 
\end{eqnarray}
where $r_d$ is the sound horizon at the baryon drag epoch. Using \eqref{eq:DM_prime} and \eqref{eq:DM_tilde}, we can rewrite \eqref{eq:w_DE_3} as
\begin{equation}
w(z) = -1 -  \frac{2(1+z)\tilde{D}'_H(z)+3\gamma(1+z)^3\,\tilde D^{3}_H(z)}{3\big[\tilde{D}_H(z)-\gamma(1+z)^3\,\tilde D^{3}_H(z)\big]} ,
\label{eq:w_DE_4}
\end{equation}
where 
\begin{equation}
\gamma = \beta r_d^2 = \frac{\alpha r_d^2}{c^2} = \frac{\Omega_{\rm m0} H_0^2 r_d^2}{c^2} \approx 1.1126 \times 10^{-7} ~ \omega_{\rm m0} \left( \frac{r_d}{{\rm Mpc}} \right)^2 .
\label{eq:defn_gamma}
\end{equation}
Note that  $\gamma$ is a dimensionless constant, unlike $\alpha$ and $\beta$.

SNIa observations use the distance modulus $\mu_B$ as an observable, which is related to the luminosity distance $d_L$ via
\begin{eqnarray}
\mu_B(z) = m_B(z)-M_B
&=& \frac{5}{\ln 10} \ln \left[ \frac{d_L(z)}{{\rm Mpc}} \right] + 25 \quad\mbox{where}~~
d_L=(1+z)D_M\,.
\label{eq:SN_dist_modulus}
\end{eqnarray}
Here  $m_B$ and $M_B$ are the observed and absolute peak magnitudes of the supernova. Then 
\begin{equation}
(1+z)D_M(z) = \exp \Big\{b\left[\mu_B(z)-25\right]\Big\} ~~{\rm Mpc}
\quad
\mbox{where}\quad b = \frac{\ln 10}{5} .
\label{eq:lum_dist_SN}
\end{equation}
It is convenient to normalise $D_M$ as
\begin{equation}
d_M(z) ={\rm e}^{b(20+M_B)}\, D_M(z) 
=\frac{ \exp \Big\{b\left[m_B(z)-5\right]\Big\}}{1+z}  ~~{\rm Mpc}.
\label{eq:SN_DM_main}
\end{equation}
Note that the constant ${\rm e}^{b(20+M_B)}$ is defined in such a way that its magnitude is of $O(1)$. Using \eqref{eq:SN_DM_main} we can rewrite \eqref{eq:w_DE_3} as
\begin{align}
w(z) = -1 -  \frac{2(1+z)d''_M(z)+3\delta(1+z)^3 d^{\prime\, 3}_M(z)}{3\big[d'_M(z)-\delta(1+z)^3\,d^{\prime\, 3}_M(z)\big]} ,
\label{eq:w_DE_6}
\end{align}
{where}
 \begin{align}
\delta = 
\beta \, {\rm e}^{-2b(20+M_B)} \approx 1.1126 \times 10^{-7} \left( \frac{\omega_{\rm m0}}{{\rm Mpc}^2} \right) {\rm e}^{-2b(20+M_B)} .
\label{eq:defn_delta}
\end{align}

\section{Observational data}
\label{sec-data}

\subsection{DESI and other BAO data}
\label{sec-bao}

We mainly focus on five different DESI DR1 measurements of anisotropic BAO observables $\tilde{D}_M$ and $\tilde{D}_H$ at five different effective redshifts,  $z_{\rm eff}$. Two other measurements which correspond to the isotropic BAO observable are excluded, because of the difficulty to include these in a model-independent analysis. The mean, standard deviation, and correlation values of $\tilde{D}_M$ and $\tilde{D}_H$ at the five  $z_{\rm eff}$ are shown in Table~\ref{table:DESI_DR1_2024}.
In Table~\ref{table:DESI_DR1_2024} and elsewhere $\Delta D$ indicates the standard deviation (1$\sigma$ confidence level) of $D$ and  $r$ is the correlation: \begin{equation}
r[D_1,D_2] = \frac{{\rm Cov}[D_1,D_2]}{\Delta D_1 \Delta D_2} .
\label{eq:norm_cov}
\end{equation}
We also consider other non-DESI anisotropic BAO data at another five different effective redshifts, as listed in Table~\ref{table:Other_BAO}.

\begin{table*}
\begin{center}
\begin{tabular}{|c|c|c|c|c|c|c|c|}
\hline &&&&&& \\
 & tracer (DESI DR1) &$z_{\rm eff}$ & $\tilde{D}_M \pm \Delta \tilde{D}_M$ & $\tilde{D}_H \pm \Delta \tilde{D}_H$ & $r[\tilde{D}_M,\tilde{D}_H]$ & Refs. \\ &&&&&& \\
\hline
1 & LRG & 0.510 & 13.62 $\pm$ 0.25 & 20.98 $\pm$ 0.61 & -0.445 & \citep{DESI:2022gle} \\
2 & LRG & 0.706 & 16.85 $\pm$ 0.32 & 20.08 $\pm$ 0.60 & -0.420 & \citep{DESI:2022gle} \\
3 & LRG+ELG & 0.930 & 21.71 $\pm$ 0.28 & 17.88 $\pm$ 0.35 & -0.389 & \citep{DESI:2024mwx} \\
4 & ELG & 1.317 & 27.79 $\pm$ 0.69 & 13.82 $\pm$ 0.42 & -0.444 & \citep{Raichoor:2022jab} \\
5 & Ly-$\alpha$ & 2.330 & 39.71 $\pm$ 0.94 & 8.52 $\pm$ 0.17 & -0.477 & \citep{DESI:2024mwx} \\
\hline
\end{tabular}
\end{center}
\caption{
Measurements of anisotropic BAO observables $\tilde{D}_M$ and $\tilde{D}_H$ and correlations between them at five different effective redshifts from DESI DR1 BAO data \citep{DESI:2024mwx}.
}
\label{table:DESI_DR1_2024}
\end{table*}

\begin{table*}
\begin{center}
\begin{tabular}{|c|c|c|c|c|c|c|}
\hline  &&&&&& \\
 & tracer (non-DESI BAO) &$z_{\rm eff}$ & $\tilde{D}_M \pm \Delta \tilde{D}_M$ & $\tilde{D}_H \pm \Delta \tilde{D}_H$ & $r[\tilde{D}_M,\tilde{D}_H]$ & Refs. \\ &&&&&& \\
\hline
1 & LRG (BOSS DR12) & 0.38 & 10.234 $\pm$ 0.151 & 24.981 $\pm$ 0.582 & -0.228 & \citep{BOSS:2016wmc} \\
2 & LRG (BOSS DR12) & 0.51 & 13.366 $\pm$ 0.179 & 22.317 $\pm$ 0.482 & -0.233 & \citep{BOSS:2016wmc} \\
3 & LRG (eBOSS DR16) & 0.698 & 17.858 $\pm$ 0.302 & 19.326 $\pm$ 0.469 & -0.239 & \citep{eBOSS:2020yzd} \\
4 & QSO (eBOSS DR16) & 1.48 & 30.688 $\pm$ 0.789 & 13.261 $\pm$ 0.469 & 0.039 & \citep{eBOSS:2020gbb} \\
5 & Ly-$\alpha$ QSO (eBOSS DR16) & 2.334 & 37.5 $\pm$ 1.2 & 8.99 $\pm$ 0.19 & -0.45 & \citep{eBOSS:2020tmo} \\
\hline
\end{tabular}
\end{center}
\caption{
Measurements of anisotropic BAO observables $\tilde{D}_M$ and $\tilde{D}_H$ and correlations between them at five different effective redshifts from non-DESI BAO data \citep{eBOSS:2020yzd}.
}
\label{table:Other_BAO}
\end{table*}

\subsection{CMB distance priors}
\label{sec-cmb}

The late-time cosmological data analysis can be done with three CMB parameters instead of the full CMB likelihood \citep{Hu:1995en,Chen:2018dbv,Zhai:2019nad,Zhai:2018vmm,Dinda:2024kjf}:
\begin{align}
\label{eq:CMB_R}
R &= \frac{1}{c}\big({\Omega_{\rm m0}H_0^2}\big)^{1/2} D_M(z_*), \\
\label{eq:CMB_lA}
l_A &= \frac{\pi D_M(z_*)}{r_s(z_*)}, \\
\label{eq:CMB_Ob0_h2}
\omega_{\rm b0} &= \Omega_{\rm b0} h^2 ,
\end{align}
where $z_*$ is the photon decoupling redshift, $R$   is the CMB shift parameter, $l_A$ is the acoustic length scale at $z_*$, $\Omega_{\rm b0}$ is the present value of the baryonic density parameter, and $r_s$ is the sound horizon. Because  $D_M(z_*)$ appears in both  $R$ and $l_A$, we can not individually use these two parameters in a model-independent data analysis. For this purpose, the ratio $\tilde{R}$ of these two parameters is used.
Then the  mean and standard deviation of the two CMB parameters, from Planck 2018 results for TT, TE, EE+lowE+lensing (considering the standard model, i.e. the base $\Lambda$CDM model) are \citep{Planck:2018vyg}
\begin{align}
\label{eq:R_by_lA}
\tilde{R} &= \frac{R}{l_A} = \frac{{\omega_{\rm m0}^{1/2}}}{3000\pi} \,\frac{r_s(z_*)}{\rm Mpc} = 0.005797 \pm 0.000013 \, , \\
\label{eq:CMB_Ob0_h2_2}
\omega_{\rm b0} &= \Omega_{\rm b0} h^2 = 0.02237 \pm 0.00015 \, , \\
\label{eq:CMB_r_Rt_wb0}
r[\tilde{R},\omega_{\rm b0}] &= -0.61\, .
\end{align}

In order to find the above values, we use the Planck 2018 data archive for the base $\Lambda$CDM model, in the chain {\sf base$_{-}$plikHM$_{-}$TTTEEE$_{-}$lowl$_{-}$lowE$_{-}$lensing.}

In standard early-time physics approximations, the sound horizon at photon decoupling $r_s(z_*)$ becomes a function of only $\omega_{\rm m0}$ and $\omega_{\rm b0}$. For details, see Appendix~\ref{sec-CMB_dist_priors}. In this case,  $\tilde{R}$ is also a function of only $\omega_{\rm m0}$ and $\omega_{\rm b0}$. Using this assumption and solving \eqref{eq:R_by_lA}--\eqref{eq:CMB_r_Rt_wb0}, we find
\begin{align}
\label{eq:wm0_PL18_standard}
\omega_{\rm m0} &= 0.1430 \pm 0.0011 ~ , \\
\label{eq:rho_wb0_wm0_PL18_standard}
 r[\omega_{\rm b0},\omega_{\rm m0}]  &= -0.42 ~ .
\end{align}
In addition, the sound horizon $r_d=r_s(z_d)$ at the baryon drag epoch $z_d$ is also a function of only $\omega_{\rm m0}$ and $\omega_{\rm b0}$. Using \eqref{eq:CMB_Ob0_h2_2}, \eqref{eq:wm0_PL18_standard}, and \eqref{eq:rho_wb0_wm0_PL18_standard}, we have
\begin{equation}
{r_d} = 146.995 \pm 0.264 ~~{{\rm Mpc}} ,
\label{eq:rd_PL18_standard}
\end{equation}
and the normalised covariance between $\omega_{\rm m0}$ and $r_d$ is
\begin{equation}
r[\omega_{\rm m0},r_d]= -0.90\,.
\label{eq:rho_wm0_rd_PL18_standard}
\end{equation}
To cross-check these values, we use a fitting formula from \citep{DESI:2024mwx} [their eq. (2.5)]: 
\begin{equation}
\frac{r_d}{{\rm Mpc}} = 147.05 \left(\frac{\omega_{\rm m0}}{0.1432}\right)^{-0.23} \left(\frac{\omega_{\rm b0}}{0.02236}\right)^{-0.13} .
\label{eq:desi_fitting_rd}
\end{equation}
Here we used the standard value $3.04$ for the effective number of relativistic species. From \eqref{eq:CMB_Ob0_h2_2}, \eqref{eq:wm0_PL18_standard}, and \eqref{eq:rho_wb0_wm0_PL18_standard} 
we obtain
\begin{equation}
{r_d} = 147.089 \pm 0.237~~{{\rm Mpc}} ,
\label{eq:desi_fitting_rd_value}
\end{equation}
which is consistent with \eqref{eq:rd_PL18_standard}. We use \eqref{eq:rd_PL18_standard} for our analysis.

From the above constraints we find constraints on $\alpha$, $\beta$ and $\gamma$:
\begin{align}
\label{eq:alpha_PL18_standard}
\alpha & \approx 1430\pm 11 ~~ \left({\rm km/s/Mpc}\right)^2\,, \\
\label{eq:beta_PL18_standard}
\beta & \approx (159.1\pm 1.2) \times 10^{-10} ~~ {\rm Mpc}^{-2} \, , \\
\label{eq:gamma_PL18_standard}
\gamma & \approx (343.8\pm 1.6) \times 10^{-6} \,.
\end{align}
These results are our CMB data.

Throughout our analysis, we consider priors on these parameters as an alternative to the full CMB likelihood. The prior values of $\alpha$, $\beta$, and $\gamma$, obtained using the base $\Lambda$CDM model from compressed CMB data, are a good approximation to, and a simplified version of, the full CMB data, provided that the true model of the Universe does not deviate significantly from $\Lambda$CDM \cite{Zhai:2019nad,Chen:2018dbv}. We also note that the values of these parameters would be slightly different for other cosmological models. Still, the main result, i.e. the reconstruction of the equation of state of dark energy, does not differ significantly when we consider other models beyond $\Lambda$CDM. We emphasize that the use of the base $\Lambda$CDM model does not significantly bias our result toward a $\Lambda$CDM value -- as shown explicitly in Appendix~\ref{sec-different_cmb_dist_prior}.

\subsection{Supernova data}
\label{sec-supernova}

We consider the Pantheon+ sample -- which we denote PantheonPlus -- for SNIa data on the apparent magnitude $m_B$. This consists of 1701 light curves for 1550 spectroscopically confirmed SNIa in the redshift range $0.00122 \leq z \leq 2.26137$ \citep{Scolnic:2021amr}.  We exclude  111 light curves at low redshifts,  $z<0.01$, in order to avoid strong peculiar velocity dependence \citep{Brout:2022vxf}. In our analysis, we define the sub-sample
\begin{align}\label{pplus}
\mbox{PantheonPlus:}~~ 1590~ \mbox{light curves in}~  0.01016 \leq z \leq 2.26137\,.
\end{align}
We consider the full statistical and systematic covariance for this data, which can be found at \url{https://github.com/PantheonPlusSH0ES/DataRelease}.


\section{Results}
\label{sec-result}

\begin{figure*}
\centering
\includegraphics[height=170pt,width=0.49\textwidth]{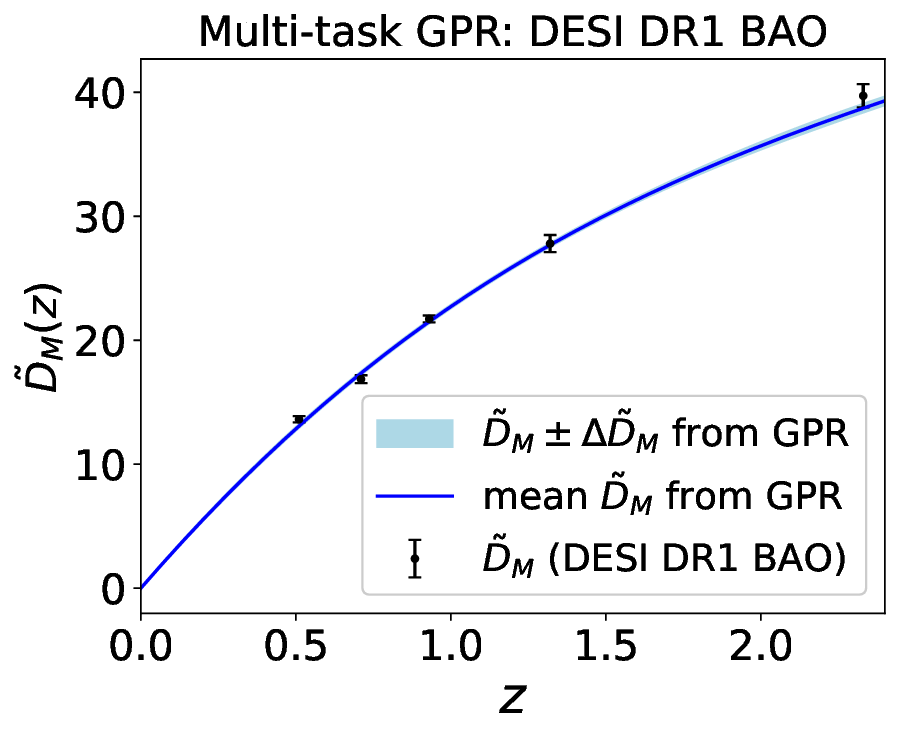}
\includegraphics[height=170pt,width=0.49\textwidth]{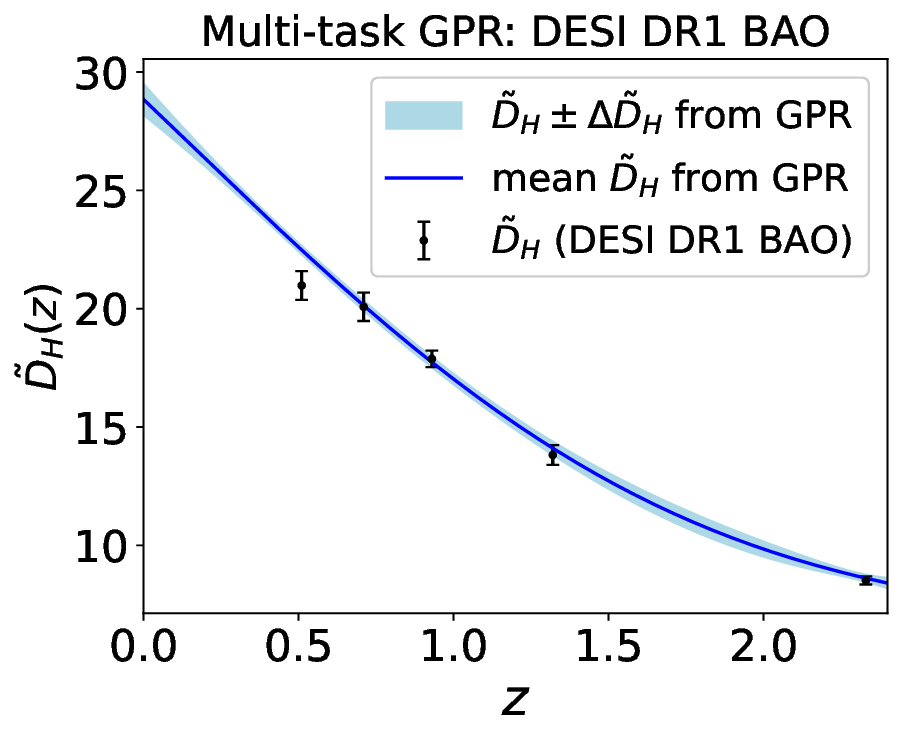} \\
\includegraphics[height=170pt,width=0.55\textwidth]{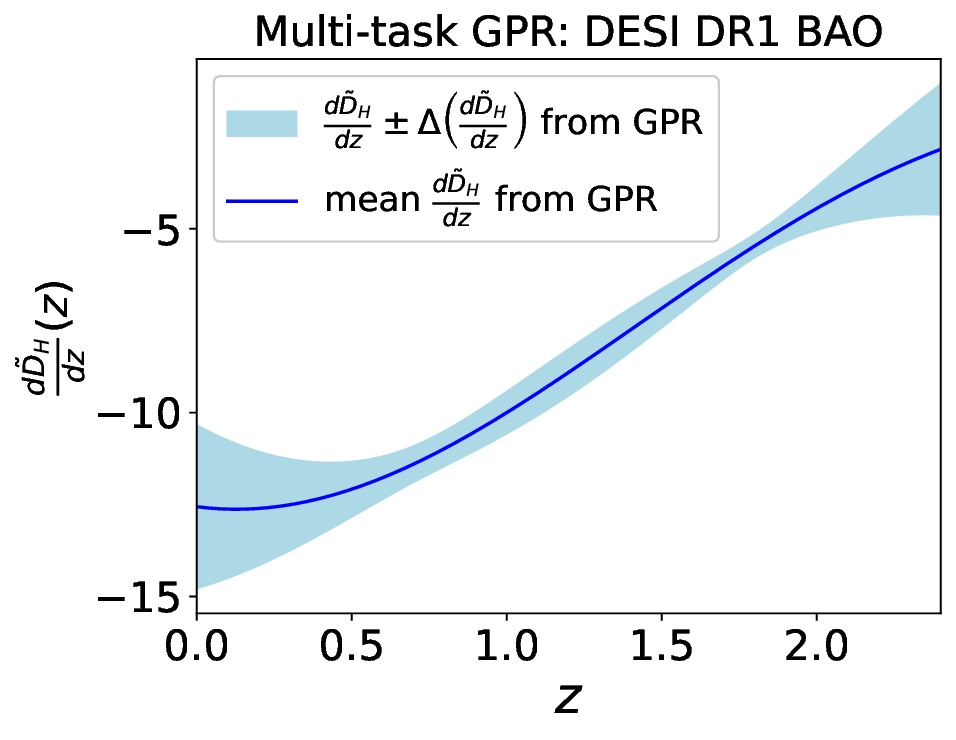}
\caption{
\label{fig:DESI_GPR_predictions}
Reconstruction of $\tilde{D}_M$ (top left), $\tilde{D}'_M=\tilde{D}_H$ (top right), and $\tilde{D}''_M=\tilde{D}'_H$ (bottom) from DESI DR1 BAO data (Table~\ref{table:DESI_DR1_2024}), using multi-task Gaussian Process regression. Blue regions are the reconstructed 1$\sigma$ confidence intervals. Blue solid lines are the best-fit values. In the top panels, black error bars correspond to the DESI DR1 BAO data.
}
\end{figure*}

In this section, we apply the GP regression (GPR) methodology, which is described in Appendix~\ref{sec-method}, to the datasets, in order to reconstruct the distances and then the dark energy (or effective dark energy) equation of state $w$.

\subsection{DESI DR1 and non-DESI BAO}

We start with  the DESI DR1 BAO data of $\tilde{D}_M$ and $\tilde{D}_H$ ($=\tilde{D}'_M$). Because $\tilde{D}_M$ and $\tilde{D}_H$ are correlated and the correlations are large, we cannot simply use the simple single-task GPR, discussed in Section~\ref{sec-single_task_GPR}, individually to $\tilde{D}_M$ and $\tilde{D}_H$. Instead, we need the multi-task GPR, which deals with the covariance between a variable and its derivative (see Section~\ref{sec-multi_task_GPR}).  In detail:
\begin{itemize}

\item
$X_1=X_2=\{z_{\rm eff}\}$ is the vector of all effective redshift points.

\item
$Y_1=\{\tilde{D}_M\}$ is the vector of all mean values of $\tilde{D}_M$.

\item
$Y'_2=\{\tilde{D}_H\}=Y_1'$ (since $X_2=X_1$) is the vector of all mean values of $\tilde{D}'_M=\tilde{D}_H$.

\item
$C_{11}=\big\{{\rm Cov}[\tilde{D}_M,\tilde{D}_M]\big\}$ is the matrix of all self-covariances of $\tilde{D}_M$.

\item
$C_{12}^{(0,1)}=\big\{{\rm Cov}[\tilde{D}_M,\tilde{D}_H]\big\}$ is matrix   of all cross covariances between $\tilde{D}_M$ and $\tilde{D}_H$.

\item
$C_{21}^{(1,0)}=\big\{{\rm Cov}[\tilde{D}_H,\tilde{D}_M]\big\}=\big( \big\{{\rm Cov}[\tilde{D}_M,\tilde{D}_H]\big\}\big)^{\rm T} = \big(C_{12}^{(0,1)}\big)^{\rm T}$ is the matrix  of all cross covariances between $\tilde{D}_H$ and $\tilde{D}_M$.

\item
$C_{22}^{(1,1)}=\big\{{\rm Cov}[\tilde{D}_H,\tilde{D}_H]\big\}$ is the matrix of all self-covariances of $\tilde{D}_H$.

\end{itemize}

We consider the most popular kernel covariance function, the squared-exponential, in which the kernel covariance between $x_1$ and $x_2$ is
\begin{equation}
k (x_1,x_2) = \sigma_f^2 \,\exp\bigg[-\frac{\left(x_1-x_2\right)^2}{l^2}\bigg] ,
\label{eq:kernel_SQ}
\end{equation}
where $\sigma_f$ and $l$ are the two kernel hyperparameters. Note that one has to choose a kernel that is differentiable at least up to the order at which we want the prediction from GPR. The squared exponential kernel is infinitely differentiable. For this reason, it can be used in any GPR task for the prediction of any order of the function.

We also need a mean function to perform a GPR task. We choose the zero mean function to avoid any cosmological model-dependent bias
\begin{equation}
\mu (x) = 0 .
\label{eq:mean_zero}
\end{equation}
Similarly to the kernel, the mean function needs to be differentiable up to the order we want the prediction. The zero mean function is infinitely differentiable and at any order, it is always the zero mean function.

Although the zero mean function is a popular choice to avoid any cosmological model-dependent bias,  if the quality of data is not good enough, then the GP reconstruction will have a bias towards the zero values. The same applies to any other mean function assumed \cite{Hwang:2022hla}. If the quality of data is good, then the reconstructed results of GP will not be significantly biased towards a chosen mean function. Fortunately, the data combinations we consider here are good enough to put reasonable constraints on the expansion history of the Universe and the zero mean function is a safe choice in our analysis. We show this in detail in Appendix~\ref{sec-gpr_different_means}.

We use the above information in \eqref{eq:tilde_Y}--\eqref{eq:double_m_log_prob_2} in order to compute the negative log marginal likelihood. This is a function of the kernel hyperparameters, including mean function parameters and other parameters (if present). We then minimise the negative of this log marginal likelihood to find the best-fit values of all the parameters.

These best-fit parameter values are used in the prediction of the functions $F_*$ (here $\tilde{D}_M$), $F'_*$ (here $\tilde{D}'_M=\tilde{D}_H$), and $F''_*$ (here $\tilde{D}''_M=\tilde{D}'_H$) at target redshift points. We want predictions for smooth functions of redshift $z$. To do so, we need to consider a large number of target points, and we use
\begin{align}
X_* =~\text{a vector of 1000 equispaced redshift points in}~0\leq z \leq 2.4\,.   
\end{align}
Finally, we obtain the mean values of $\tilde{D}_M$ (corresponding to $\bar{F}_*$) using \eqref{eq:predict_f_multi} and the corresponding 1$\sigma$ confidence region from the prediction of the square root of self-covariances (variances at each target point) using  \eqref{eq:predict_V_f_f_multi}. We plot these values in the top left panel of Fig.~\ref{fig:DESI_GPR_predictions}. The mean function is represented by the solid blue line and the 1$\sigma$ region is represented by the blue shading. To compare the predicted outcome of the GPR with the actual DESI DR1 BAO data, we also plot the $\tilde{D}_M$ data with black error bars. 

We see that the predictions are quite consistent with the data. Similarly, we get the mean and standard deviation values of $\tilde{D}'_M=\tilde{D}_H$ (corresponding to $\bar{F}'_*$) using  \eqref{eq:predict_fp_multi} and~\eqref{eq:predict_V_fp_fp_multi}, shown in the top right panel of Fig.~\ref{fig:DESI_GPR_predictions}. We compare this prediction with the  DESI DR1 BAO data by plotting the $\tilde{D}_H$ data with black error bars. Here the predictions are also consistent with the data. 

Predictions for the mean and standard deviation of $\tilde{D}''_M=\tilde{D}'_H$ (corresponding to $\bar{F}''_*$), using  \eqref{eq:predict_fpp_multi} and~\eqref{eq:predict_V_fpp_fpp_multi}, are displayed in the bottom panel of Fig.~\ref{fig:DESI_GPR_predictions}. Note that for this second-order derivative prediction, there is no data for comparison with the predicted outcomes.

\begin{figure*}
\centering
\includegraphics[height=170pt,width=0.49\textwidth]{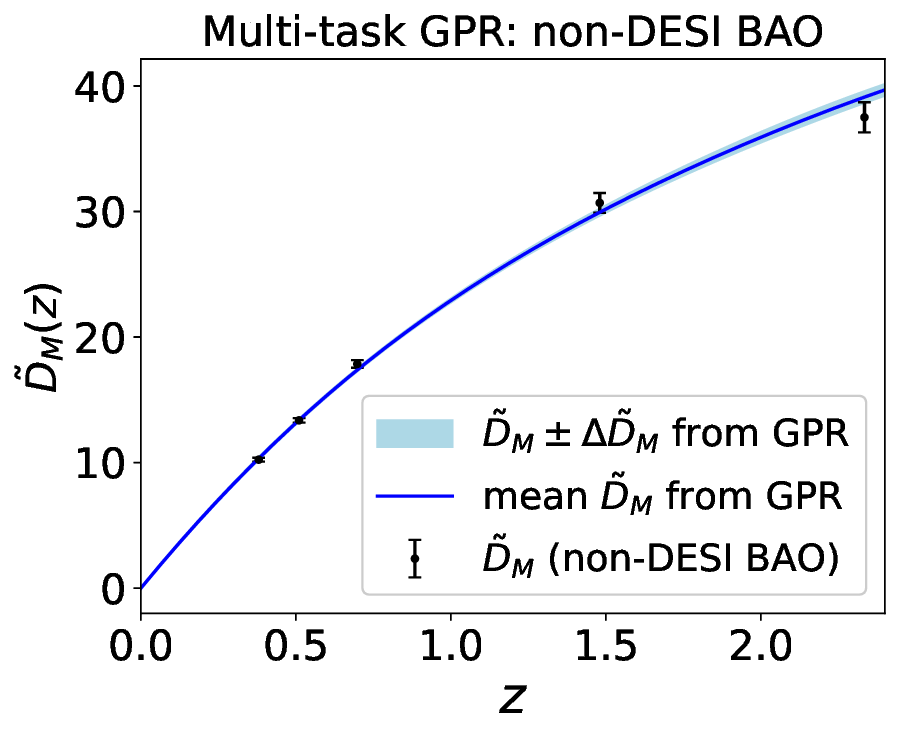}
\includegraphics[height=170pt,width=0.49\textwidth]{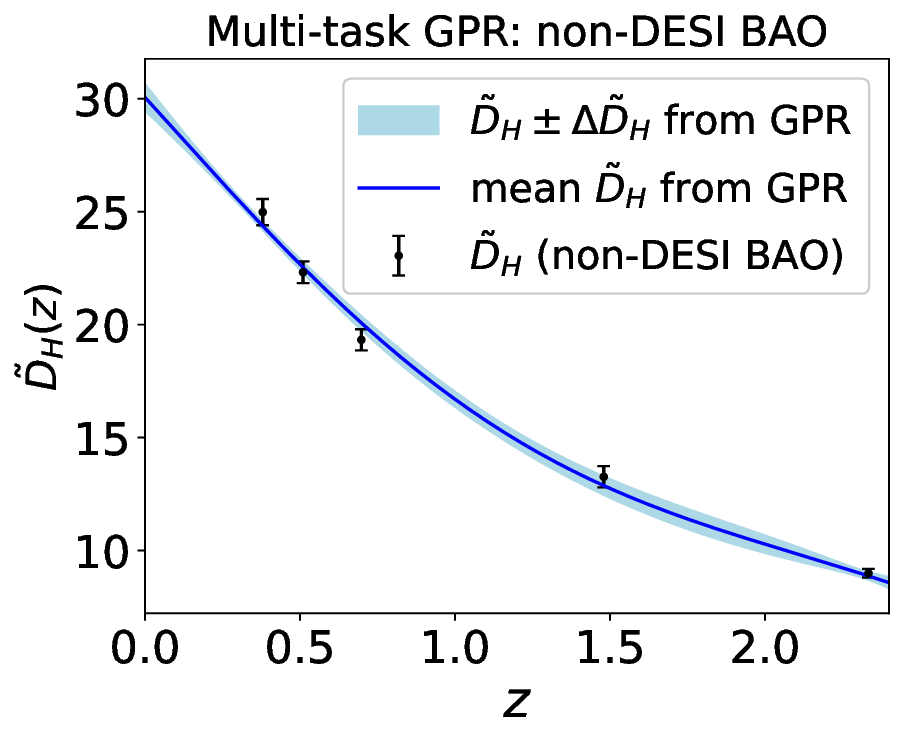} \\
\includegraphics[height=170pt,width=0.55\textwidth]{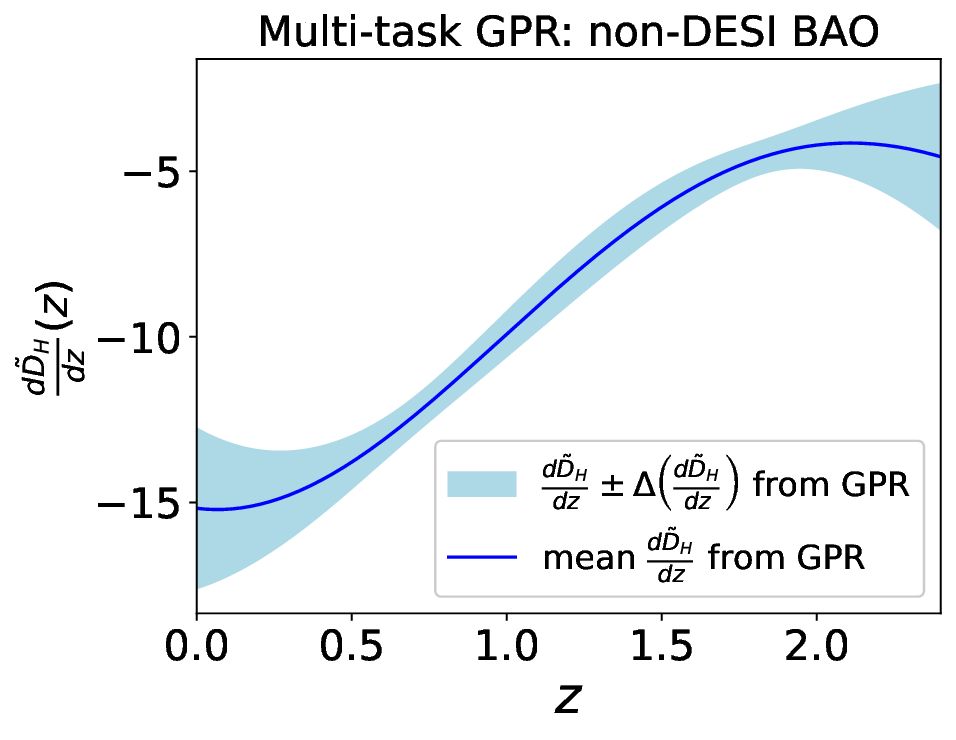}
\caption{
\label{fig:non_DESI_GPR_predictions}
As in Fig.~\ref{fig:DESI_GPR_predictions}, for
non-DESI BAO data  (Table~\ref{table:Other_BAO}).
}
\end{figure*}

For the non-DESI BAO, we follow the same procedure to compute the predicted mean functions of $\tilde{D}_M$, $\tilde{D}'_M=\tilde{D}_H$, and $\tilde{D}''_M=\tilde{D}'_H$ and the corresponding 1$\sigma$ regions. These are shown in  Fig.~\ref{fig:non_DESI_GPR_predictions} with the same color codes. The non-DESI BAO data are shown as black points with error bars. We find again that the predicted values are consistent with the data.

\begin{figure*}
\centering
\includegraphics[width=0.7\textwidth]{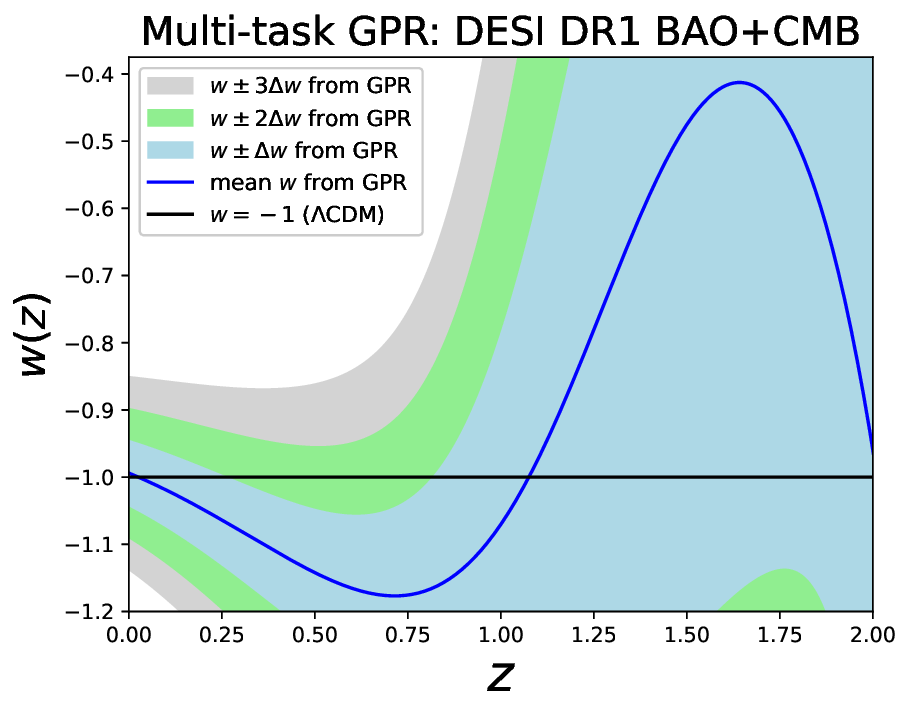}
\caption{
\label{fig:CMB_DESI_w_reconstruction}
Reconstructed redshift evolution of equation of state of (effective) dark energy ($w$), obtained via \eqref{eq:w_DE_4} from the reconstructed $\tilde{D}'_M$ and $\tilde{D}''_M$ (from DESI DR1 BAO with multi-task GPR),  with the constraints on $\gamma$ obtained from \eqref{eq:gamma_PL18_standard} (Planck 2018 CMB). Blue, green, and grey shadings are the  1$\sigma$, 2$\sigma$, and 3$\sigma$ confidence intervals of $w$. The blue line is the best-fit value of $w$ and the black line is $w=-1$  ($\Lambda$CDM).
}
\end{figure*}

\begin{figure*}
\centering
\includegraphics[width=0.7\textwidth]{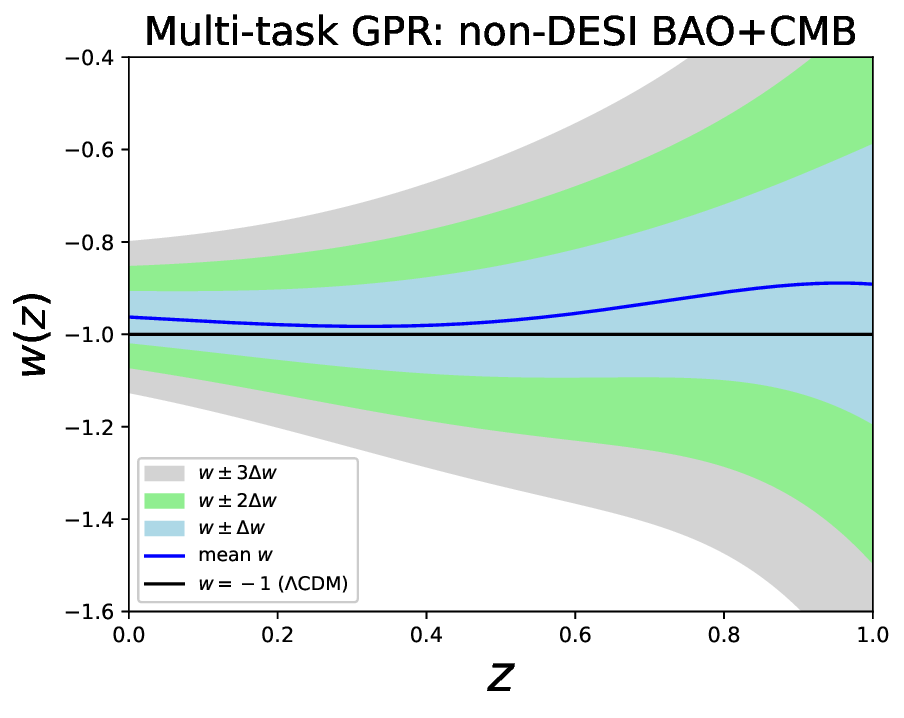}
\caption{
\label{fig:CMB_non_DESI_w_reconstruction}
As in Fig.~\ref{fig:CMB_DESI_w_reconstruction}, for CMB+non-DESI BAO (multi-task GPR).
}
\end{figure*}

\subsection{CMB+DESI DR1 BAO and CMB+non-DESI BAO}

Next, from the reconstructed values of $\tilde{D}'_M$, and $\tilde{D}''_M$, we compute the redshift evolution of $w$ using  \eqref{eq:w_DE_4}. Note that in order to compute $w$, we also need the value of $\gamma$ given in \eqref{eq:gamma_PL18_standard}, which we find from Planck 2018 CMB results (considering standard early time physics) for TT, TE, EE+lowE+lensing. Thus computation of $w$ requires a combination of DESI DR1 BAO data and CMB data, denoted as CMB+DESI DR1 BAO. Also note that when we compute the errors in $w$ using the propagation of uncertainty, we need values of the cross covariances between $\tilde{D}'_M$ and $\tilde{D}''_M$ along with their self-covariances (variances). We compute these cross covariances using  \eqref{eq:predict_C_fp_fpp_multi}. The reconstructed mean function of $w$ (blue line) and the 1$\sigma$ (blue), 2$\sigma$ (green + blue) and 3$\sigma$ (grey + green + blue) confidence regions are displayed in Fig.~\ref{fig:CMB_DESI_w_reconstruction}.

It is evident that for the redshift range  $0\lesssim z\lesssim 1$, the best-fit equation of state is phantom and the $\Lambda$CDM model is  $\sim 1-1.5\sigma$ away. Around $z=1.05$, there is a phantom crossing and after that  $w$ is non-phantom. However, this phantom crossing is not that significant, because the $\Lambda$CDM model is within the 1$\sigma$ limit at $z\gtrsim 0.8$.

For the CMB+non-DESI BAO data, we follow the same procedure to compute $w$ and the associated errors. The results are shown in Fig.~\ref{fig:CMB_non_DESI_w_reconstruction} with the same color codes as in Fig.~\ref{fig:CMB_DESI_w_reconstruction}. In the non-DESI case, the $\Lambda$CDM model is well within the 1$\sigma$ region for the entire redshift range.\footnote{Note that our results are based on the BAO data, which may change if either the primordial or late-time model or both are far from the fiducial model used to obtain the BAO data \cite{Heinesen:2019phg,Bernal:2020vbb,Pan:2023zgb,PhysRevD.107.123506,PhysRevD.101.083517}.}

\begin{figure*}
\centering
\includegraphics[height=170pt,width=0.46\textwidth]{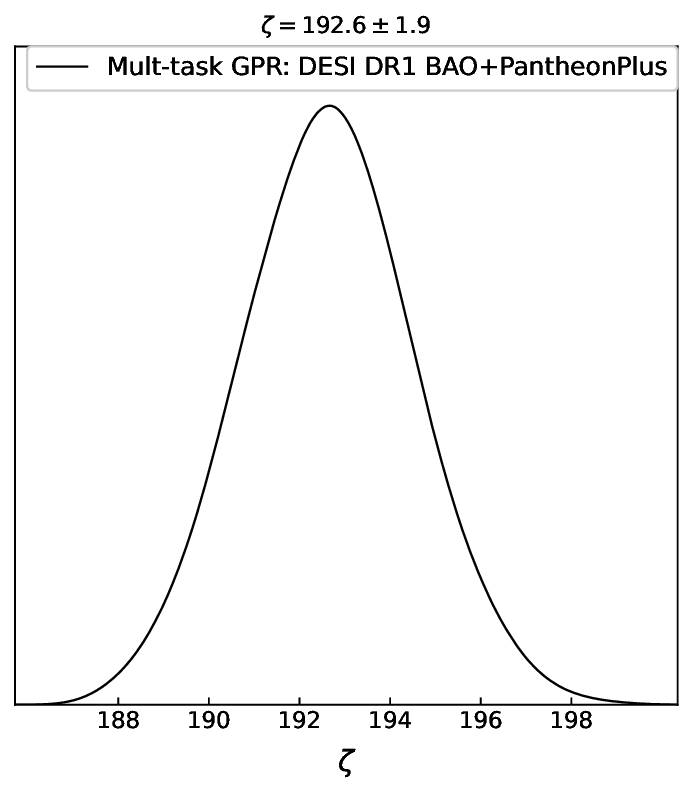}
\includegraphics[height=170pt,width=0.52\textwidth]{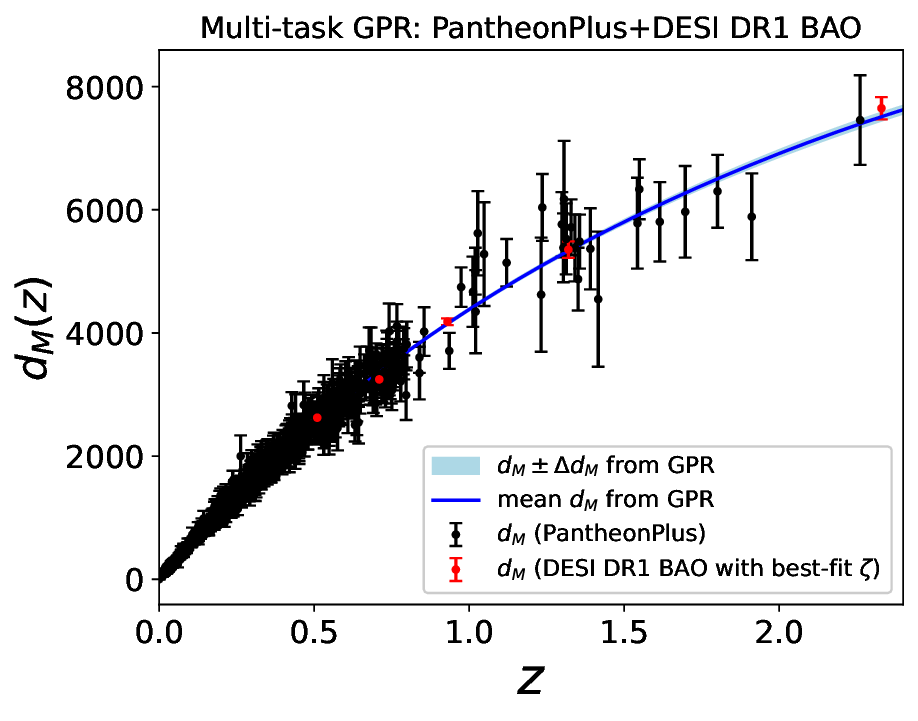} \\
\includegraphics[height=170pt,width=0.49\textwidth]{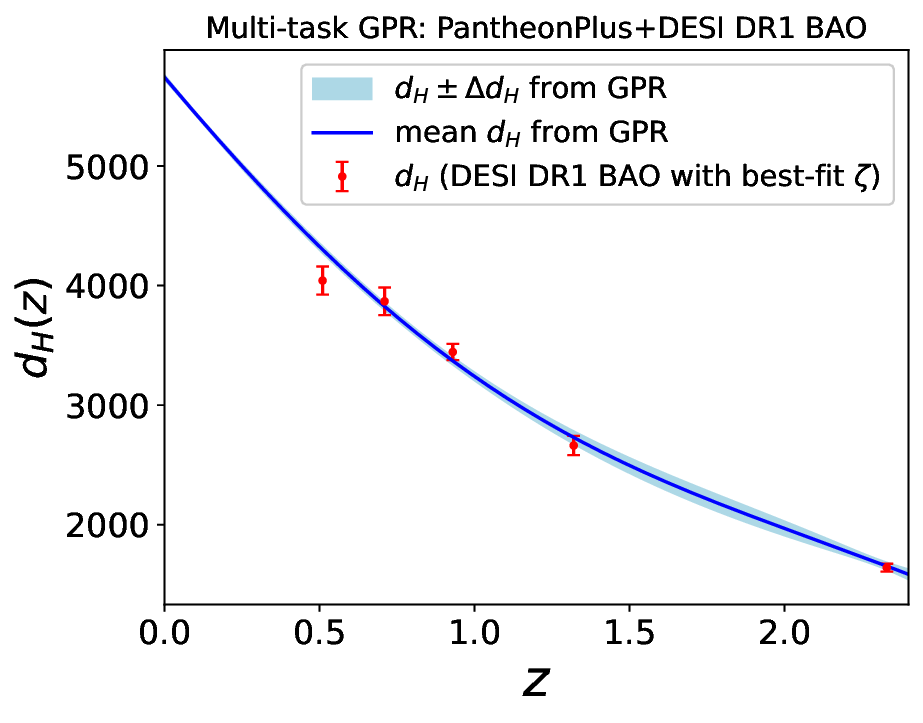}
\includegraphics[height=170pt,width=0.49\textwidth]{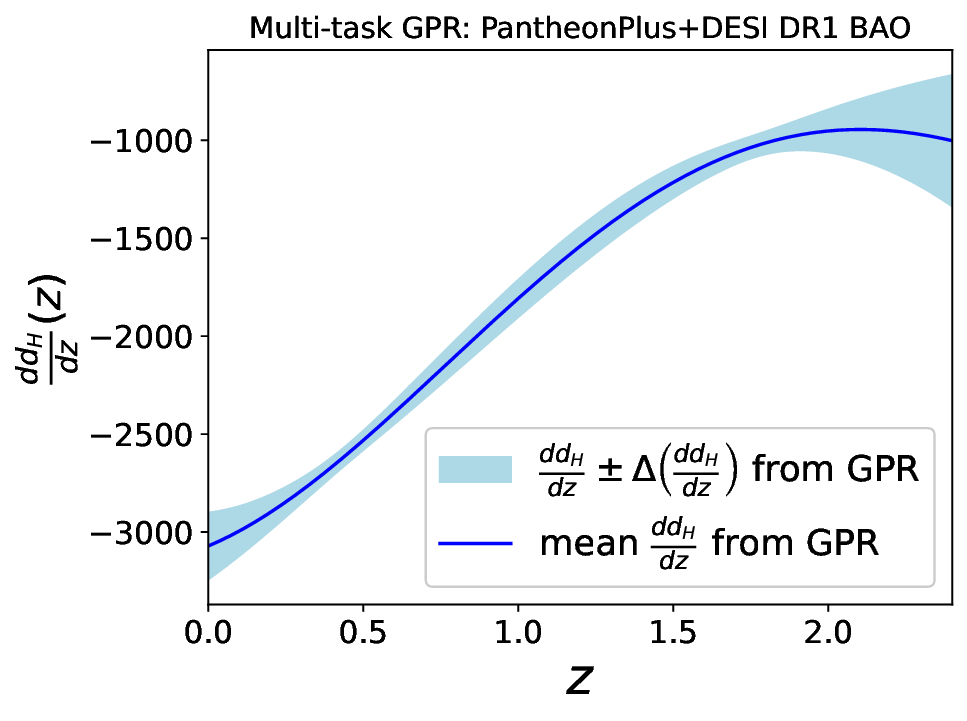}
\caption{
\label{fig:CMB_DESI_SNPP_1}
Multi-task GPR analysis of the combination DESI DR1 BAO+PantheonPlus.
Marginalised probability distribution of  $\zeta$ parameter (top left). Black error bars (top right) show $d_M \pm \Delta d_M$ from observed $m_B$ of PantheonPlus data, using  \eqref{eq:SN_DM_main} (as in Fig.~\ref{fig:CMB_SNPP_MB_w}). Red error bars show $d_M \pm \Delta d_M$ for DESI DR1 BAO from $\tilde{D}_M$ (top right) or $d_H \pm \Delta d_H$ from $\tilde{D}_H$  (bottom left), using the constraint on $\zeta$. Blue solid lines and shadings in the top right, bottom left, and bottom right panels correspond to reconstructed $d_M$, $d'_M=d_H$, $d''_M=d'_H$, and associated 1$\sigma$ confidence regions.
}
\end{figure*}

\subsection{DESI DR1 BAO+PantheonPlus}

Next, we combine DESI DR1 data with PantheonPlus data. We do not consider any $M_B$ priors, since the combination of these datasets can in principle constrain the $M_B$ parameter using the CMB constraint on $r_d$. This can be seen through the connection between variables $d_M$ and $\tilde{D}_M$: from \eqref{eq:DM_tilde} and~\eqref{eq:SN_DM_main}, we find 
\begin{align}
d_M (z) &= \zeta\, \tilde{D}_M(z) ,
\label{eq:dM_DMtilde_relation}
\\
\zeta &= r_d\, {\rm e}^{b(20+M_B)} 
\label{eq:zeta}
\end{align}
Similarly, using \eqref{eq:DM_prime} and~\eqref{eq:dM_DMtilde_relation}, we get
\begin{eqnarray}
d'_M (z) &=& d_H(z) = \zeta \tilde{D}'_M(z) = \zeta \tilde{D}_H(z) ,
\label{eq:dH_DHtilde_relation} \\
d''_M (z) &=& d'_H(z) = \zeta \tilde{D}''_M(z) = \zeta \tilde{D}'_H(z) .
\label{eq:dHp_DHptilde_relation}
\end{eqnarray}
These equations suggest that a combination of SNIa and BAO observations constrains the parameter $\zeta$. We use $d_M$ from BAO as a function of $\zeta$ via \eqref{eq:dM_DMtilde_relation}. We also use the PantheonPlus $d_M$. We then apply multi-task GPR, where the first derivative information is provided by $d_M'$ ($=d_H$) from DESI DR1 BAO as a function of $\zeta$ using \eqref{eq:dH_DHtilde_relation}. This corresponds to
\begin{align}
X_1 &=
\begin{bmatrix} 
\{z_{\rm eff}\} ~ (\text{DESI DR1 BAO}) \\
\{z\} ~ (\text{PantheonPlus})
\end{bmatrix}
,  \\  
X_2 &= \{z_{\rm eff}\} ~ (\text{DESI DR1 BAO}) ,    
\\ 
Y_1 &=
\begin{bmatrix} 
\{d_M(\zeta)\} ~ (\text{DESI DR1 BAO}) \\
\{d_M\} ~ (\text{PantheonPlus})
\end{bmatrix}
,   
\\
Y'_2 &= \{d'_M(\zeta)=d_H(\zeta)\} ~ (\text{DESI DR1 BAO}) .    
\end{align}
Then we perform a multi-task GPR analysis.  Minimisation of the negative log marginal likelihood provides constraints on kernel hyperparameters as well as on  $\zeta$:
\begin{equation}
\text{DESI DR1 BAO+PantheonPlus:} ~~~ \zeta = 192.6 \pm 1.9 ~~ \text{Mpc} .
\label{eq:desi_PP_zeta}
\end{equation}
The marginalised probability of $\zeta$ obtained from the minimisation is shown in the top left panel of Fig.~\ref{fig:CMB_DESI_SNPP_1}.

In the top right panel of Fig.~\ref{fig:CMB_DESI_SNPP_1}, we plot the reconstructed $d_M$ and associated 1$\sigma$ confidence (solid blue line and shading). The black error bars (top right) show $d_M \pm \Delta d_M$ obtained directly from PantheonPlus data of $m_B$ using \eqref{eq:SN_DM_main}. The red error bars also correspond to $d_M \pm \Delta d_M$ but obtained from DESI DR1 BAO data of $\tilde{D}_M$, using the constraint on $\zeta$ in \eqref{eq:desi_PP_zeta}. We see that the obtained reconstructed function of $d_M$ from multi-task GPR is consistent with the $d_M$ obtained from both PantheonPlus and DESI DR1 BAO data.

The predicted  $d'_M=d_H$ and associated 1$\sigma$ confidence interval are shown in the bottom left panel of Fig.~\ref{fig:CMB_DESI_SNPP_1} in blue. In the same panel, the red error bars correspond to $d_H \pm \Delta d_H$ obtained from DESI DR1 BAO data of $\tilde{D}_M$ and constraints on $\zeta$ using  \eqref{eq:desi_PP_zeta}. It is evident that the predicted $d_H$ is consistent with the $d_H$ obtained from DESI DR1 BAO (with $\zeta$ value).

Similarly,  the predicted second derivative  $d''_M=d'_H$ and the associated 1$\sigma$ confidence region are shown in the bottom right panel of Fig.~\ref{fig:CMB_DESI_SNPP_1}. In this case, there is no data to compare the predicted outcome from multi-task GPR.

\begin{figure*}
\centering
\includegraphics[height=170pt,width=0.46\textwidth]{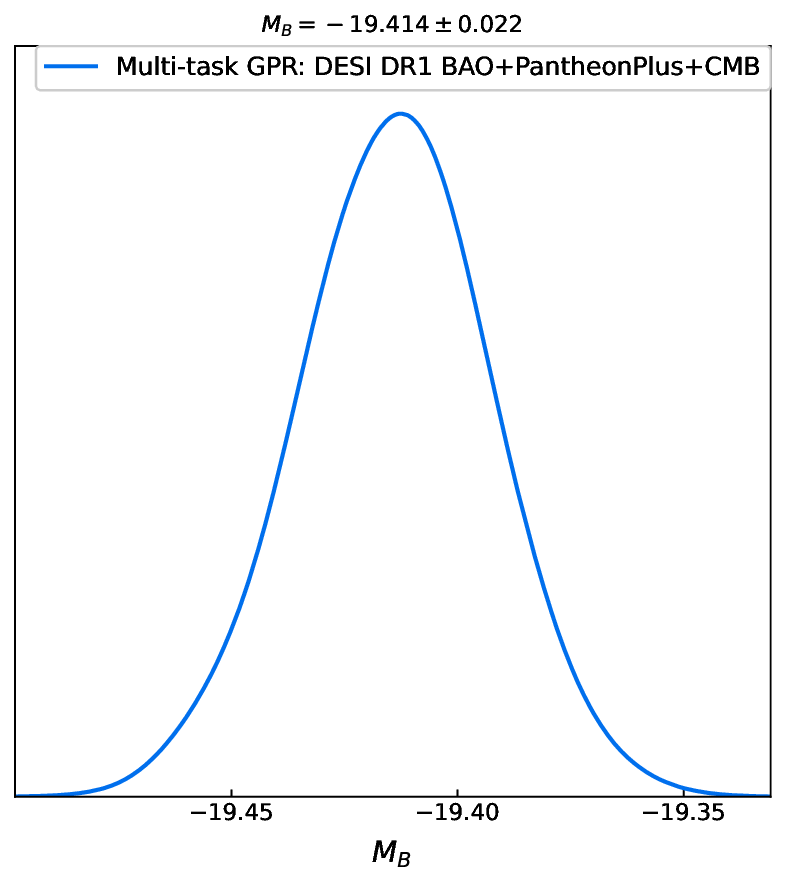}
\includegraphics[height=170pt,width=0.52\textwidth]{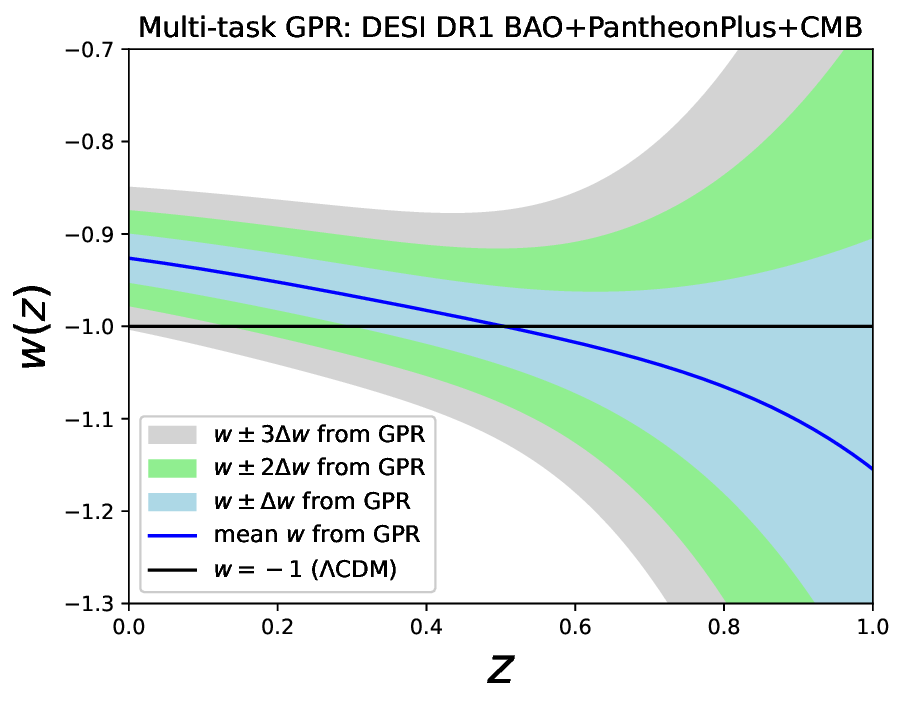}
\caption{
\label{fig:CMB_DESI_SNPP_2}
For the combination CMB+DESI DR1 BAO+PantheonPlus (multi-task GPR),
the left panel shows the marginalised probability of the $M_B$ parameter, obtained from constraints on $\zeta$ and from the $r_d$ Gaussian prior in  \eqref{eq:rd_PL18_standard}. The reconstructed $w$ and associated confidence regions are displayed in the right panel, using  \eqref{eq:w_DE_6}, \eqref{eq:defn_delta} and~\eqref{eq:beta_PL18_standard} with the obtained $M_B$ constraints.
}
\end{figure*}

\subsection{CMB+DESI DR1 BAO+PantheonPlus}

In the left panel of Fig.~\ref{fig:CMB_DESI_SNPP_2}, we plot the marginalised probability of the $M_B$ parameter. This is obtained using the $\zeta$ constraint and the $r_d$ Gaussian prior mentioned in  \eqref{eq:desi_PP_zeta} and~\eqref{eq:rd_PL18_standard} respectively. Because we include $r_d$ prior from CMB to break the degeneracy between $r_d$ and $M_B$, this corresponds to the inclusion of CMB data and the constraint on $M_B$ from CMB + DESI DR1 BAO + PantheonPlus is 
\begin{equation}
\text{CMB+DESI DR1 BAO+PantheonPlus:} ~~~ M_B = -19.414 \pm 0.022 ~ {\rm mag} .
\label{eq:cmb_desi_PP_MB}
\end{equation}
Note that this value of $M_B$ is smaller than the SHOES value.

From the reconstructed $d'_M$, $d''_M$, constraint on $M_B$ from  \eqref{eq:cmb_desi_PP_MB}, and constraint on $\beta$ from  \eqref{eq:beta_PL18_standard}, we compute $w$ and the associated errors using \eqref{eq:w_DE_6} and~\eqref{eq:defn_delta}. These are plotted in the right panel of Fig.~\ref{fig:CMB_DESI_SNPP_2}. The solid blue line, the blue region, the green region (including the blue region), and the gray region (including blue and green regions) correspond to the mean function of $w$, associated 1$\sigma$, 2$\sigma$, 3$\sigma$ confidence regions respectively. We see that at lower redshifts $0<z<0.5$ non-phantom regions are preferable for the CMB + DESI DR1 BAO + PantheonPlus combination of data. In this redshift region, the $\Lambda$CDM model is around more than 2$\sigma$ to less than 1$\sigma$ away (gradually decreasing with increasing redshift) from the reconstructed mean of $w$ and in other redshift regions, it is well within the 1$\sigma$ confidence regions.

\begin{figure*}
\centering
\centering
\includegraphics[height=170pt,width=0.49\textwidth]{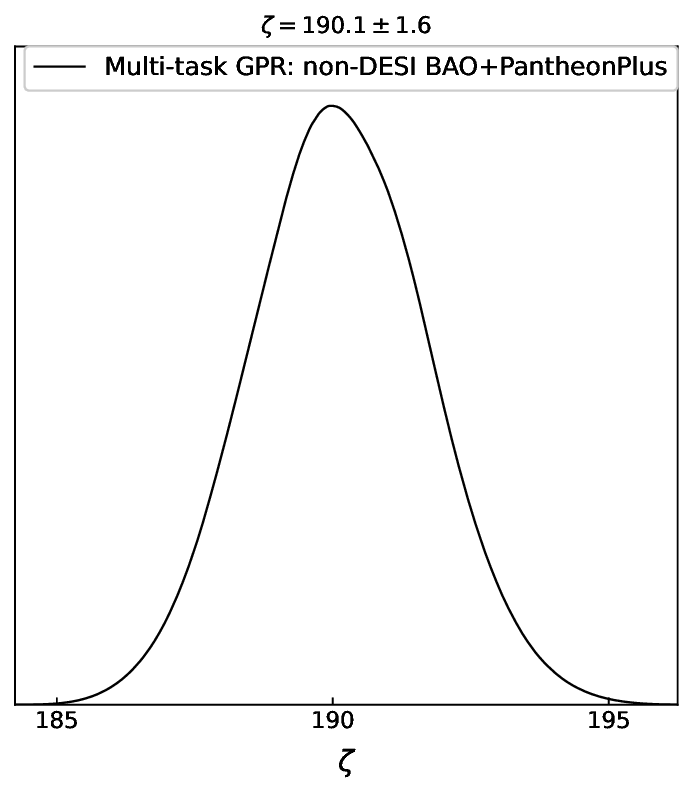}
\includegraphics[height=170pt,width=0.49\textwidth]{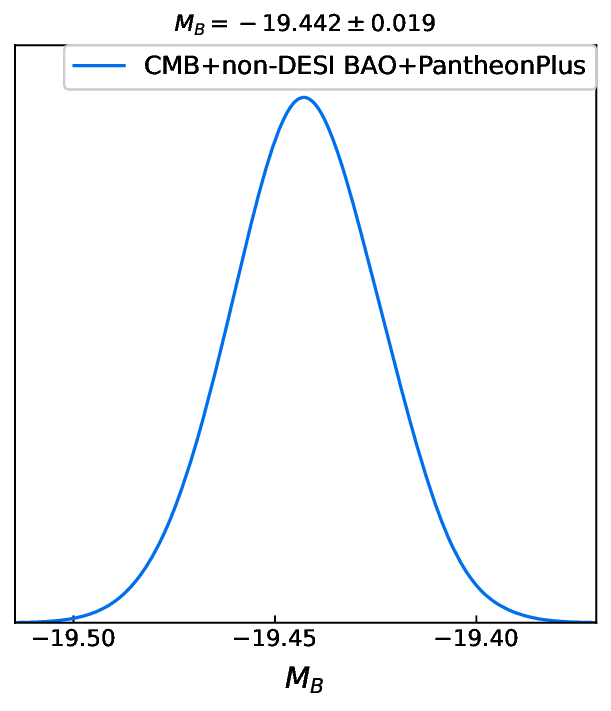} \\
\includegraphics[width=0.55\textwidth]{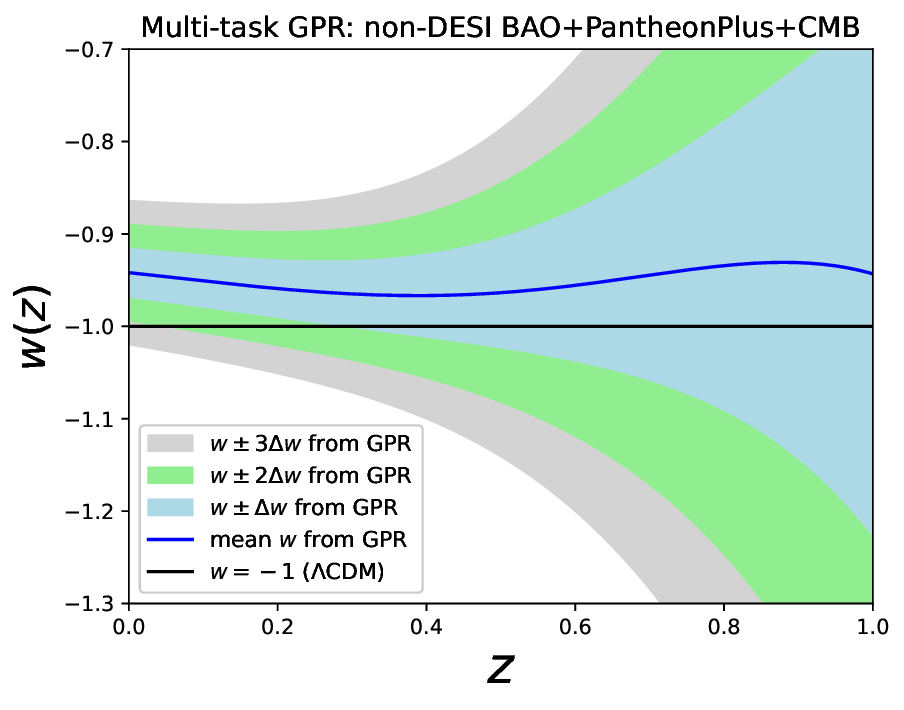}
\caption{
\label{fig:CMB_nondesi_SNPP}
Non-DESI BAO+PantheonPlus and CMB+non-DESI BAO+PantheonPlus combinations, using multi-task GPR.
Top panels: 
marginalised probabilities  of $\zeta$, without CMB (left) and of $M_B$, with CMB (right). Reconstructed $w$ and associated confidence regions are shown in the bottom panel for CMB+non-DESI BAO+PantheonPlus (multi-task GPR) using the same methodology as for Figs.~\ref{fig:CMB_DESI_SNPP_1} and~\ref{fig:CMB_DESI_SNPP_2}.
}
\end{figure*}

\subsection{CMB+Non-DESI BAO+PantheonPlus}

For the non-DESI BAO data, we start with the multi-task GPR combination with PantheonPlus, without using CMB information.  We find a constraint on  $\zeta$, 
\begin{equation}
\text{non-DESI BAO+PantheonPlus:} ~~~ \zeta = 190.1 \pm 1.6 ~~ \text{Mpc} .
\label{eq:nondesi_pp_zeta}
\end{equation}

Introducing CMB data via $r_d$ constraints,  we obtain constraints on the $M_B$ parameter for CMB+non-DESI BAO+PantheonPlus:
\begin{equation}
\text{CMB+non-DESI BAO+PantheonPlus:} ~~~ M_B = -19.442 \pm 0.019 ~~ {\rm mag} .
\label{eq:cmb_nondesi_pp_MB}
\end{equation}
Note that this value is slightly smaller than the one obtained from CMB + DESI DR1 BAO + PantheonPlus in \eqref{eq:cmb_desi_PP_MB}. The corresponding marginalised probabilities of $\zeta$ and $M_B$  are shown in the top left and top right panels of Fig.~\ref{fig:CMB_nondesi_SNPP}.

Following the same methodology as in the previous case, we reconstruct $w$ and the associated confidence regions. These are displayed in the bottom panel of Fig.~\ref{fig:CMB_nondesi_SNPP}. We find that in the redshift range $0<z\lesssim 0.35$, the reconstructed $w$ is preferably in the non-phantom region at more than 1$\sigma$ confidence. In the same redshift range, the $\Lambda$CDM model is around 2$\sigma$ to 1$\sigma$ away and in other regions, it is well within the 1$\sigma$ region.

\subsection{Reconstructed constants and consistency check of $\Lambda$CDM}

Once we have all the corresponding reconstructed functions and consequently their values at present ($z=0$), we obtain constraints on the constants used in the cosmological analysis. We compute these constants using combinations of data sets as follows:
\begingroup
\allowdisplaybreaks
\begin{align}
\label{eq:bao_H0rd}
 \text{BAO:} ~~~ H_0 r_d &= \frac{c}{\tilde{D}_H(z=0)} , \\
\label{eq:bao_sn_H0rd}
 \text{BAO+SNIa:} ~~~ H_0 r_d &= \frac{c \zeta}{d_H(z=0)} , \\
\label{eq:cmb_bao_H0_Om0}
 \text{CMB+BAO:} ~~~ H_0 r_d &= \frac{c}{\tilde{D}_H(z=0)} ~ (\text{same as for only BAO}) , \nonumber\\
  H_0 &= \frac{c}{r_d \tilde{D}_H(z=0)} , ~~~ \Omega_{\rm m0} = \frac{\omega_{\rm m0}}{h^2} , \\
\label{eq:cmb_bao_sn_H0_MB_Om0}
 \text{CMB+BAO+SNIa:} ~~~ H_0 r_d &= \frac{c \zeta}{d_H(z=0)} ~ (\text{same as for BAO+SNIa}) , \nonumber\\
  H_0 &= \frac{c \zeta}{r_d d_H(z=0)}, ~~~ \Omega_{\rm m0} = \frac{\omega_{\rm m0}}{h^2}, ~~~ M_B = \frac{1}{b} \ln{\left(\frac{\zeta}{r_d}\right)}-20 .
\end{align}
\endgroup
\begin{table*}
\begin{center}
\begin{tabular}{|c|c|c|c|c|c|c|c|}
\hline &&&&\\
Data combination & $H_0r_d$ [100 km/s] & $H_0$ [km/s/Mpc] & $\Omega_{\rm m0}$ & $M_B$ [mag] \\
&&&&\\
\hline 
DESI & $104.02 \pm 2.34$ & - & - & - \\
\hline
non-DESI & $99.78 \pm 1.94$ & - & - & - \\
\hline
DESI+PP & $100.65 \pm 1.03$ & - & - & - \\
\hline
non-DESI+PP & $99.30 \pm 0.87$ & - & - & - \\
\hline
CMB+DESI & $104.02 \pm 2.34$ & $70.74 \pm 1.60$ & $0.286 \pm 0.013$ & - \\
\hline
CMB+non-DESI & $99.78 \pm 1.94$ & $67.86 \pm 1.33$ & $0.310 \pm 0.013$ & - \\
\hline
CMB+DESI+PP & $100.65 \pm 1.03$ & $68.45 \pm 0.71$ & $0.305 \pm 0.007$ & $-19.414 \pm 0.022$ \\
\hline
CMB+non-DESI+PP & $99.30 \pm 0.87$ & $67.53 \pm 0.61$ & $0.313 \pm 0.006$ & $-19.442 \pm 0.019$ \\
\hline
\end{tabular}
\end{center}
\caption{
Reconstructed values of $H_0r_d$, $H_0$, $\Omega_{\rm m0}$ and $M_B$. PP denotes PantheonPlus data \eqref{pplus}.
}
\label{table:constants}
\end{table*}
%
The mean and standard deviation values are listed in Table~\ref{table:constants}.

\begin{figure*}
\centering
\centering
\includegraphics[height=170pt,width=0.49\textwidth]{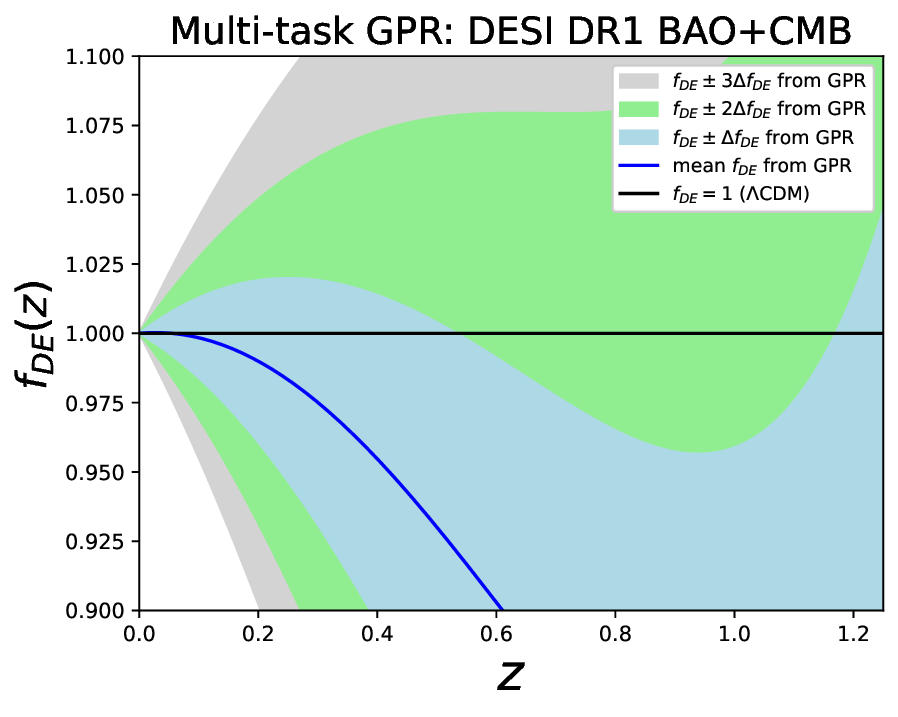}
\includegraphics[height=170pt,width=0.49\textwidth]{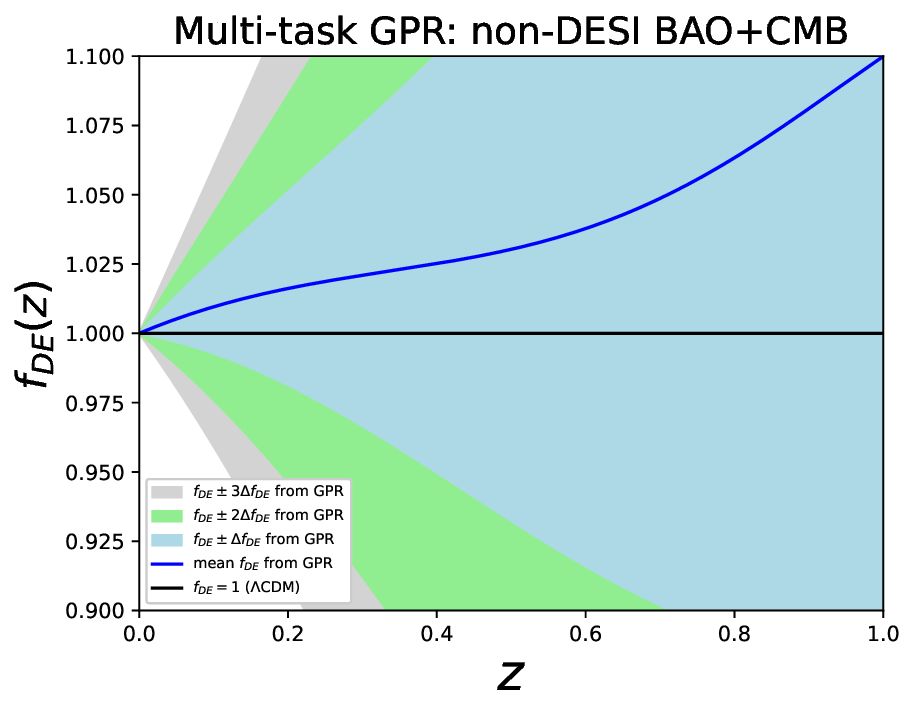} \\
\includegraphics[height=170pt,width=0.49\textwidth]{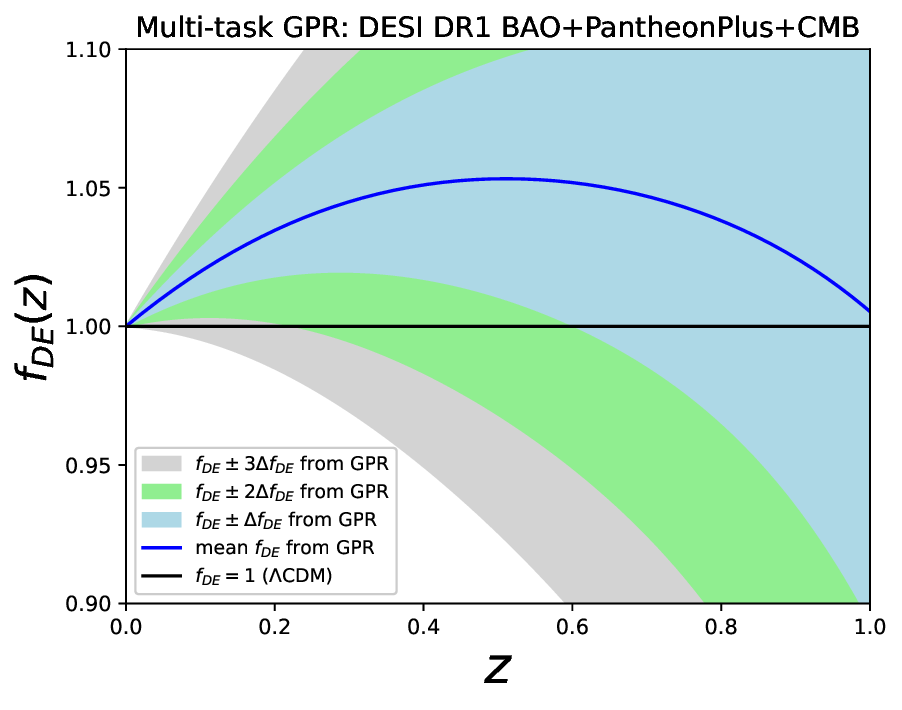}
\includegraphics[height=170pt,width=0.49\textwidth]{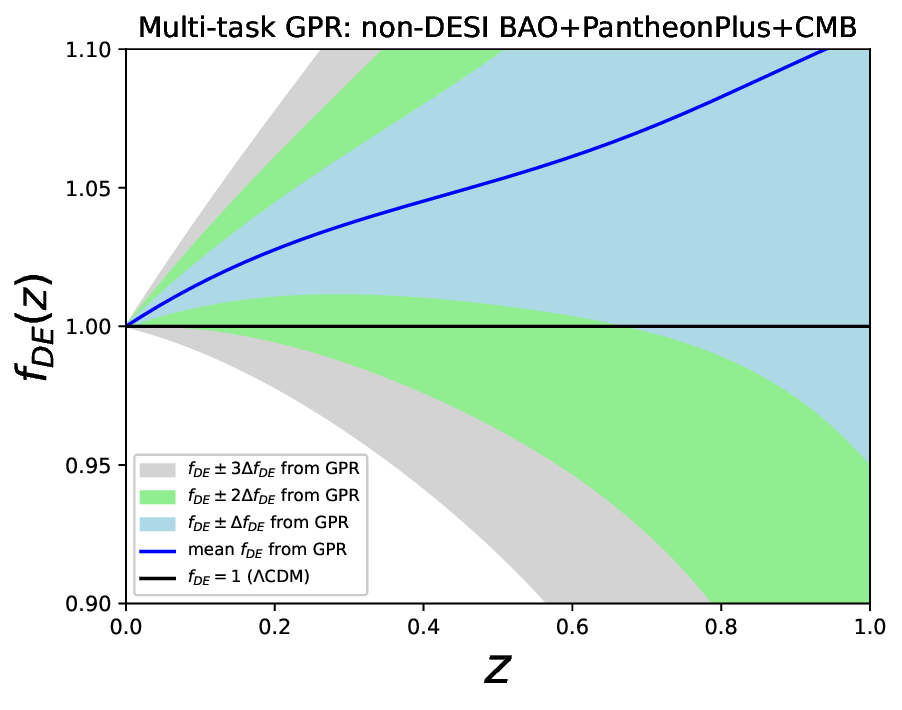}
\caption{
\label{fig:all_fDE}
Reconstruction of $f_{\rm DE}$ using  \eqref{eq:fde_wrt_H} using CMB+DESI DR1 BAO, CMB+non-DESI BAO, CMB+DESI DR1 BAO+PantheonPlus and CMB+non-DESI BAO+PantheonPlus  data. Blue lines are the mean values of $f_{\rm DE}$ obtained from multi-task GPR analysis. Confidence regions correspond to 1$\sigma$, 2$\sigma$, and 3$\sigma$. Horizontal black lines correspond to $\Lambda$CDM.
}
\end{figure*}

Once we know  $H_0$ and $\Omega_{\rm m0}$ separately, we can check the consistency of the $\Lambda$CDM model with different combinations of data in another way through the parameter
\begin{equation}
    f_{\rm DE}(z) = \exp{ \left[ 3 \int_{0}^{z} \frac{1+w(\tilde{z})}{1+\tilde{z}} d\tilde{z} \right] } .
    \label{eq:defn_fde}
\end{equation}
For the $\Lambda$CDM model, 
$f_{\rm DE, \Lambda\text{CDM}} = 1$.
Using \eqref{eq:flat_hubble} and \eqref{eq:alpha}, 
\begin{equation}
f_{\rm DE}(z) = \frac{H^2(z)-\alpha (1+z)^3}{H_0^2-\alpha} .
\label{eq:fde_wrt_H}
\end{equation}
This allows us to rewrite  $f_{\rm DE}$ in terms of the other variables used in our analysis.
Figure~\ref{fig:all_fDE} shows the reconstructed function $f_{\rm DE}$ for the datasets

CMB+DESI DR1 BAO, CMB+non-DESI BAO, 

CMB+DESI DR1 BAO+PantheonPlus,  CMB+non-DESI BAO+PantheonPlus. \\
Solid blue lines give the reconstructed mean $f_{\rm DE}$. The blue, green (including blue), and gray (including blue and green) regions correspond to 1$\sigma$, 2$\sigma$, and 3$\sigma$ confidence regions. Horizontal black lines represent $f_{\rm DE}=1$,  corresponding to the standard $\Lambda$CDM model.
\begin{itemize}
    \item 
For the CMB+DESI DR1 BAO combination of data, we find that the $\Lambda$CDM model is little more than 1$\sigma$ away in the redshift range  $0.6\lesssim z\lesssim 1.15$. In other redshift regions, it is well within the 1$\sigma$ region. This means there are no significant deviations from the $\Lambda$CDM model in the CMB+DESI DR1 BAO data combination. 
\item
For the CMB+DESI DR1 BAO+PantheonPlus combination of data, we find that in  $z\lesssim 0.2$ the $\Lambda$CDM model is little more than 2$\sigma$ away. In  $0.2\lesssim z\lesssim 0.6$, the $\Lambda$CDM model is around 2$\sigma$ to 1$\sigma$ away (gradually decreasing with increasing redshift). In other redshift regions, it is within the 1$\sigma$ limit. This means that the deviations are low to moderate, but not very significant.
\item
For the CMB+non-DESI BAO combination of data, the $\Lambda$CDM model is well within the 1$\sigma$ region in the entire redshift region. This means there is no evidence for the deviations from the $\Lambda$CDM model.
\item
For the CMB+non-DESI BAO+PantheonPlus combination of data, the $\Lambda$CDM model is little more than 1$\sigma$ away in the redshift range around $0<z<0.6$ and it is well within the 1$\sigma$ region in the other redshift regions. This means for this combination of data the deviations from the $\Lambda$CDM model are mild.

\end{itemize}

\subsection{Hubble tension, $M_B$ tension and their connection to $w$}

Table~\ref{table:constants} reveals interesting correlations between $H_0$, $M_B$, and $\Omega_{\rm m0}$ which bear the Hubble tension and the related $M_B$ tension. This tension is the disagreement between the direct low-redshift measurement from SNIa and the indirect measurement from the CMB.

\begin{itemize}
    
    \item The reconstructed $H_0$ values, where no SHOES data is involved or the corresponding $M_B$ value ($\approx -19.25$) is not considered, are lower than the $H_0\approx 73$ obtained via the SHOES data from the calibration of SNIa and observations of Cepheid variables. This is the well-established Hubble tension \citep{DiValentino:2021izs,Vagnozzi:2019ezj,Krishnan:2021dyb}.
    
    \item
    A similar conclusion is applicable for $M_B$, corresponding to the $M_B$ tension \citep{Camarena:2021jlr,Efstathiou:2021ocp,Dinda:2021ffa}.
    
    \item We see that the higher the value of $H_0$, the higher the value of $M_B$, and vice-versa when SNIa data is involved. It is also apparent in Table~\ref{table:extended_constants} in Appendix~\ref{sec-extended_constants}. This is a well-known fact in cosmology \citep{Camarena:2021jlr,Efstathiou:2021ocp,Dinda:2021ffa}.

	\item
    We do not find any immediate connection between $H_0$ and $w$ when SNIa data are not involved. However, when SNIa (PantheonPlus) data is involved, we see a higher $H_0$, corresponding to the higher value of $w$ at lower redshift. The same applies to the correlation between $M_B$ and $w$. Similarly, the inverse relation applies between $\Omega_{\rm m0}$ and $w$ at lower redshift when SNIa data is involved.    
    
    \item There is an inverse correlation between $H_0$ and $\Omega_{\rm m0}$, when CMB data is involved (with or without SNIa)  \citep{Sakr:2023hrl,Dinda:2023xqx}. The same inverse correlation exists between $M_B$ and $\Omega_{\rm m0}$ when SNIa data is involved. This can also be seen in Table~\ref{table:extended_constants} in Appendix~\ref{sec-extended_constants}. This is also a well-known fact in cosmology.

\end{itemize}

\section{Conclusions}
\label{sec-conclusion}

Recent results from DESI DR1 BAO data \citep{DESI:2024mwx} suggest evidence for dynamical dark energy, with a preference for phantom behaviour in significant redshift ranges. These results rely on a 
phenomenological parametrised equation of state $w$ for dark energy (or effective dark energy). Given the limitations of imposing such an equation of state, we 
used a model-agnostic approach to reconstruct $w(z)$ directly from the data.

In addition to the DESI DR1 BAO data, we considered other non-DESI BAO data, CMB distance data from Planck 2018, and SNIa data from PantheonPlus. The model-agnostic methodology is based on the simple (single-task) posterior Gaussian Process regression as well as the multi-task GP regression. Multi-task GP is required to correctly incorporate the correlation between anisotropic BAO observables parallel and perpendicular to the line of sight.

From the CMB+DESI DR1 BAO combination data (see Fig.~\ref{fig:CMB_DESI_w_reconstruction}), we find that the reconstructed mean function $w$ is on the phantom side; however, this evidence is not significant. To be more specific, in the redshift range  $0.3\lesssim z\lesssim 0.8$, the $\Lambda$CDM model is little more than 1$\sigma$ away, while in other redshift interval it is well within 1$\sigma$ limit.

In Fig.~\ref{fig:CMB_DESI_SNPP_2}, when we consider SNIa data from PantheonPlus, the low redshift reconstruction of $w$ is dominated by this data, which has a preference for non-phantom behaviour. The CMB+DESI DR1 BAO+PantheonPlus data leads to a reconstructed $w$ that is not in tension with $\Lambda$CDM model -- which is $\lesssim 2\sigma$ to 1$\sigma$ away in the redshift range $z\lesssim 0.35$.

When we replace the DESI DR1 BAO data with the previous non-DESI BAO data, using the combinations CMB+non-DESI BAO (Fig.~\ref{fig:CMB_non_DESI_w_reconstruction}) and CMB+non-DESI BAO+PantheonPlus (Fig.~\ref{fig:CMB_nondesi_SNPP}), 
we find no evidence for deviation from the $\Lambda$CDM model, which are within the 1$\sigma$ and 2$\sigma$ limits respectively.

We also compute constraints on different constants involved in a cosmological analysis, such as the Hubble constant and the present value of the matter energy density parameter. 
These constraints reflect the Hubble tension and the corresponding $M_B$ tension. When SNIa data are used, we note a higher $H_0$ together with a corresponding higher $w$ at low redshift.

Finally, we reconstruct the normalised energy density parameter of dark energy, $f_{\rm DE}$, as a probe to check the consistency of the $\Lambda$CDM model against all these observations in Fig.~\ref{fig:all_fDE}. We find a similar conclusion:  the deviations from the $\Lambda$CDM value ($f_{\rm DE}=1$) are not very significant.

Note that all these results use CMB distance priors, which are obtained from the standard base $\Lambda$CDM model instead of the full CMB likelihood. The results would be similar for other models of cosmology, provided that the considered model does not differ significantly from the base $\Lambda$CDM model.

In summary, our model-agnostic approach suggests significantly low evidence for dynamical dark energy  and this evidence is not strong enough to firmly conclude that there is evidence for dynamical dark energy.

\acknowledgments
The authors are supported by the South African Radio Astronomy Observatory and the National Research Foundation (Grant No. 75415).

\clearpage

\appendix

\section{Computation of $r_s(z_*)$ and $r_d$ from CMB distance priors}
\label{sec-CMB_dist_priors}

The sound horizon  is 
\begin{equation}
r_s (z) = \int_z^{\infty} \frac{c_s(\tilde{z}) }{H(\tilde{z})}\,d\tilde{z} = \int_0^{1/(1+z)} \frac{c_s(a) }{a^2H(a)}\,da .
\label{eq:defn_sound_horizon}
\end{equation}
For early times 
\begin{equation}
c_s^2 (a) = \frac{c^2}{3\big[1+{3}\,\omega_{\rm b0}/(4\,\omega_{\rm \gamma 0})a\big]} \,,
\label{eq:early_sound_speed}
\end{equation}
where $\omega_{\rm \gamma 0}=\Omega_{\rm \gamma 0}h^2$ and $\Omega_{\rm \gamma 0}$ is the present value of the photon energy density parameter. At sufficiently early times of our interest, the Hubble parameter can be approximated by
\begin{equation}
H^2(z) \approx H_0^2 \big[{\Omega_{\rm m0}(1+z)^3+\Omega_{\rm r0}(1+z)^4}\big] ,
\label{eq:H_early_appx}
\end{equation}
where $\Omega_{\rm r0}$ is the present value of the radiation energy density parameter. In this approximation, the sound horizon is \citep{Dinda:2024kjf}
\begin{equation}
r_s (z) = \tilde{\delta} ~ {\rm log} \left\{ \frac{{2}+(1+z)(\tilde{\alpha}+\tilde{\gamma})+2\big[(1+z)^2{\tilde{\alpha}\tilde{\gamma}+ (1+z)({\tilde{\alpha}+\tilde{\gamma}})+1}\big]^{1/2}}{(1+z)\big[{\tilde{\alpha}+\tilde{\gamma}}+2({\tilde{\alpha}\tilde{\gamma}})^{1/2}\big]} \right\} ,
\label{eq:sound_horizon_appx}
\end{equation}
where \citep{Hu:1995en,Chen:2018dbv,Zhai:2019nad,Zhai:2018vmm,Dinda:2024kjf}
\begin{align}
\tilde{\alpha} &= \frac{\tilde{\beta}}{\omega_{\rm b0}} ,
\label{eq:tilde_alpha}\\
\tilde{\beta} &= \frac{4\omega_{\rm \gamma 0}}{3} = \frac{\theta^{4}}{31500} ,
\label{eq:tilde_beta}\\
\theta &= \frac{T_{\rm cmb}}{2.7~\rm K} ,
\label{eq:theta}\\
\tilde{\gamma} &= \frac{\Omega_{\rm r0}}{\Omega_{\rm m0}} = \frac{1}{1+z_{\rm eq}} ,
\label{eq:gamma_tilde}\\
\tilde{\delta} &= \frac{3000 \,{\tilde{\beta}}^{1/2} }{\big({3 \omega_{\rm m0} \omega_{\rm b0}}\big)^{1/2}}~~ {\rm Mpc} .
\label{eq:delta_tilde}
\end{align}
Here $T_{\rm cmb}=2.7255\,$K and 
$z_{\rm eq}$ is at radiation-matter equality \citep{Hu:1995en,Chen:2018dbv,Zhai:2019nad,Zhai:2018vmm,Dinda:2024kjf}:
\begin{equation}
z_{\rm eq} = 25000~\omega_{\rm m0} \,\theta^{-4} \approx 24077.44059~\omega_{\rm m0} .
\label{eq:z_eq_rad_mat}
\end{equation}

The redshift of the photon decoupling epoch $z_*$ has an approximate expression for standard early time physics  \citep{Hu:1995en,Chen:2018dbv,Zhai:2019nad,Zhai:2018vmm,Dinda:2024kjf}:
\begin{equation}
z_* = 1048 \left( 1+0.00124 \omega_{\rm b0}^{-0.738} \right) \big( 1+g_1 \omega_{\rm m0}^{g_2} \big) ,
\label{eq:z_photon_decoupling}
\end{equation}
where 
\begin{eqnarray}
g_1 &=& \frac{ 0.0783\, \omega_{\rm b0}^{-0.238} }{ 1+39.5\, \omega_{\rm b0}^{0.763} } ,
\label{eq:z_cmb_g1} \\
g_2 &=& \frac{0.560}{ 1+21.1\, \omega_{\rm b0}^{1.81} } .
\label{eq:z_cmb_g_2}
\end{eqnarray}

Similarly, the redshift of the baryon drag epoch $z_d$ has an approximate expression in standard early time physics  \citep{Hu:1995en}:
\begin{equation}
z_d = \frac{1345\, \omega_{\rm m0}^{0.251}\big(1+b_1\,\omega_{\rm b0}^{b_2}\big)}{1+0.659\,\omega_{\rm m0}^{0.828}} ,
\label{eq:z_baryon_drag}
\end{equation}
where 
\begin{eqnarray}
b_1 &=& 0.313\, \omega_{\rm m0}^{-0.419}\left(1+0.607\,\omega_{\rm m0}^{0.674}\right) ,
\label{eq:z_d_b1} \\
b_2 &=& 0.238\,\omega_{\rm m0}^{0.223} .
\label{eq:z_d_b2}
\end{eqnarray}

Considering all the above expressions, we find that both $r_s(z_*)$ and $r_d$ are functions of only $\omega_{\rm m0}$ and $\omega_{\rm b0}$.

\section{Gaussian Process Regression}
\label{sec-method}

\subsection{Single-task Gaussian Process regression up to second order derivative}
\label{sec-single_task_GPR}

Gaussian process regression (GPR) analysis is useful to predict a smooth function from a given data set. Let us denote vectors of data points and the observed values of any observable by $X_1$ and $Y_1=Y(X_1)$ respectively with the observational errors denoted by a matrix $C_{11}$ corresponding to all the variances and covariances. GPR is only useful if these errors are Gaussian. With this condition, in GPR analysis, data sets are considered to be a multivariate Gaussian (normal) distribution with a specified kernel covariance function $K$ and a mean function $\mu$: 
\begin{equation}
Y_1 = Y(X_1) \thicksim \mathcal{N} \left( \mu_1, \Sigma_{11} \right) ,
\label{eq:single_norm}
\end{equation}
where $\mathcal{N}$ is a joint multivariate Gaussian distribution, $\mu_1$ is the mean vector corresponding to the chosen mean function at $X_1$ and $\Sigma_{11}$ is the total covariance matrix: 
\begin{eqnarray}
\mu_1 &=& \mu(X_1) ,
\label{eq:single_mu_1} \\
\Sigma_{11} &=& K_{11}+C_{11} ,
\label{eq:single_Sigma} \\
K_{11} &=& K(X_1,X_1) ,
\label{eq:K_11} \\
C_{11} &=& {\rm Cov} [Y_1,Y_1] = {\rm Cov} [Y(X_1),Y(X_1)] ~ ({\rm from~observational~errors}) .
\label{eq:C_11}
\end{eqnarray}
The exact vector form of $\mu_1$ and the matrix forms of $C_{11}$ and $K_{11}$ are given in Appendix~\ref{sec-GPR_short_notations}.

\subsubsection{Training the single-task GPR}

Before any prediction from GPR, in a posterior approach, it is trained by minimizing the negative of the logarithmic marginal likelihood
\begin{equation}
- \log P(Y_1|X_1) = \frac{1}{2} (Y_1-\mu_1)^{\rm T} \Sigma_{11}^{-1} (Y_1-\mu_1)+\frac{1}{2} \log |\Sigma_{11}|+\frac{n_1}{2} \log (2\pi) ,
\label{eq:single_m_log_prob}
\end{equation}
where  $n_1$ is the total number of data points, and the notation $|A|$ represents the determinant of any matrix $A$. We use this notation for the determinant throughout this analysis. The kernel covariance function is described by the kernel hyper-parameters, and best-fit values of these parameters are obtained from the minimisation of the negative log marginal likelihood, mentioned above.

\subsubsection{Prediction from single-task GPR}

The best-fit values, obtained from the minimization of the negative log marginal likelihood, mentioned in \eqref{eq:single_m_log_prob}, are used for the prediction of the function or the function values at a specific target set of points. The predicted values of the function and its derivatives can be considered to be a joint multivariate Gaussian distribution with the observed data. The distribution up to the second-order derivative prediction of the function is 
\begin{equation}
\begin{bmatrix} 
Y_1 \\
F_* \\
F'_* \\
F''_*
\end{bmatrix}
\sim \mathcal{N} \left(
\begin{bmatrix} 
\mu_1 \\
\mu_* \\
\mu'_* \\
\mu''_*
\end{bmatrix},
\begin{bmatrix} 
\Sigma_{11} & K_{1*} & K_{\rm 1*}^{(0,1)} & K_{\rm 1*}^{(0,2)} \\
K_{\rm *1} & K_{**} & K^{(0,1)}_{**} & K^{(0,2)}_{**} \\
K_{\rm *1}^{(1,0)} & K^{(1,0)}_{**} & K^{(1,1)}_{**} & K^{(1,2)}_{**} \\
K_{\rm *1}^{(2,0)} & K^{(2,0)}_{**} & K^{(2,1)}_{**} & K^{(2,2)}_{**}
\end{bmatrix}
\right) ,
\label{eq:joint_distribution_GPR_single}
\end{equation}
where subscript '$*$' corresponds to the target points $X_*$ where GPR predicts the values of a function and its derivatives. All the short notations used above are standard. To have ideas of these short notations, we list this in Appendix~\ref{sec-GPR_short_notations}. The predictions of GPR are based on the conditional distributions corresponding to the joint probability distribution of $Y_*^{(p)}$ (for $F_*=Y(X_*)$, $F'_*=Y'(X_*)$, and $F''_*=Y''(X_*)$) given $Y_1$ and $\Sigma_{11}$. These can be easily seen if we rewrite the  joint distribution \eqref{eq:joint_distribution_GPR_single} as
\begin{equation}
\begin{bmatrix} 
Y_1 \\
Y_*^{(p)}
\end{bmatrix}
\sim \mathcal{N} \left(
\begin{bmatrix} 
\mu_1 \\
\mu_*^{(p)}
\end{bmatrix},
\begin{bmatrix} 
\Sigma_{11} & \Sigma_{1*}^{(0,p)} \\
\Sigma_{*1}^{(p,0)} & \Sigma_{**}^{(p,p)}
\end{bmatrix}
\right) ,
\label{eq:joint_distribution_GPR_single_2}
\end{equation}
where 
\begin{align}
\label{eq:conditional_predict_no_change_1}
Y_*^{(p)} &=
\begin{bmatrix}
F_* = Y(X_*) \\
F'_* = Y'(X_*) \\
F''_* = Y''(X_*)
\end{bmatrix}
, ~~~~~
\mu_*^{(p)} =
\begin{bmatrix}
\mu_* = \mu(X_*) \\
\mu'_* = \mu'(X_*) \\
\mu''_* = \mu''(X_*)
\end{bmatrix}
, \\
\label{eq:conditional_predict_change_1}
\Sigma_{\rm 1*}^{(0,p)} &=
\begin{bmatrix}
K_{\rm 1*} & K_{\rm 1*}^{(0,1)} & K_{\rm 1*}^{(0,2)}
\end{bmatrix}
, \\
\label{eq:conditional_predict_change_2}
\Sigma_{\rm *1}^{(p,0)} &= 
\begin{bmatrix} 
K_{\rm *1} \\
K_{\rm *1}^{(0,1)} \\
K_{\rm *1}^{(0,2)}
\end{bmatrix}
= 
\begin{bmatrix} 
K_{\rm *1}^{\rm T} & \left(K_{\rm *1}^{(0,1)}\right)^{\rm T} & \left(K_{\rm *1}^{(0,2)}\right)^{\rm T}
\end{bmatrix}
^{\rm T} \nonumber\\
&= \left(
\begin{bmatrix} 
K_{\rm 1*} & K_{\rm 1*}^{(1,0)} & K_{\rm 1*}^{(2,0)}
\end{bmatrix}
\right)^{\rm T} = \left(\Sigma_{\rm 1*}^{(0,p)}\right)^{\rm T} , \\
\label{eq:conditional_predict_no_change_2}
\Sigma_{\rm **}^{(p,p)} &=
\begin{bmatrix} 
K_{**} & K^{(0,1)}_{**} & K^{(0,2)}_{**} \\
K^{(1,0)}_{**} & K^{(1,1)}_{**} & K^{(1,2)}_{**} \\
K^{(2,0)}_{**} & K^{(2,1)}_{**} & K^{(2,2)}_{**}
\end{bmatrix}
.
\end{align}

Note that one can consider $F_*$, $F'_*$, and $F''_*$ separately for a conditional distribution to find mean values and self-covariances of each quantity, but this can not give a prediction for cross covariances. The results would be the same for mean and self-covariances. The conditional distribution for the prediction of $Y_*^p$ given $Y_1$ corresponding to the Gaussian joint distribution in the form given in  \eqref{eq:joint_distribution_GPR_single_2} becomes
\begin{align}
\label{eq:conditional_mean}
\bar{Y}_*^{(p)} &= \mu_*^{(p)}+\Sigma_{\rm *1}^{(p,0)} \Sigma_{11}^{-1} (Y_1-\mu_1) , \\
\label{eq:conditional_covariance}
{\rm Cov}[Y_*^{(p)},Y_*^{(p)}] &= \Sigma_{\rm **}^{(p,p)} - \Sigma_{\rm *1}^{(p,0)} \Sigma_{11}^{-1} \Sigma_{\rm 1*}^{(0,p)} ,
\end{align}
where $\bar{Y}_*^{(p)}$ and ${\rm Cov}[Y_*^{(p)},Y_*^{(p)}]$ are predicted mean (vector) and covariance (matrix) respectively. Throughout this analysis, the overbar on any quantity denotes the mean of the quantity. Putting the corresponding expressions of  \eqref{eq:conditional_predict_no_change_1} and~\eqref{eq:conditional_predict_change_2} in  \eqref{eq:conditional_mean}, we get the predicted mean of the function and its derivatives at target points:
\begin{align}
\label{eq:predict_f}
\bar{F}_* &= \mu_*+K_{\rm *1} \Sigma_{11}^{-1} (Y_1-\mu_1) , \\
\label{eq:predict_fp}
\bar{F}'_* &= \mu'_*+K_{\rm *1}^{(1,0)} \Sigma_{11}^{-1} (Y_1-\mu_1) , \\
\label{eq:predict_fpp}
\bar{F}''_* &= \mu''_*+K_{\rm *1}^{(2,0)} \Sigma_{11}^{-1} (Y_1-\mu_1) .
\end{align}

The above three equations can be combined as 
\begin{equation}
\bar{F}_*^{(u)} = \mu_*^{(u)}+K_{\rm *1}^{(u,0)} \Sigma_{11}^{-1} (Y_1-\mu_1) ,
\label{eq:single_predict_general_mean}
\end{equation}
where superscript $u$ is the order of the derivative. The above general expression can also be used for the prediction of mean values of the third and higher-order derivatives. Putting the corresponding expressions of  \eqref{eq:conditional_predict_change_1},~\eqref{eq:conditional_predict_change_2}, and~\eqref{eq:conditional_predict_no_change_2} in  \eqref{eq:conditional_covariance}, we get the predicted self-covariances,
\begin{align}
\label{eq:predict_V_f_f}
{\rm Cov}[F_*,F_*] &= K_{**}-K_{\rm *1} \Sigma_{11}^{-1} K_{\rm 1*} , \\
\label{eq:predict_V_fp_fp}
{\rm Cov}[F'_*,F'_*] &= K^{(1,1)}_{**}-K_{\rm *1}^{(1,0)} \Sigma_{11}^{-1} K_{\rm 1*}^{(0,1)} , \\
\label{eq:predict_V_fpp_fpp}
{\rm Cov}[F''_*,F''_*] &= K^{(2,2)}_{**}-K_{\rm *1}^{(2,0)} \Sigma_{11}^{-1} K_{\rm 1*}^{(0,2)} .
\end{align}
Similarly, using  \eqref{eq:conditional_predict_change_1}--\eqref{eq:conditional_predict_no_change_2} in  \eqref{eq:conditional_covariance}, we find the cross covariances:
\begin{align}
\label{eq:predict_C_f_fp}
{\rm Cov}[F_*,F'_*] &= K^{(0,1)}_{**}-K_{\rm *1} \Sigma_{11}^{-1} K_{\rm 1*}^{(0,1)} , \\
\label{eq:predict_C_f_fpp}
{\rm Cov}[F_*,F''_*] &= K^{(0,2)}_{**}-K_{\rm *1} \Sigma_{11}^{-1} K_{\rm 1*}^{(0,2)} , \\
\label{eq:predict_C_fp_fpp}
{\rm Cov}[F'_*,F''_*] &= K^{(1,2)}_{**}-K_{\rm *1}^{(1,0)} \Sigma_{11}^{-1} K_{\rm 1*}^{(0,2)} .
\end{align}
The above six equations can be written as
\begin{equation}
{\rm Cov}[F_*^{(u)},F_*^{(v)}] = K^{(u,v)}_{**}-K_{\rm *1}^{(u,0)} \Sigma_{11}^{-1} K_{\rm 1*}^{(0,v)} ,
\label{eq:single_predict_general_cov}
\end{equation}
This general expression can be used for third and higher-order order derivatives as well.

\subsection{Multi-task GP regression up to second-order derivative}
\label{sec-multi_task_GPR}

The main aim of this analysis is to use GPR in BAO observations. Because there are large covariances between $\tilde{D}_M$ and $\tilde{D}_H$, we should not apply simple GPR to each of them separately. For this purpose, we consider multi-task GPR in which the information of function ($Y_1=Y(X_1)$) and its first derivative ($Y'_2=Y'(X_2)$) is considered simultaneously at data points $X_1$ and $X_2$ respectively. Note that, in the BAO data set the function and its derivative are in the same data points, but here for generality of the methodology, we consider $X_1$ and $X_2$ to be different in general (the subset can be the same data points). In this case, we assume that $Y_1=Y(X_1)$ and $Y'_2=Y'(X_2)$ follow a joint-multivariate Gaussian distribution
\begin{equation}
\tilde{Y}_1 \thicksim \mathcal{N} \left( \tilde{\mu}_1, \tilde{\Sigma}_{11} \right) ,
\label{eq:double_norm}
\end{equation}
where $\tilde{Y}_1$ is the joint vector consisting of vector $Y_1$ followed by vector $Y'_2$, $\tilde{\mu}_1$ is the joint vector consisting of vector $\mu_1$ followed by vector $\mu'_2$, and $\tilde{\Sigma}_{11}$ is the joint block matrix consisting of matrices $\Sigma_{11}$, $\Sigma_{12}^{(0,1)}$, $\Sigma_{21}^{(1,0)}$, and $\Sigma_{22}^{(1,1)}$:
\begin{align}
\label{eq:tilde_Y}
\tilde{Y}_1 &=
\begin{bmatrix} 
Y_1 \\
Y'_2
\end{bmatrix}
=
\begin{bmatrix} 
Y(X_1) \\
Y'(X_2)
\end{bmatrix}
, \\
\label{eq:tilde_mu}
\tilde{\mu}_1 &=
\begin{bmatrix} 
\mu_1 \\
\mu'_2
\end{bmatrix}
=
\begin{bmatrix} 
\mu(X_1) \\
\mu'(X_2)
\end{bmatrix}
, \\
\label{eq:tilde_Sigma}
\tilde{\Sigma}_{11} &=
\begin{bmatrix} 
\Sigma_{11} & \Sigma_{12}^{(0,1)} \\
\Sigma_{21}^{(1,0)} & \Sigma_{22}^{(1,1)}
\end{bmatrix}
,
\end{align}
In  \eqref{eq:tilde_Sigma}, $\Sigma_{11}$, $\Sigma_{12}^{(0,1)}$, $\Sigma_{21}^{(1,0)}$, and $\Sigma_{22}^{(1,1)}$ correspond to the total self covariances of $Y_1$, total cross covariances between $Y_1$ and $Y'_2$, total cross covariances between $Y'_2$ and $Y_1$, and total self covariances of $Y'_2$:
\begin{align}
\label{eq:Sigma}
\Sigma_{11} &= K_{11}+C_{11} ~~~ \text{-- as in  \eqref{eq:single_Sigma}}, \\
\label{eq:Sigma_p}
\Sigma_{12}^{(0,1)} &= K_{12}^{(0,1)}+C_{12}^{(0,1)} , \\
\label{eq:Sigma_p_T}
\Sigma_{21}^{(1,0)} &= K_{21}^{(1,0)}+C_{21}^{(1,0)} = \left(K_{12}^{(0,1)}\right)^{\rm T}+\left(C_{12}^{(0,1)}\right)^{\rm T} = \left(K_{12}^{(0,1)}+C_{12}^{(0,1)}\right)^{\rm T} = \left(\Sigma_{12}^{(0,1)}\right)^{\rm T} , \\
\label{eq:Sigma_p_p}
\Sigma_{22}^{(1,1)} &= K_{22}^{(1,1)}+C_{22}^{(1,1)} .
\end{align}
Here $C_{11}$, $C_{12}^{(0,1)}$, $C_{21}^{(1,0)}$, and $C_{22}^{(1,1)}$ are the self covariances of $Y_1$, cross covariances between $Y_1$ and $Y'_2$, cross covariances between $Y'_2$ and $Y_1$, and self covariances of $Y'_2$ respectively, corresponding to the contribution only from the observational errors. The actual matrix forms of these matrices are given in  Appendix~\ref{sec-GPR_short_notations}.

\subsubsection{Training the multi-task GPR}

Similar to the previous case, here also the GPR is trained by minimizing
\begin{equation}
- \log P(\tilde{Y}_1|\tilde{X}_1) = \frac{1}{2} (\tilde{Y}_1-\tilde{\mu}_1)^{\rm T} \tilde{\Sigma}_{11}^{-1} (\tilde{Y}_1-\tilde{\mu}_1)+\frac{1}{2} \log |\tilde{\Sigma}_{11}|+ \frac{\tilde{n}_1}{2} \log (2\pi) ,
\label{eq:double_m_log_prob}
\end{equation}
where $\tilde{X}_1$ is the joint data points vector consisting of vector $X_1$ followed by the vector $X_2$ and $\tilde{n}_1$ is the total number of data points in the observations corresponding to the function and its first derivative. Then the dimension of the vector $\tilde{X}_1$ is
\begin{equation}
\tilde{n}_1 = n_1+n_2 \,.
\label{eq:n_tilde}
\end{equation}

The inverse of block matrix $\tilde{\Sigma}_{11}$ can be represented by its constituent matrices:
\begin{equation}
\tilde{\Sigma}_{11}^{-1} =
\begin{bmatrix} 
A_{11} & A_{12} \\
A_{21} & A_{22}
\end{bmatrix}
,
\label{eq:Sigma_tilde_inverse}
\end{equation}
where 
\begin{align}
\label{eq:A11}
A_{11} &= \Sigma_{11}^{-1}+\Sigma_{11}^{-1}\Sigma_{12}^{(0,1)}S\Sigma_{21}^{(1,0)}\Sigma_{11}^{-1} , \\
\label{eq:A12}
A_{12} &= -\Sigma_{11}^{-1}\Sigma_{12}^{(0,1)}S , \\
\label{eq:A21}
A_{21} &= -S\Sigma_{21}^{(1,0)}\Sigma_{11}^{-1} = A_{12}^{\rm T} , \\
\label{eq:A22}
A_{22} &= S = \left( \Sigma_{22}^{(1,1)}-\Sigma_{21}^{(1,0)}\Sigma_{11}^{-1}\Sigma_{12}^{(0,1)} \right)^{-1} .
\end{align}
The combination $\tilde{\Sigma}_{11}^{-1}(\tilde{Y}_1-\tilde{\mu}_1)$ is a vector:
\begin{align}
\tilde{\Sigma}_{11}^{-1}(\tilde{Y}_1-\tilde{\mu}_1) &=
\begin{bmatrix} 
B_1 \\
B_2
\end{bmatrix}
,
\label{eq:Sigma_Y_m_mu_tilde}\\
\label{eq:B1}
B_1 &= A_{11}(Y_1-\mu_1)+A_{12}(Y'_2-\mu'_2) , \\
\label{eq:B2}
B_2 &= A_{21}(Y_1-\mu_1)+A_{22}(Y'_2-\mu'_2) .
\end{align}
The combination $(\tilde{Y}_1-\tilde{\mu}_1)^{\rm T}\tilde{\Sigma}_{11}^{-1}(\tilde{Y}_1-\tilde{\mu}_1)$ is a scalar:
\begin{equation}
(\tilde{Y}_1-\tilde{\mu}_1)^{\rm T}\tilde{\Sigma}_{11}^{-1}(\tilde{Y}_1-\tilde{\mu}_1) = (Y_1-\mu_1)^{\rm T}B_1+(Y'_2-\mu'_2)^{\rm T}B_2 \equiv Q \, .
\label{eq:Y_m_mu_T_Sigma_Y_m_mu_tilde}
\end{equation}

The determinant of the block matrix $\tilde{\Sigma}_{11}$ can be expressed by the determinants of its constituent matrices,
\begin{equation}
|\tilde{\Sigma}_{11}| = |\Sigma_{11}||\Sigma_{22}^{(1,1)}-\Sigma_{21}^{(1,0)}\Sigma_{11}^{-1}\Sigma_{12}^{(0,1)}| = |\Sigma_{11}||S^{-1}| = \frac{|\Sigma_{11}|}{|S|} .
\label{eq:det_Sigma_tilde}
\end{equation}
Putting  \eqref{eq:n_tilde}, \eqref{eq:Y_m_mu_T_Sigma_Y_m_mu_tilde} and~\eqref{eq:det_Sigma_tilde} into  \eqref{eq:double_m_log_prob}, we have
\begin{equation}
- \log P(\tilde{Y}|\tilde{X}) = \frac{1}{2} \Big[Q + \log |\Sigma_{11}| - \log |S| + (n_1+n_2) \log (2\pi)\Big] .
\label{eq:double_m_log_prob_2}
\end{equation}

\subsubsection{Prediction from multi-task GPR}

The observed data of the main function $Y_1=Y(X_1)$ and its derivative $Y'_2=Y'(X_2)$ and the predicted values of the function and its derivatives can be represented by a joint-multivariate Gaussian distribution:
\begin{equation}
\begin{bmatrix} 
Y_1 \\
Y'_2 \\
F_* \\
F'_* \\
F''_*
\end{bmatrix}
\sim \mathcal{N} \left(
\begin{bmatrix} 
\mu_1 \\
\mu'_2 \\
\mu_* \\
\mu'_* \\
\mu''_*
\end{bmatrix},
\begin{bmatrix} 
\Sigma_{11} & \Sigma_{12}^{(0,1)} & K_{\rm 1*} & K_{\rm 1*}^{(0,1)} & K_{\rm 1*}^{(0,2)} \\
\Sigma_{21}^{(1,0)} & \Sigma_{22}^{(1,1)} & K_{\rm 2*}^{(1,0)} & K_{\rm 2*}^{(1,1)} & K_{\rm 2*}^{(1,2)} \\
K_{\rm *1} & K_{\rm *2}^{(0,1)} & K_{**} & K^{(0,1)}_{**} & K^{(0,2)}_{**} \\
K_{\rm *1}^{(1,0)} & K_{\rm *2}^{(1,1)} & K^{(1,0)}_{**} & K^{(1,1)}_{**} & K^{(1,2)}_{**} \\
K_{\rm *1}^{(2,0)} & K_{\rm *2}^{(2,1)} & K^{(2,0)}_{**} & K^{(2,1)}_{**} & K^{(2,2)}_{**}
\end{bmatrix}
\right)
.
\label{eq:joint_distribution_GPR_multi}
\end{equation}
Here the notations are standard for all the vectors and matrices and these are given in Appendix~\ref{sec-GPR_short_notations}. This distribution can be rewritten in a form equivalent to the form in  \eqref{eq:joint_distribution_GPR_single} using definitions of $\tilde{Y}_1$, $\tilde{\mu}_1$, and $\tilde{\Sigma}_{11}$ from \eqref{eq:tilde_Y}--\eqref{eq:tilde_Sigma}:
\begin{equation}
\begin{bmatrix} 
\tilde{Y}_1 \\
F_* \\
F'_* \\
F''_*
\end{bmatrix}
\sim \mathcal{N} \left(
\begin{bmatrix} 
\tilde{\mu}_1 \\
\mu_* \\
\mu'_* \\
\mu''_*
\end{bmatrix},
\begin{bmatrix} 
\tilde{\Sigma}_{11} & \tilde{K}_{1*} & \tilde{K}_{\rm 1*}^{(0,1)} & \tilde{K}_{\rm 1*}^{(0,2)} \\
\tilde{K}_{\rm *1} & K_{**} & K^{(0,1)}_{**} & K^{(0,2)}_{**} \\
\tilde{K}_{\rm *1}^{(1,0)} & K^{(1,0)}_{**} & K^{(1,1)}_{**} & K^{(1,2)}_{**} \\
\tilde{K}_{\rm *1}^{(2,0)} & K^{(2,0)}_{**} & K^{(2,1)}_{**} & K^{(2,2)}_{**}
\end{bmatrix}
\right) .
\label{eq:joint_distribution_GPR_multi_2}
\end{equation}
Here 
\begin{align}
\label{eq:K_tilde_star}
\tilde{K}_{1*} &=
\begin{bmatrix} 
K_{1*} \\
K_{2*}^{(1,0)}
\end{bmatrix}
, \\
\label{eq:K_tilde_star_p}
\tilde{K}_{1*}^{(0,1)} &=
\begin{bmatrix} 
K_{1*}^{(0,1)} \\
K_{2*}^{(1,1)}
\end{bmatrix}
, \\
\label{eq:K_tilde_star_p_p}
\tilde{K}_{1*}^{(0,2)} &=
\begin{bmatrix} 
K_{1*}^{(0,2)} \\
K_{2*}^{(1,2)}
\end{bmatrix}
.
\end{align}
Similarly, 
\begin{align}
\label{eq:K_tilde_star_T}
\tilde{K}_{*1} &=
\begin{bmatrix} 
K_{*1} & K_{*2}^{(0,1)}
\end{bmatrix}
= 
\begin{bmatrix} 
K_{*1}^{\rm T} \\
\left(K_{*2}^{(0,1)}\right)^{\rm T}
\end{bmatrix}
^{\rm T} =
\begin{bmatrix} 
K_{1*} \\
K_{2*}^{(1,0)}
\end{bmatrix}
^{\rm T} =
\tilde{K}_{1*}^{\rm T} , \\
\label{eq:K_tilde_star_p_T}
\tilde{K}_{*1}^{(1,0)} &=
\begin{bmatrix} 
K_{*1}^{(1,0)} & K_{*2}^{(1,1)}
\end{bmatrix}
= 
\begin{bmatrix} 
\left(K_{*1}^{(1,0)}\right)^{\rm T} \\
\left(K_{*2}^{(1,1)}\right)^{\rm T}
\end{bmatrix}
^{\rm T}
= 
\begin{bmatrix} 
K_{1*}^{(0,1)} \\
K_{2*}^{(1,1)}
\end{bmatrix}
^{\rm T} = \left(\tilde{K}_{1*}^{(0,1)}\right)^{\rm T} , \\
\label{eq:K_tilde_star_p_p_T}
\tilde{K}_{*1}^{(2,0)} &=
\begin{bmatrix} 
K_{*1}^{(2,0)} & K_{*2}^{(2,1)}
\end{bmatrix}
= 
\begin{bmatrix} 
\left(K_{*1}^{(2,0)}\right)^{\rm T} \\
\left(K_{*2}^{(2,1)}\right)^{\rm T}
\end{bmatrix}
^{\rm T}
= 
\begin{bmatrix} 
K_{1*}^{(0,2)} \\
K_{2*}^{(1,2)}
\end{bmatrix}
^{\rm T}
= \left(\tilde{K}_{1*}^{(0,2)}\right)^{\rm T} .
\end{align}

Comparing the same structure between  \eqref{eq:joint_distribution_GPR_single} and  \eqref{eq:joint_distribution_GPR_multi_2}, we find predictions with the similar structure. For the mean values of the functions and their derivatives:
\begin{align}
\label{eq:predict_f_multi}
\bar{F}_* &= \mu_*+\tilde{K}_{*1} \tilde{\Sigma}_{11}^{-1} (\tilde{Y}_1-\tilde{\mu}_1) = \mu_*+K_{*1}B_1+K_{*2}^{(0,1)}B_2 , \\
\label{eq:predict_fp_multi}
\bar{F}'_* &= \mu'_*+\tilde{K}_{*1}^{(1,0)} \tilde{\Sigma}_{11}^{-1} (\tilde{Y}_1-\tilde{\mu}_1) = \mu'_*+K_{*1}^{(1,0)}B_1+K_{*2}^{(1,1)}B_2 , \\
\label{eq:predict_fpp_multi}
\bar{F}''_* &= \mu''_*+\tilde{K}_{*1}^{(2,0)} \tilde{\Sigma}_{11}^{-1} (\tilde{Y}_1-\tilde{\mu}_1) = \mu''_*+K_{*1}^{(2,0)}B_1+K_{*2}^{(2,1)}B_2 .
\end{align}
These equations can be written  as
\begin{equation}
\bar{F}_*^{(u)} = \mu_*^{(u)}+K_{\rm *1}^{(u,0)}B_1+K_{*2}^{(u,1)}B_2 ,
\label{eq:multi_predict_general_mean}
\end{equation}
which can be used for the prediction of mean values of the third and higher-order derivatives too. Similarly, we get the prediction for the self-covariances
\begin{align}
\label{eq:predict_V_f_f_multi}
&{\rm Cov}[F_*,F_*] = K_{**}-\tilde{K}_{*1} \tilde{\Sigma}_{11}^{-1} \tilde{K}_{1*} , \nonumber\\
&~~~~~= K_{**}-K_{*1}\left(A_{11}K_{1*}+A_{12}K_{2*}^{(1,0)}\right)-K_{*2}^{(0,1)}\left(A_{21}K_{1*}+A_{22}K_{2*}^{(1,0)}\right) , \\
\label{eq:predict_V_fp_fp_multi}
&{\rm Cov}[F'_*,F'_*] = K^{(1,1)}_{**}-\tilde{K}_{*1}^{(1,0)} \tilde{\Sigma}_{11}^{-1} \tilde{K}_{1*}^{(0,1)} , \nonumber\\
&~~~~~= K^{(1,1)}_{**}-K_{*1}^{(1,0)}\left(A_{11}K_{1*}^{(0,1)}+A_{12}K_{2*}^{(1,1)}\right)-K_{*2}^{(1,1)}\left(A_{21}K_{1*}^{(0,1)}+A_{22}K_{2*}^{(1,1)}\right) , \\
\label{eq:predict_V_fpp_fpp_multi}
&{\rm Cov}[F''_*,F''_*] = K^{(2,2)}_{**}-\tilde{K}_{*1}^{(2,0)} \tilde{\Sigma}_{11}^{-1} \tilde{K}_{1*}^{(0,2)} \nonumber\\
&~~~~~= K^{(2,2)}_{**}-K_{*1}^{(2,0)}\left(A_{11}K_{1*}^{(0,2)}+A_{12}K_{2*}^{(1,2)}\right)-K_{*2}^{(2,1)}\left(A_{21}K_{1*}^{(0,2)}+A_{22}K_{2*}^{(1,2)}\right) ,
\end{align}
and for the cross covariances,
\begin{align}
\label{eq:predict_C_f_fp_multi}
&{\rm Cov}[F_*,F'_*] = {\rm Cov}[F'_*,F_*]^{\rm T} = K^{(0,1)}_{**}-\tilde{K}_{*1} \tilde{\Sigma}_{11}^{-1} \tilde{K}_{1*}^{(0,1)} , \nonumber\\
&~~~~~= K^{(0,1)}_{**}-K_{*1}\left(A_{11}K_{1*}^{(0,1)}+A_{12}K_{2*}^{(1,1)}\right)-K_{*2}^{(0,1)}\left(A_{21}K_{1*}^{(0,1)}+A_{22}K_{2*}^{(1,1)}\right) , \\
\label{eq:predict_C_f_fpp_multi}
&{\rm Cov}[F_*,F''_*] = {\rm Cov}[F''_*,F_*]^{\rm T} = K^{(0,2)}_{**}-\tilde{K}_{*1} \tilde{\Sigma}_{11}^{-1} \tilde{K}_{1*}^{(0,2)} , \nonumber\\
&~~~~~= K^{(0,2)}_{**}-K_{*1}\left(A_{11}K_{1*}^{(0,2)}+A_{12}K_{2*}^{(1,2)}\right)-K_{*2}^{(0,1)}\left(A_{21}K_{1*}^{(0,2)}+A_{22}K_{2*}^{(1,2)}\right) , \\
\label{eq:predict_C_fp_fpp_multi}
&{\rm Cov}[F'_*,F''_*] = {\rm Cov}[F''_*,F'_*]^{\rm T} = K^{(1,2)}_{**}-\tilde{K}_{*1}^{(1,0)} \tilde{\Sigma}_{11}^{-1} \tilde{K}_{1*}^{(0,2)} \nonumber\\
&~~~~~= K^{(1,2)}_{**}-K_{*1}^{(1,0)}\left(A_{11}K_{1*}^{(0,2)}+A_{12}K_{2*}^{(1,2)}\right)-K_{*2}^{(1,1)}\left(A_{21}K_{1*}^{(0,2)}+A_{22}K_{2*}^{(1,2)}\right) .
\end{align}
The above six equations can be written in the form \begin{eqnarray}
{\rm Cov}[F_*^{(u)},F_*^{(v)}] &= K^{(u,v)}_{**}-K_{\rm *1}^{(u,0)} \left(A_{11}K_{1*}^{(0,v)}+A_{12}K_{2*}^{(1,v)}\right)
\nonumber\\&\qquad\qquad{}
-K_{*2}^{(u,1)}\left(A_{21}K_{1*}^{(0,v)}+A_{22}K_{2*}^{(1,v)}\right) .
\label{eq:multi_predict_general_cov}
\end{eqnarray}
This general expression can also be used for third and higher-order derivatives.

\subsection{Recovering single-task GPR from multi-task GPR}

We can recover the single-task GPR results if we do not have any information from the derivative of the main function $Y'_2=Y'(X_2)$. This can either be done by removing all the quantities which are made up of data in the first derivative of the function or it can done by quantifying the standard deviation in the derivative of the function corresponding to infinite observational errors, i.e.,
\begin{equation}
\text{each diagonal element of}~C_{22}^{(1,1)} \rightarrow \infty , \\
\label{eq:multi_to_single_condition}
\end{equation}
Using this in \eqref{eq:A22}, we get the zero matrix:
\begin{equation}
S = \{0\} \,.
\label{eq:main_condition}
\end{equation}
From \eqref{eq:A11}--\eqref{eq:B2},
\begin{align}
A_{11}& = \Sigma_{11}^{-1}, ~~~~~ A_{12} = \{0\}, ~~~~~ A_{21} = \{0\}, ~~~~~ A_{22} = \{0\} ,
\label{eq:condition_set_1}\\
B_{1} &= \Sigma_{11}^{-1}(Y_1-\mu_1), ~~~~~ B_{2} = \{0\} .
\label{eq:condition_set_2}
\end{align}
Then \eqref{eq:Y_m_mu_T_Sigma_Y_m_mu_tilde} gives
\begin{equation}
Q = (Y-\mu)^{\rm T}\Sigma_{11}^{-1}(Y_1-\mu_1) .
\label{eq:condition_3}
\end{equation}
Using \eqref{eq:main_condition} and~\eqref{eq:condition_3} in \eqref{eq:double_m_log_prob_2}, the negative log marginal likelihood becomes
\begin{equation}
- \log P(\tilde{Y}|\tilde{X}) \rightarrow - \log P(Y_1|X_1) + \left[ \infty + \frac{n_2}{2} \log (2\pi) \right] ,
\label{eq:n_log_mrg_lk_m_s}
\end{equation}
showing that no constraints come from the derivative information.

Now we can see that if we use \eqref{eq:condition_set_2} in  \eqref{eq:predict_f_multi}--\eqref{eq:multi_predict_general_mean}, we recover the standard results of  \eqref{eq:predict_f}--\eqref{eq:single_predict_general_mean}, where there is no information on the derivative of the function.
Similarly, putting conditions of \eqref{eq:condition_set_1} in \eqref{eq:predict_V_f_f_multi}--\eqref{eq:multi_predict_general_cov}, we recover the standard results  \eqref{eq:predict_V_f_f}--\eqref{eq:single_predict_general_cov}.

\section{Notation used in the GPR analysis}
\label{sec-GPR_short_notations}

Here we list all the short notations which were used in the GPR analysis in the main text. The short notations for vectors are as follows:
\begin{align}
X_1 &=
\begin{bmatrix} 
x_1^1 \\
x_1^2 \\
. \\
. \\
. \\
x_1^{n_1}
\end{bmatrix}
, ~ Y_1 = Y(X_1) =
\begin{bmatrix} 
y_1^1 = y(x_1^1) \\
y_1^2 = y(x_1^2) \\
. \\
. \\
. \\
y_1^{n_1} = y(x_1^{n_1})
\end{bmatrix}
, ~ X_2 =
\begin{bmatrix} 
x_2^1 \\
x_2^2 \\
. \\
. \\
. \\
x_2^{n_2}
\end{bmatrix}
, ~ Y'_2 = Y'(X_2) =
\begin{bmatrix} 
y'_2{}^1 = y'(x_2^1) \\
y'_2{}^2 = y'(x_2^2) \\
. \\
. \\
. \\
y'_2{}^{n_2} = y'(x_2^{n_2})
\end{bmatrix}
, \nonumber\\
\nonumber\\
X_* &=
\begin{bmatrix} 
x_*^1 \\
x_*^2 \\
. \\
. \\
. \\
x_*^m
\end{bmatrix}
, ~ F_* = Y(X_*) =
\begin{bmatrix} 
f_*^1 = y(x_*^1) \\
f_*^2 = y(x_*^2) \\
. \\
. \\
. \\
f_*^m = y(x_*^m)
\end{bmatrix}
, ~ F'_* = Y'(X_*) =
\begin{bmatrix} 
f'{}_*^1=y'(x_*^1) \\
f'{}_*^2=y'(x_*^2) \\
. \\
. \\
. \\
f'{}_*^m=y'(x_*^m)
\end{bmatrix}
,
\label{eq:vectors_1}
\end{align}
where $m$ is the number of points, where we want the predictions from GPR for the function and its derivatives. Similar to the notations of vectors in  \eqref{eq:vectors_1}, the same rule applies for the notations of vectors $F''_*=Y''(X_*)$, $\bar{F}_*=\bar{Y}(X_*)$, $\bar{F}'_*=\bar{Y}'(X_*)$, and $\bar{F}''_*=\bar{Y}''(X_*)$ of dimension $m$.

The covariance matrix $C_{11}$ corresponding to the self covariances of the main function $Y_1=Y(X_1)$ has the matrix form given as
\begin{align}
\label{eq:data_self_cov_f}
C_{11} &= {\rm Cov} [Y_1,Y_1] = {\rm Cov} [Y(X_1),Y(X_1)] = \left\{ c_{11}^{ij} \right\} = \left\{ {\rm Cov}[y_1^i,y_1^j] \right\} \nonumber\\
&=
\begin{bmatrix} 
{\rm Cov}[y_1^1,y_1^1] & {\rm Cov}[y_1^1,y_1^2] & ... & {\rm Cov}[y_1^1,y_1^{n_1}] \\
{\rm Cov}[y_1^2,y_1^1] & {\rm Cov}[y_1^2,y_1^2] & ... & {\rm Cov}[y_1^2,y_1^{n_1}] \\
. & . & ... & . \\
. & . & ... & . \\
. & . & ... & . \\
{\rm Cov}[y_1^{n_1},y_1^1] & {\rm Cov}[y_1^{n_1},y_1^2] & ... & {\rm Cov}[y_1^{n_1},y_1^{n_1}]
\end{bmatrix}
,
\end{align}
where $i$ and $j$ run from $1$ to $n_1$. We shall also consider indices $p$ and $q$ which run from $1$ to $n_2$ and indices $r$ and $s$ which run from $1$ to $m$:
\begin{align}
\label{eq:indices}
i,j \in [1,2,...,n_1] , ~~
p,q  \in [1,2,...,n_2] , ~~
r,s  \in [1,2,...,m]  .
\end{align}

The covariance matrix $C_{12}^{(0,1)}$ corresponding to the cross covariances of the main function $Y_1=Y(X_1)$ and its derivative $Y'_2=Y'(X_2)$ has the matrix form 
\begin{align}
\label{eq:data_self_cov_f_2}
C_{12}^{(0,1)} &= {\rm Cov} [Y_1,Y'_2] = {\rm Cov} [Y(X_1),Y'(X_2)] = \left\{ c_{12}^{(0,1)ip} \right\} = \left\{ {\rm Cov}[y_1^i,y'_2{}^p] \right\} \nonumber\\
&=
\begin{bmatrix} 
{\rm Cov}[y_1^1,y'_2{}^1] & {\rm Cov}[y_1^1,y'_2{}^2] & ... & {\rm Cov}[y_1^1,y'_2{}^{n_2}] \\
{\rm Cov}[y_1^2,y'_2{}^1] & {\rm Cov}[y_1^2,y'_2{}^2] & ... & {\rm Cov}[y_1^2,y'_2{}^{n_2}] \\
. & . & ... & . \\
. & . & ... & . \\
. & . & ... & . \\
{\rm Cov}[y_1^{n_1},y'_2{}^1] & {\rm Cov}[y_1^{n_1},y'_2{}^2] & ... & {\rm Cov}[y_1^{n_1},y'_2{}^{n_2}]
\end{bmatrix}
= \left(C_{21}^{(1,0)}\right)^{\rm T} .
\end{align}

The covariance matrix $C_{22}^{(1,1)}$ corresponding to the self covariances of the derivative of the function $Y'_2=Y'(X_2)$ has the matrix form 
\begin{align}
\label{eq:data_self_cov_f_3}
C_{22}^{(1,1)} &= {\rm Cov} [Y'_2,Y'_2] = {\rm Cov} [Y'(X_2),Y'(X_2)] = \left\{ c_{22}^{(1,1)pq} \right\} = \left\{ {\rm Cov}[y'_2{}^p,y'_2{}^q] \right\} \nonumber\\
&=
\begin{bmatrix} 
{\rm Cov}[y'_2{}^1,y'_2{}^1] & {\rm Cov}[y'_2{}^1,y'_2{}^2] & ... & {\rm Cov}[y'_2{}^1,y'_2{}^{n_2}] \\
{\rm Cov}[y'_2{}^2,y'_2{}^1] & {\rm Cov}[y'_2{}^2,y'_2{}^2] & ... & {\rm Cov}[y'_2{}^2,y'_2{}^{n_2}] \\
. & . & ... & . \\
. & . & ... & . \\
. & . & ... & . \\
{\rm Cov}[y'_2{}^{n_2},y'_2{}^1] & {\rm Cov}[y'_2{}^{n_2},y'_2{}^2] & ... & {\rm Cov}[y'_2{}^{n_2},y'_2{}^{n_2}]
\end{bmatrix}
.
\end{align}

The short notations for the other matrices are 
\begin{align}
K_{11} &= K(X_1,X_1) = \Big\{ k(x_1^i,x_1^j) \Big\} =
\begin{bmatrix} 
k(x_1^1,x_1^1) & k(x_1^1,x_1^2) & ... & k(x_1^1,x_1^{n_1}) \\
k(x_1^2,x_1^1) & k(x_1^2,x_1^2) & ... & k(x_1^2,x_1^{n_1}) \\
. & . & ... & . \\
. & . & ... & . \\
. & . & ... & . \\
k(x_1^{n_1},x_1^1) & k(x_1^{n_1},x_1^2) & ... & k(x_1^{n_1},x_1^{n_1}) \\
\end{bmatrix}
,
\label{eq:K_matrices_11}
\\
K_{1*} &= K(X_1,X_*) = \Big\{ k(x_1^i,x_*^r) \Big\} =
\begin{bmatrix} 
k(x_1^1,x_*^1) & k(x_1^1,x_*^2) & ... & k(x_1^1,x_*^m) \\
k(x_1^2,x_*^1) & k(x_1^2,x_*^2) & ... & k(x_1^2,x_*^m) \\
. & . & ... & . \\
. & . & ... & . \\
. & . & ... & . \\
k(x_1^{n_1},x_*^1) & k(x_1^{n_1},x_*^2) & ... & k(x_1^{n_1},x_*^m) \\
\end{bmatrix}
= K_{*1}^{\rm T} .
\label{eq:K_matrices_12}
\end{align}

The same matrix notations are applied to other matrices with the same rules. Some relevant matrix notations are listed below without showing the actual matrix:
\begin{align}
\label{eq:K_matrices_2}
K_{**} &= K(X_*,X_*) = \left\{ k(x_*^r,x_*^s) \right\} , \\
K_{11}^{(u,v)} &= K^{(u,v)}(X_1,X_1) = \left\{ k^{(u,v)}(x_1^i,x_1^j) \right\} = \left\{ \dfrac{\partial^{(u+v)} k(x_1^i,x_1^j)}{\partial x_1^i{}^u \partial x_1^j{}^v} \right\} , \\
K_{1*}^{(u,v)} &= K^{(u,v)}(X_1,X_*) = \left\{ k^{(u,v)}(x_1^i,x^r_*) \right\} = \left\{ \dfrac{\partial^{(u+v)} k(x_1^i,x^r_*)}{\partial x_1^i{}^u \partial x^r_*{}^v} \right\} , \\
K_{**}^{(u,v)} &= K^{(u,v)}(X_*,X_*) = \left\{ k^{(u,v)}(x_*^r,x_*^s) \right\} = \left\{ \dfrac{\partial^{(u+v)} k(x_*^r,x_*^s)}{\partial x_*^r{}^u \partial x_*^s{}^v} \right\} ,
\end{align}
and so on, where $u$ and $v$ are the orders of the differentiation of the kernel with respect to the first and second arguments.


\section{Application of single-task GPR to PantheonPlus data and role of $M_B$} 
\label{sec-pantheon_plus_data_related}

\begin{figure*}
\centering
\includegraphics[height=170pt,width=0.49\textwidth]{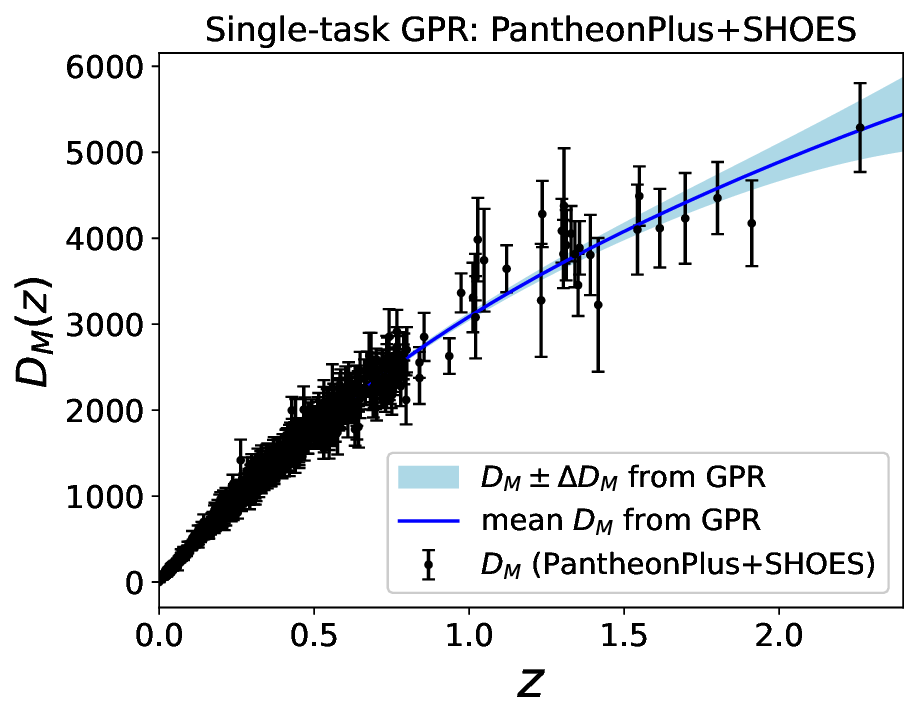}
\includegraphics[height=170pt,width=0.49\textwidth]{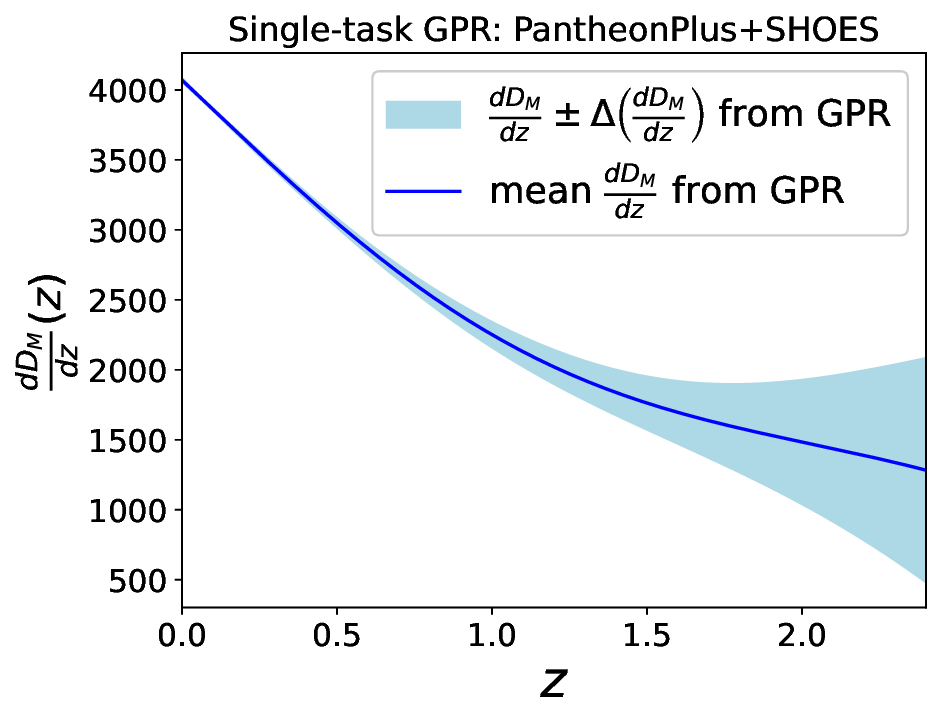} \\
\includegraphics[height=170pt,width=0.49\textwidth]{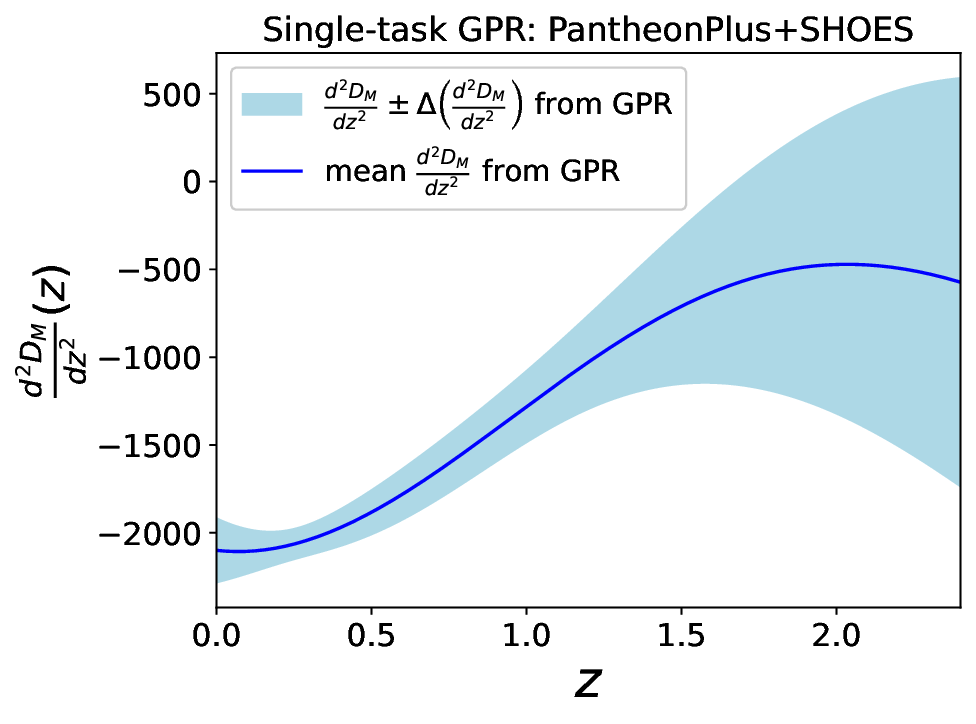}
\includegraphics[height=170pt,width=0.49\textwidth]{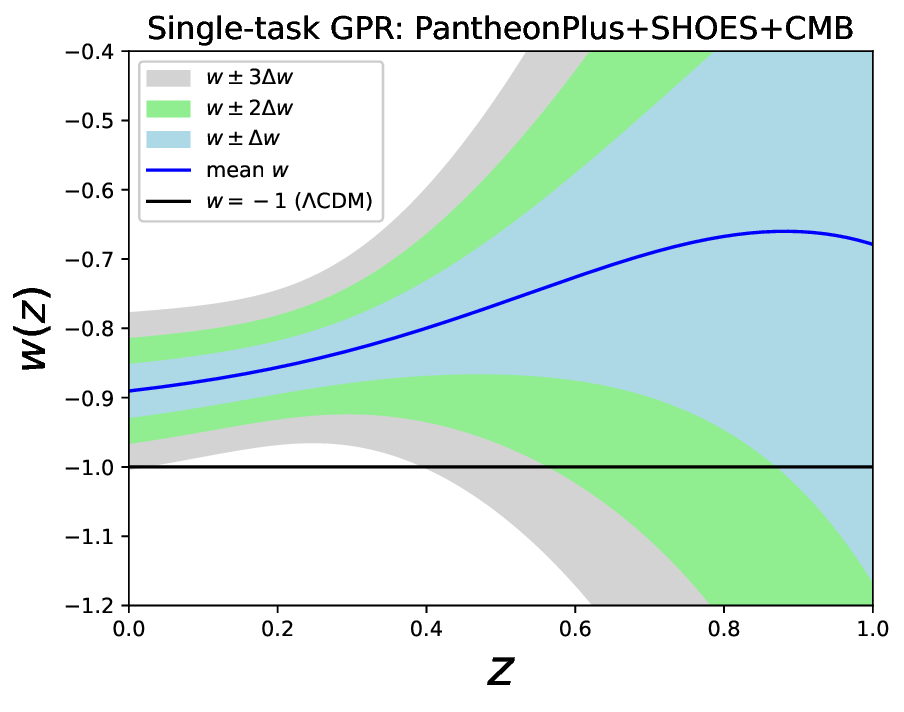}
\caption{
\label{fig:all_pp_shoes}
Black error bars in the top left panel give $D_M \pm \Delta D_M$ obtained from observed $\mu_B$ of PantheonPlus+SHOES, using \eqref{eq:lum_dist_SN}. Solid blue lines and blue shadings in the top left, top right, and bottom left panels correspond to the reconstructed $D_M$, $D'_M$, and $D''_M$, with associated 1$\sigma$ confidence regions, using single-task GPR. The bottom right panel shows the reconstructed $w$ and associated confidence regions for CMB+PantheonPlus+SHOES (single-task GPR) using  \eqref{eq:w_DE_3} and~\eqref{eq:beta_PL18_standard}.
}
\end{figure*}

\subsection{PantheonPlus+SHOES and CMB+PantheonPlus+SHOES}

The PantheonPlus dataset \eqref{pplus} provides apparent magnitude $m_B$ and also data on distance modulus $\mu_B$, obtained from the calibration with local Cepheid measurements from SHOES observations. When we consider $m_B$ data, we are using only PantheonPlus data. On the other hand, when we consider $\mu_B$ data,  we are using PantheonPlus and SHOES data together. 

We first consider the $\mu_B$ data and the associated errors. For this, we consider the full covariance coming from both statistics and systematics. Then we compute $D_M$ using \eqref{eq:lum_dist_SN} and the corresponding error by propagation of uncertainty. These are shown in the top left panel of Fig.~\ref{fig:all_pp_shoes} with black error bars. We apply the GPR analysis to these $D_M$ data. Unlike BAO data, these data correspond to only one function $D_M$. So, in this case, single-task GPR is applicable. Applying single-task GPR, we compute predictions for mean values of $D_M$, its first derivative, and its second derivative using  \eqref{eq:predict_f}--\eqref{eq:predict_fpp}. We also compute errors  using  \eqref{eq:predict_V_f_f}--\eqref{eq:predict_V_fpp_fpp}.  The predicted $D_M$ and the associated 1$\sigma$ confidence interval are shown in the top left panel of Fig.~\ref{fig:all_pp_shoes} as a blue line and blue shading.

\begin{figure*}
\centering
\includegraphics[height=170pt,width=0.49\textwidth]{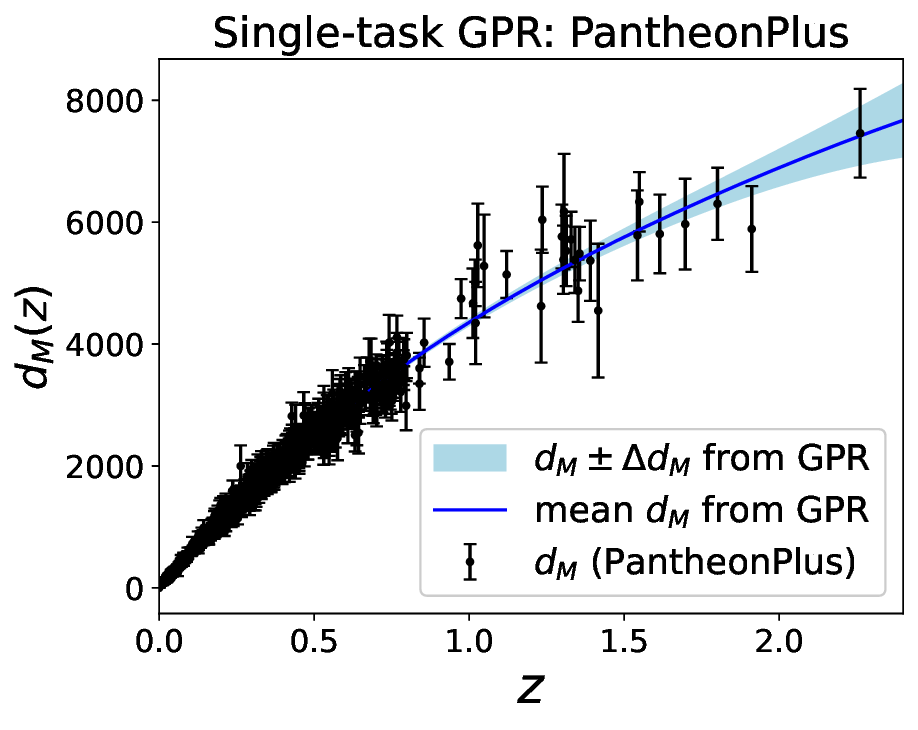}
\includegraphics[height=170pt,width=0.49\textwidth]{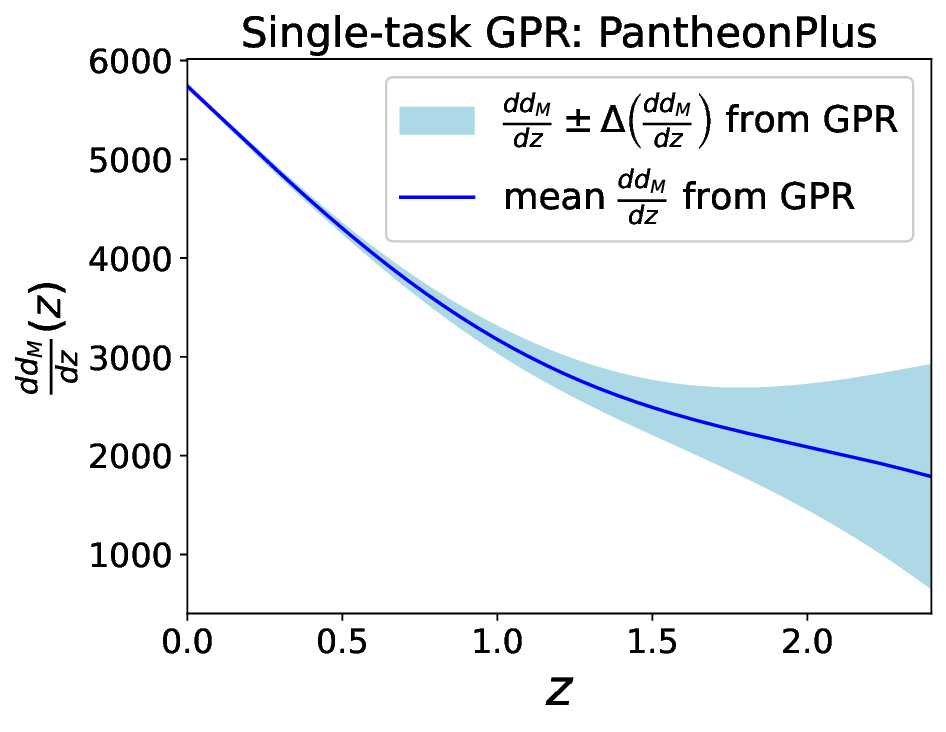} \\
\includegraphics[height=170pt,width=0.49\textwidth]{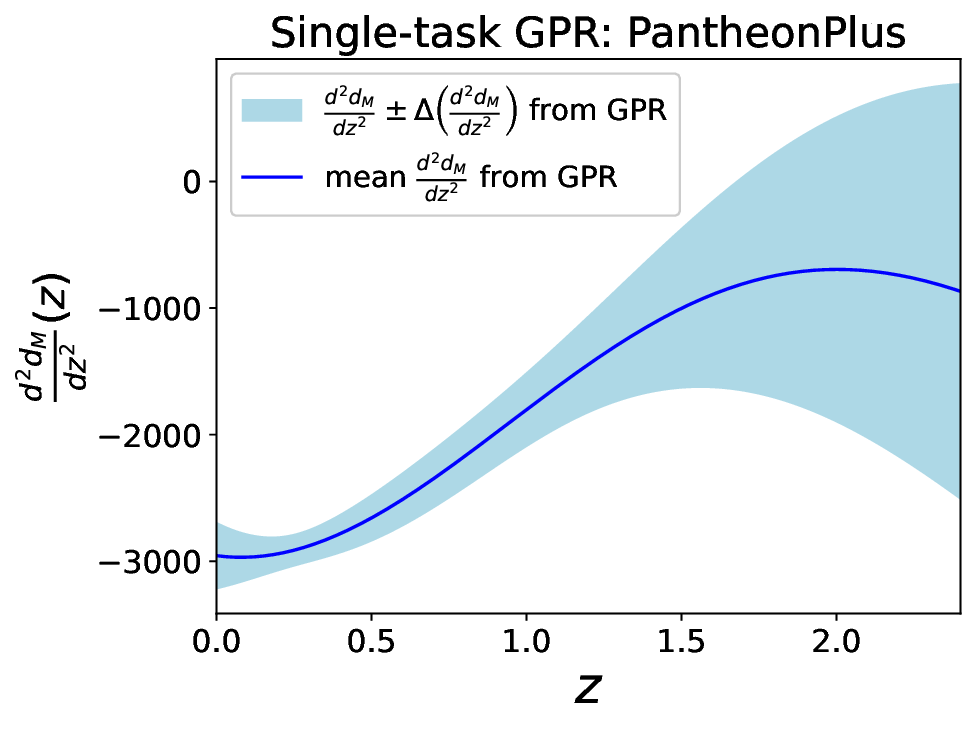}
\includegraphics[height=170pt,width=0.49\textwidth]{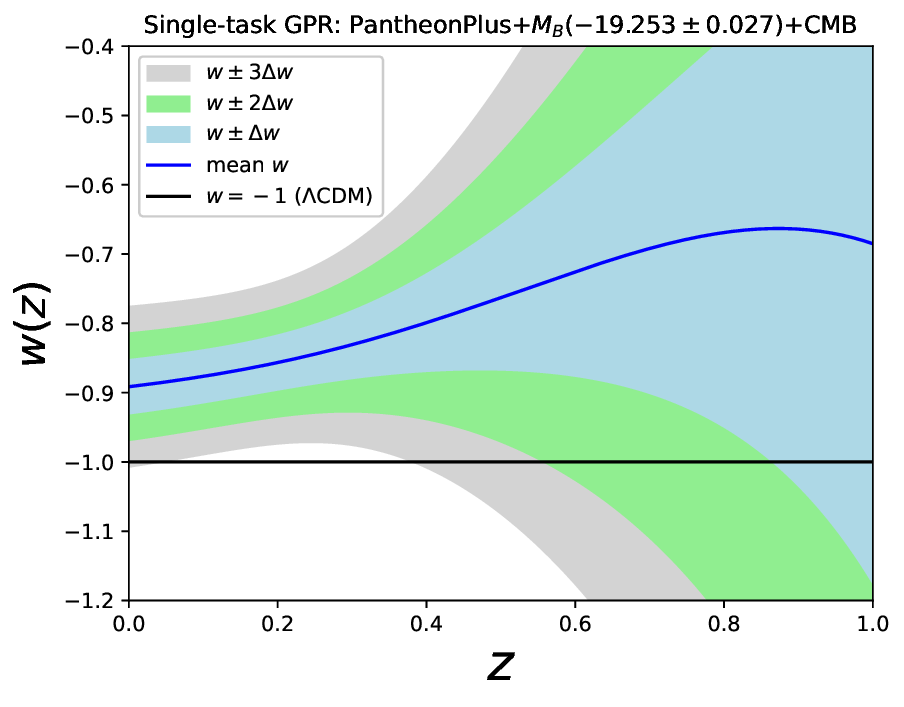} \\
\includegraphics[height=170pt,width=0.49\textwidth]{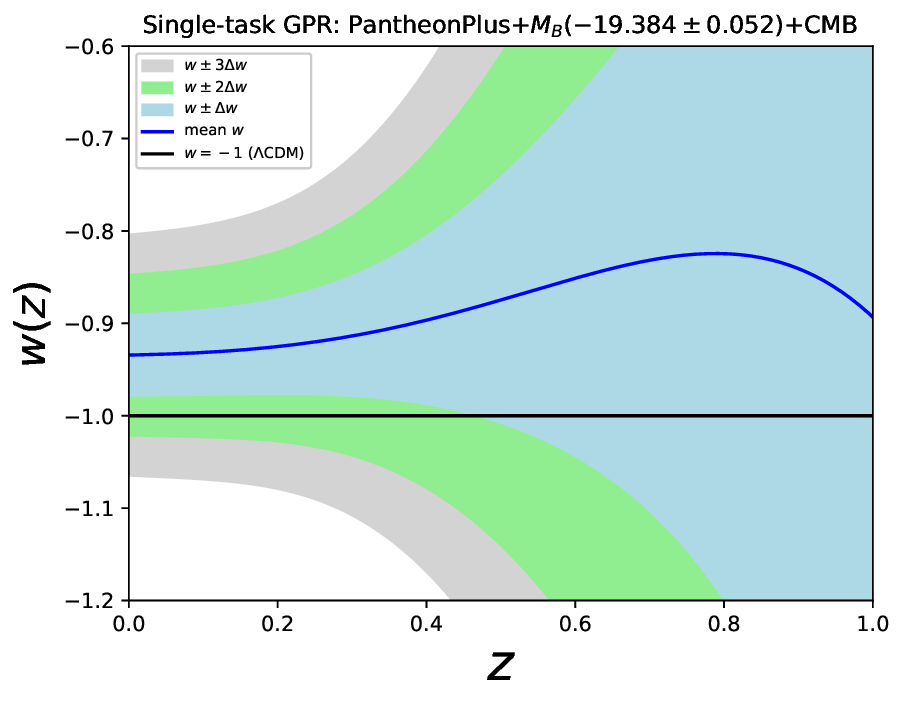}
\includegraphics[height=170pt,width=0.49\textwidth]{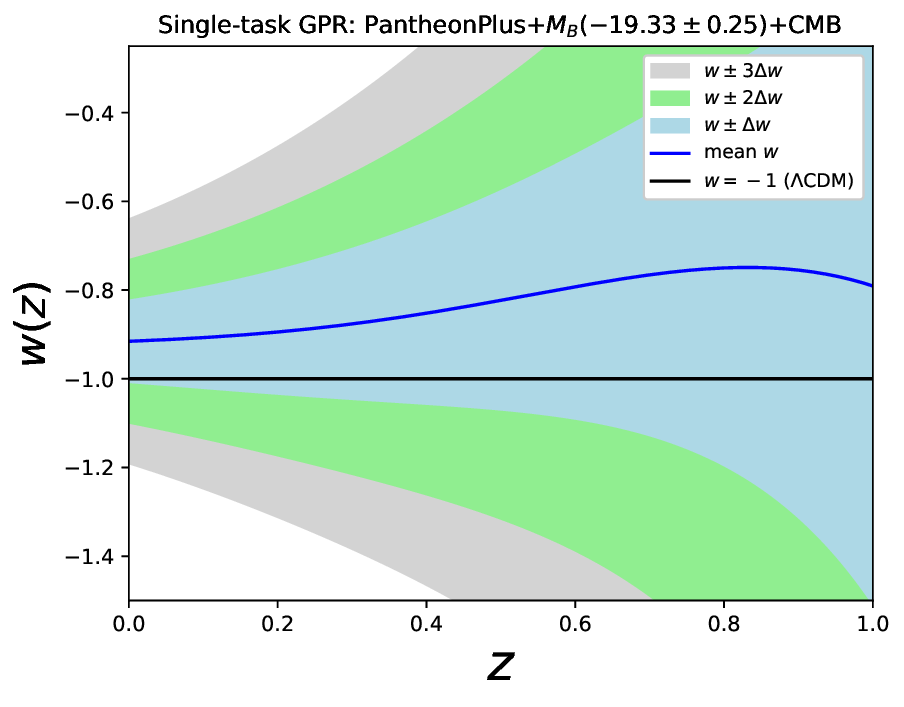}
\caption{
\label{fig:CMB_SNPP_MB_w}
The top panels and the middle left panel show the reconstructed $d_M$, $d'_M$, and $d''_M$  and associated 1$\sigma$ confidence regions respectively using single-task GPR.
Black error bars in the top left panel are from the observed $m_B$ of PantheonPlus data, using \eqref{eq:SN_DM_main}. The remaining three panels show the reconstructed $w$ and associated confidence regions for CMB+PantheonPlus (single-task GPR) with three different $M_B$ Gaussian priors \eqref{pr1}--\eqref{pr3},  using \eqref{eq:w_DE_6},~\eqref{eq:defn_delta} and~\eqref{eq:beta_PL18_standard}.
}
\end{figure*}

Similarly, we show $D'_M$ and $D''_M$, together with the associated 1$\sigma$ confidence intervals, in the top right and bottom left panels of Fig.~\ref{fig:all_pp_shoes}. We then compute cross covariances between $D'_M$ and $D''_M$ using  \eqref{eq:predict_C_fp_fpp}.

Using the reconstructed $D'_M$ and $D''_M$, their self-covariances and cross-covariances, we compute the redshift evolution of $w$ from  \eqref{eq:w_DE_3},  with the additional CMB information of $\beta$  in  \eqref{eq:beta_PL18_standard}. The reconstructed $w$ and its errors are displayed in the bottom right panel of Fig.~\ref{fig:all_pp_shoes}, with the same color codes as in Fig.~\ref{fig:CMB_DESI_w_reconstruction}. 
From this plot, we see that the $\Lambda$CDM model is more than 3$\sigma$ away in the redshift range $0\lesssim z\lesssim 0.4$,  for the CMB+PantheonPlus+SHOES combination of data sets. It is 3$\sigma$ to 1$\sigma$ away in the redshift range around $0.4\lesssim z\lesssim 0.9$. For $z\gtrsim 0.9$ it is well within the 1$\sigma$ limit. In most of the lower redshift range, the reconstructed $w$ is in the non-phantom region.

\subsection{PantheonPlus, PantheonPlus+$M_B$ and CMB+PantheonPlus+$M_B$}

It is not essential that we calibrate PantheonPlus data with the SHOES observations. To keep other options open, we combine PantheonPlus data with other data sets. Without calibration with SHOES, we cannot use $\mu_B$ data from PantheonPlus. For this case, we consider instead $m_B$ data. Here, also we consider full covariance including systematic errors from PantheonPlus data. From this $m_B$ data, we compute $d_M$ using \eqref{eq:SN_DM_main} and the associated error using propagation of uncertainty. The result is shown in the top left panel of Fig.~\ref{fig:CMB_SNPP_MB_w} with black error bars. We apply single-task GPR to the $d_M$ data in order to predict the smooth function of $d_M$ and its first and second derivatives, and the associated errors (the blue line and shading). We see that the predicted  $d_M(z)$ is consistent with the  $d_M$ data. The predicted $d'_M$ and $d''_M$ and the associated 1$\sigma$ confidence intervals are plotted in the top right and middle left panels of Fig.~\ref{fig:CMB_SNPP_MB_w}.

In order to reconstruct $w$ from the predicted $d'_M$ and $d''_M$ using  \eqref{eq:w_DE_6}, we need values of the parameter $\delta$ in  \eqref{eq:defn_delta}, which involves parameters $\beta$ and $M_B$. We use the $\beta$ value from  CMB data, as in \eqref{eq:beta_PL18_standard}.  Additional information on the $M_B$ parameter is needed. For this, we use three significantly different Gaussian priors on $M_B$ \citep{Chen:2024gnu}:
\begin{align}\label{pr1}
& M_B~\text{Prior I:} ~~~~~ M_B=-19.253\pm0.027 ,\\
& M_B ~\text{Prior II:} ~~~~ M_B=-19.384\pm0.052 ,\\
& M_B ~\text{Prior III:} ~~~ M_B=-19.33~\,\pm0.25 \,. \label{pr3}
\end{align}
Prior I is from SHOES, used to cross-check the consistency with the previous result \citep{Riess:2021jrx}. Prior II is obtained from the calibration of the Pantheon sample with cosmic chronometer observations of the Hubble parameter \citep{Dinda:2022jih}. Prior III is from 18 SNIa \citep{Wang:1999bz} obtained using multicolor light curve shapes \citep{SupernovaSearchTeam:1998fmf}. 

With $M_B$ Prior I, we reconstruct $w$ and the associated error regions, shown in the middle right panel of Fig.~\ref{fig:CMB_SNPP_MB_w}.  Comparing with the bottom right panel of Fig.~\ref{fig:all_pp_shoes}, we see that the results are consistent and the same conclusion is applicable here. 

Using $M_B$ Prior II, we reconstruct $w$ and associated confidence regions, shown in the bottom left panel of Fig.~\ref{fig:CMB_SNPP_MB_w}. In this case, the $\Lambda$CDM model is 1.5$\sigma$ to 1$\sigma$ away in the redshift range $0\lesssim z\lesssim 0.45$ and well within 1$\sigma$ region in the redshift range $z\gtrsim 0.45$. 

The $M_B$ Prior III leads to the bottom right panel of Fig.~\ref{fig:CMB_SNPP_MB_w}. Here we see that $\Lambda$CDM  is within the 1$\sigma$ region in the entire redshift range. Note that, at lower redshifts, the reconstructed mean function of $w$ lies within the mean functions corresponding to $M_B$ Prior I and $M_B$ Prior II. This is because the mean value of $M_B$ in Prior III lies within the mean values of Priors I and II. The higher the $M_B$ prior value, the higher the mean value of $w$ in the non-phantom region. The confidence region is largest for $M_B$ Prior III since the error in $M_B$ is largest in this case.

\subsection{Extended list of reconstructed constants}
\label{sec-extended_constants}
\begingroup
\allowdisplaybreaks
\begin{align}
\label{eq:sn_MB}
 \text{SNIa+$M_B$:} ~~~ H_0 &= \frac{c\, {\rm e}^{b(20+M_B)}}{d_H(z=0)} , ~~~ M_B ~ (\text{provided}) , \\
\label{eq:sn_MB_cmb}
 \text{CMB+SNIa+$M_B$:} ~~~ H_0 &= \frac{c\, {\rm e}^{b(20+M_B)}}{d_H(z=0)} ~ (\text{same as for SNIa+$M_B$}) , ~~~ H_0 r_d = H_0 \times r_d , \nonumber\\
\Omega_{\rm m0} &= \frac{\omega_{\rm m0}}{h^2} , ~~~ M_B ~ (\text{provided}) .
\end{align}
\endgroup
\begin{table*}[!h]
\begin{center}
\begin{tabular}{|c|c|c|c|c|c|c|c|}
\hline &&&&\\
Data combination & $H_0r_d$ [100 km/s] & $H_0$ [km/s/Mpc] & $\Omega_{\rm m0}$ & $M_B$ [mag] \\
&&&&\\
\hline 
PP+$M_B$ Prior I & - & $73.72\pm0.95$ & - & $-19.253\pm0.027$ \\
\hline
PP+$M_B$ Prior II & - & $69.41\pm1.68$ & - & $-19.384\pm0.052$ \\
\hline
PP+$M_B$ Prior III & - & $71.16\pm8.20$ & - & $-19.33~\,\pm0.25$ \\
\hline
CMB+PP+$M_B$ Prior I & $108.40\pm1.42$ & $73.72\pm0.95$ & $0.263\pm0.007$ & $-19.253\pm0.027$ \\
\hline
CMB+PP+$M_B$ Prior II & $102.06\pm2.48$ & $69.41\pm1.68$ & $0.297\pm0.015$ & $-19.384\pm0.052$ \\
\hline
CMB+PP+$M_B$ Prior III & $104.63\pm12.05$ & $71.16\pm8.20$ & $0.282\pm0.065$ & $-19.33~\,\pm0.25$ \\
\hline
\end{tabular}
\end{center}
\caption{
Extended list of reconstructed values of $H_0r_d$, $H_0$, $\Omega_{\rm m0}$ and $M_B$.
}
\label{table:extended_constants}
\end{table*}
%

\section{Dependence on different CMB distance priors}
\label{sec-different_cmb_dist_prior}


To compute CMB distance priors we used  {\sf base$_{-}$plikHM$_{-}$TTTEEE$_{-}$lowl$_{-}$lowE$_{-}$lensing} from the Planck data archive with standard $\Lambda$CDM \cite{Zhai:2018vmm}. We denote the resulting CMB distance prior  CMB(P18 I). Here we check how the results change if we change the CMB distance prior \cite{Poulin:2024ken,Pedrotti:2024kpn}. For the second CMB distance prior, CMB(P18 II), we exclude the constraints from lensing, i.e. we use Planck 2018 TT,TE,EE+lowE, and we assume $w$CDM  instead of $\Lambda$CDM. The corresponding CMB distance priors use the chain {\sf base$_{-}$w$_{-}$plikHM$_{-}$TTTEEE$_{-}$lowl$_{-}$lowE} with the $w$CDM model \cite{Chen:2018dbv}. For the third CMB distance prior, CMB(P18 III), we generalize the early physics of $\Lambda$CDM  to the varying sum of neutrino masses and varying effective number of relativistic species, in order to get comparatively broader standard deviations in distance priors. For this, we use the chain {\sf base$_{-}$nnu$_{-}$mnu$_{-}$plikHM$_{-}$TT$_{-}$lowl$_{-}$lowE$_{-}$post$_{-}$lensing} \cite{Zhai:2019nad}.  We obtain the $\gamma$ values
\begin{align}
\label{eq:gamma_Pl18_I}
\gamma &= (343.8\pm 1.6) \times 10^{-6} \, \quad \text{CMB(P18 I)} , \\
\label{eq:gamma_Pl18_II}
\gamma &= (344.4\pm 1.9) \times 10^{-6} \, \quad \text{CMB(P18 II)} , \\
\label{eq:gamma_Pl18_III}
\gamma &= (342.5\pm 3.1) \times 10^{-6} \, \quad \text{CMB(P18 III)} .
\end{align}
\begin{figure*}
\centering
\centering
\includegraphics[height=180pt,width=0.8\textwidth]{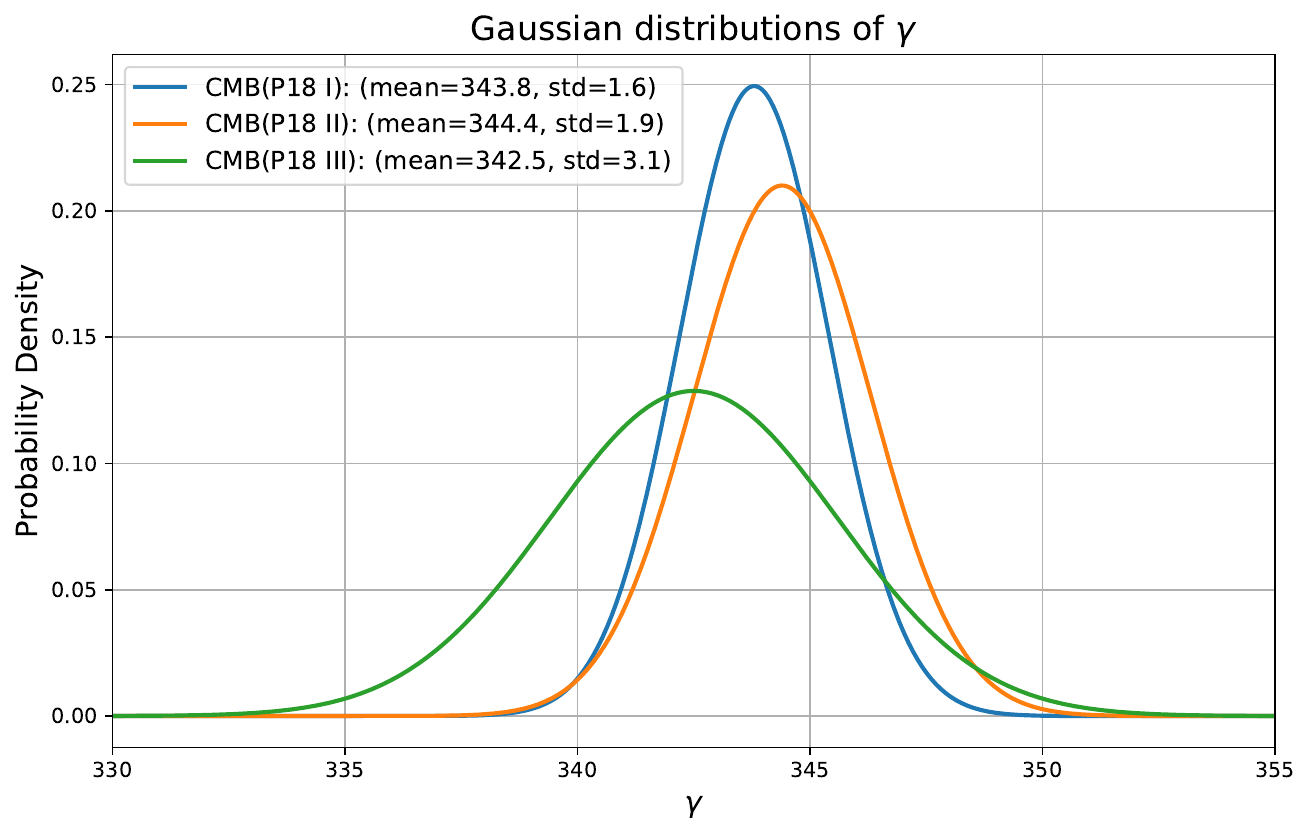}
\caption{
\label{fig:probability_gamma}
Probability distribution of $\gamma$ obtained from three CMB distance priors.
}
\end{figure*}
The mean values are slightly different in the three cases.
CMB(P18 II) has a slightly larger standard deviation than  CMB(P18 I), whereas CMB(P18 III) has a significantly larger standard deviation. Figure~\ref{fig:probability_gamma} shows the Gaussian distributions of $\gamma$.

\begin{figure*}
\centering
\centering
\includegraphics[height=130pt,width=0.32\textwidth]{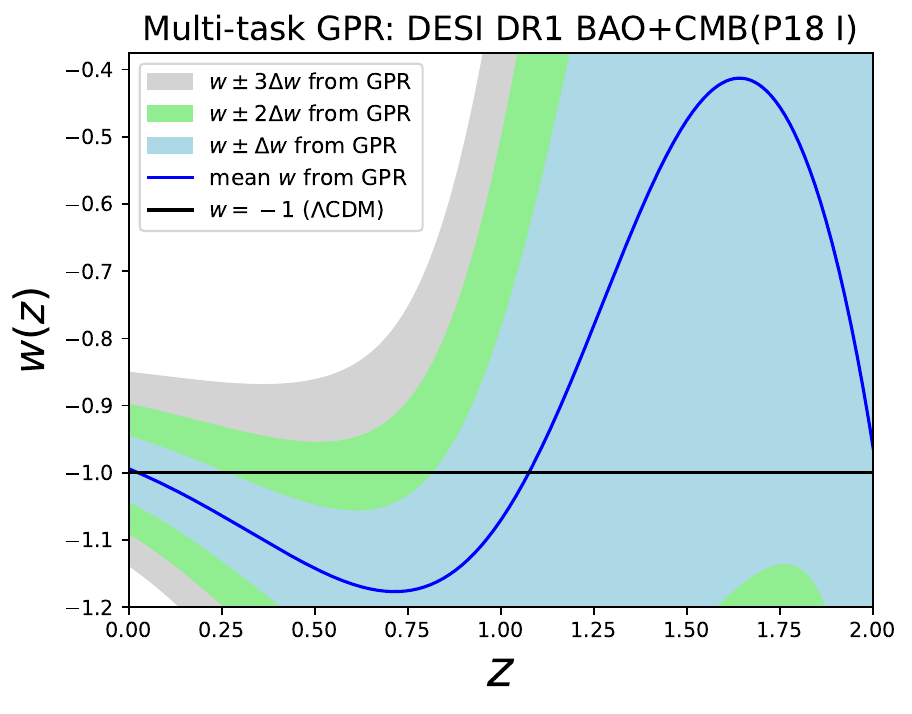}
\includegraphics[height=130pt,width=0.32\textwidth]{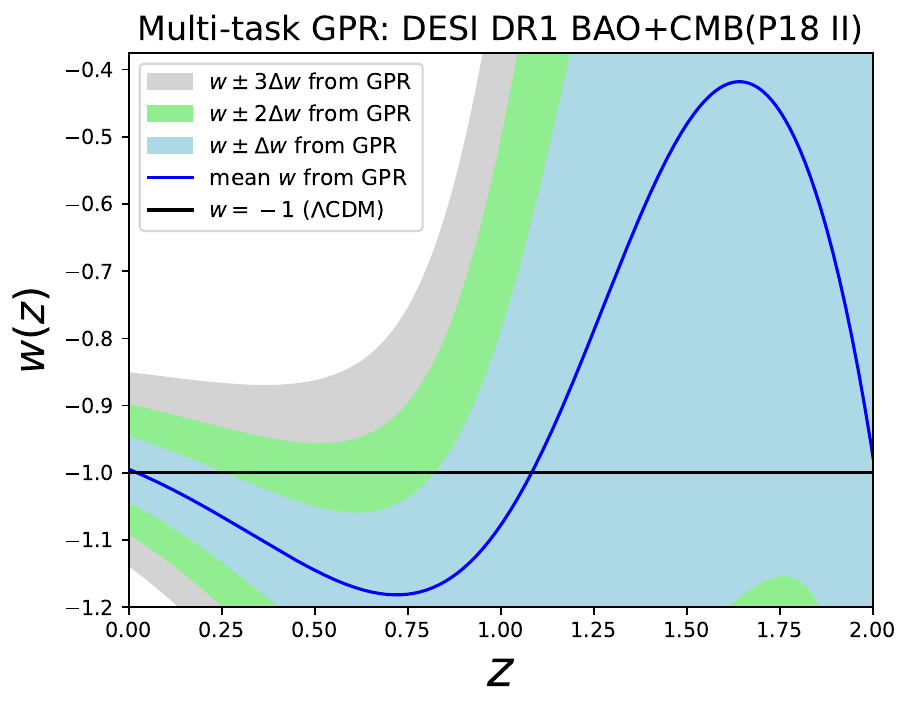}
\includegraphics[height=130pt,width=0.32\textwidth]{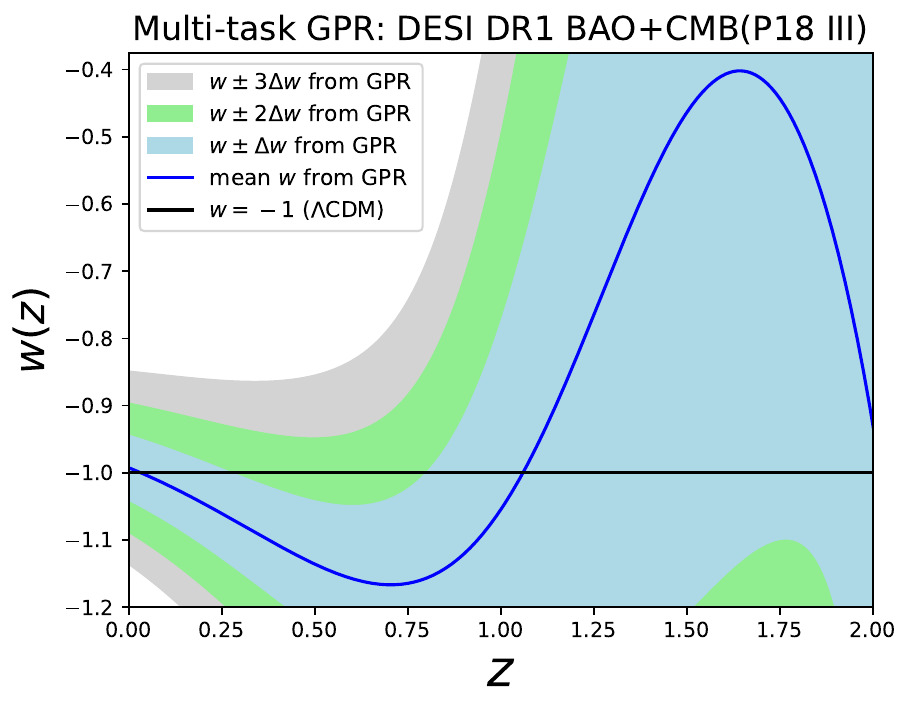}
\caption{
\label{fig:DESI_diff_cmb_priors_eos}
Dark energy equation of state for DESI DR1 BAO combined with three different CMB distance priors.
}
\end{figure*}

In Fig.~\ref{fig:DESI_diff_cmb_priors_eos}, we plot the dark energy equation of state for DESI DR1 BAO combined with three different CMB distance priors. It is evident that there are no significant differences in $w$ for the three different $\gamma$.

\begin{figure*}
\centering
\centering
\includegraphics[height=130pt,width=0.32\textwidth]{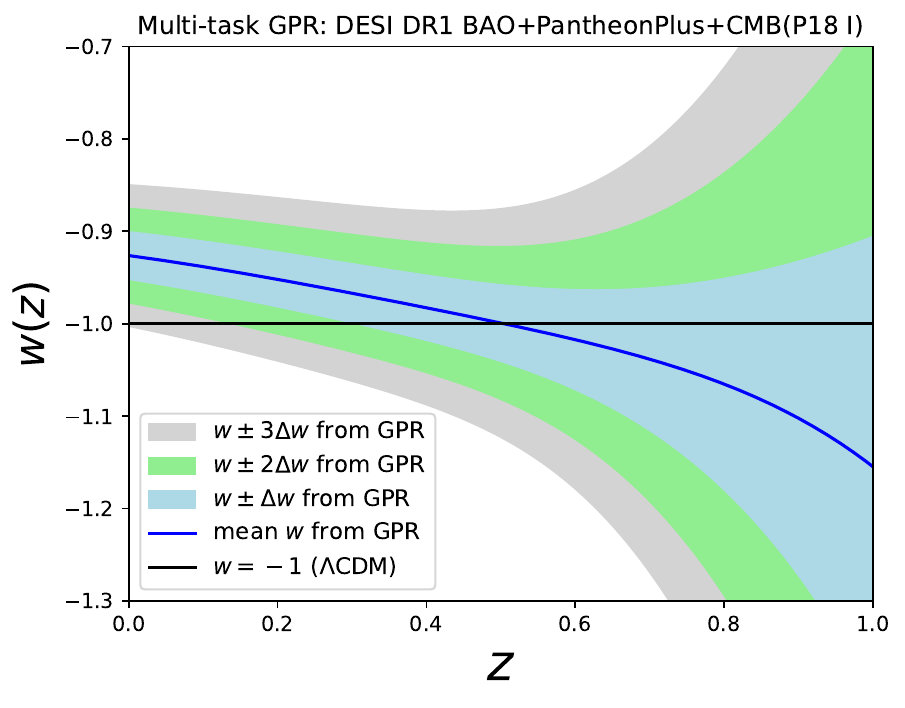}
\includegraphics[height=130pt,width=0.32\textwidth]{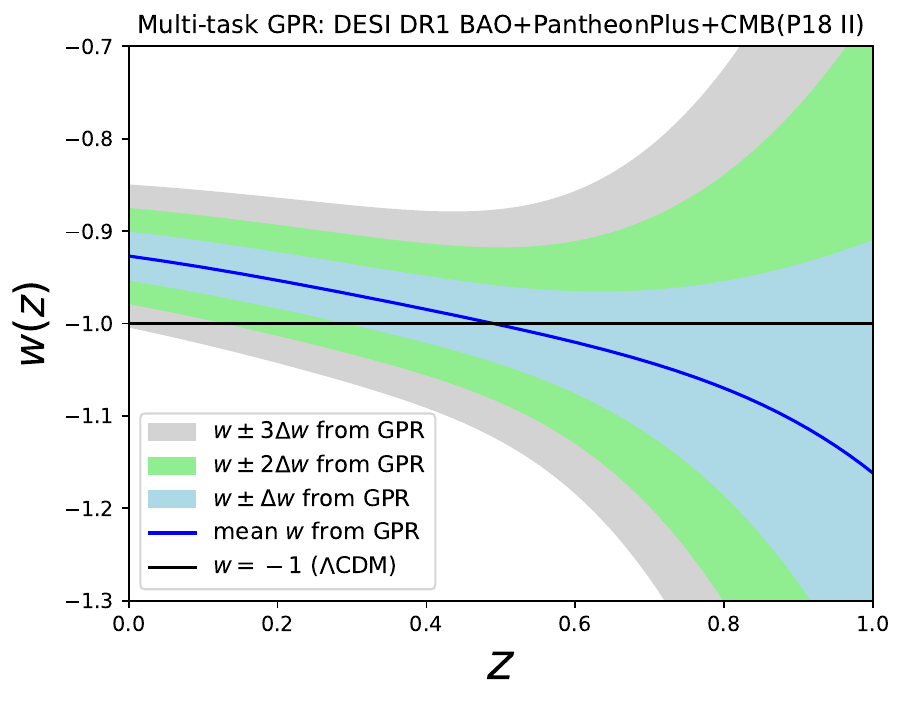}
\includegraphics[height=130pt,width=0.32\textwidth]{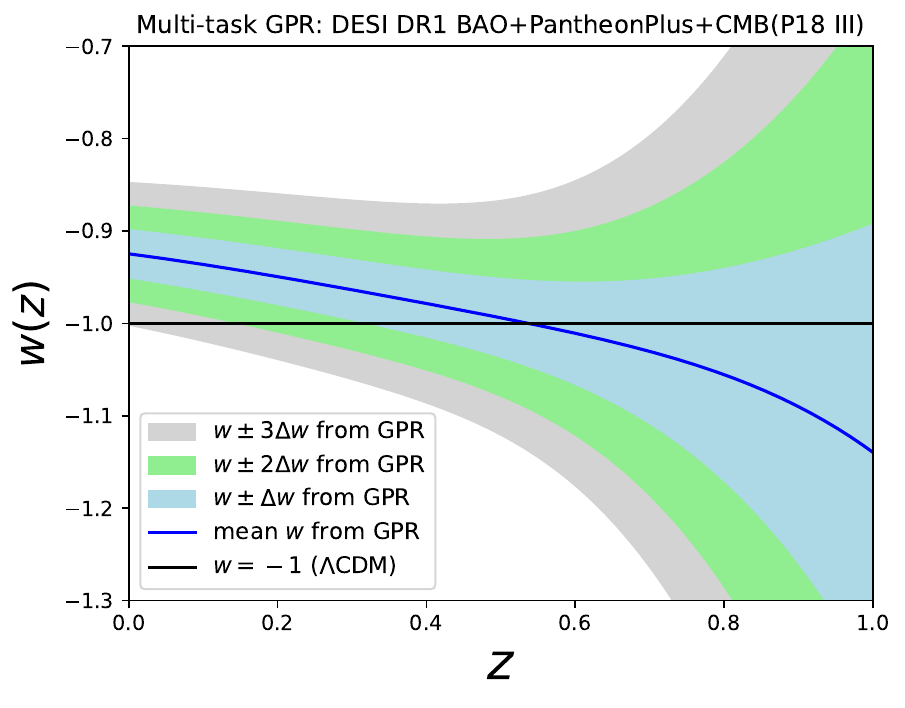}
\caption{
\label{fig:DESI_PP_diff_cmb_priors_eos}
Dark energy equation of state for DESI DR1 BAO+PantheonPlus combined with three different CMB distance priors.
}
\end{figure*}

Figure~\ref{fig:DESI_PP_diff_cmb_priors_eos} repeats the plots in Fig.~\ref{fig:DESI_diff_cmb_priors_eos}, but with the addition of the PantheonPlus data, i.e. for DESI DR1 BAO+PantheonPlus combined with three different CMB distance priors. Again, we see that the results are similar.

\begin{figure*}
\centering
\centering
\includegraphics[height=130pt,width=0.32\textwidth]{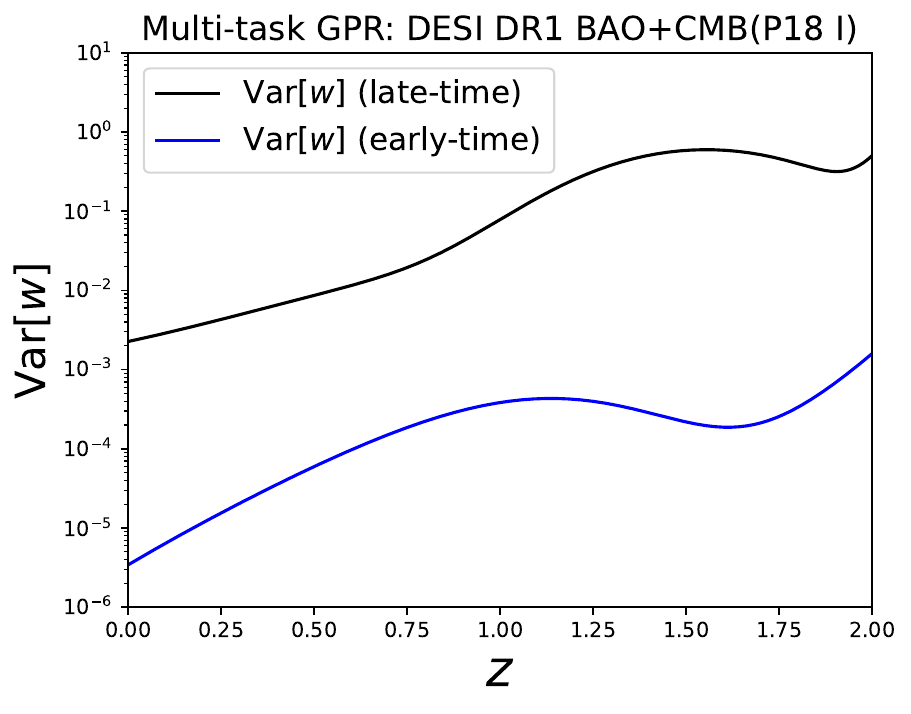}
\includegraphics[height=130pt,width=0.32\textwidth]{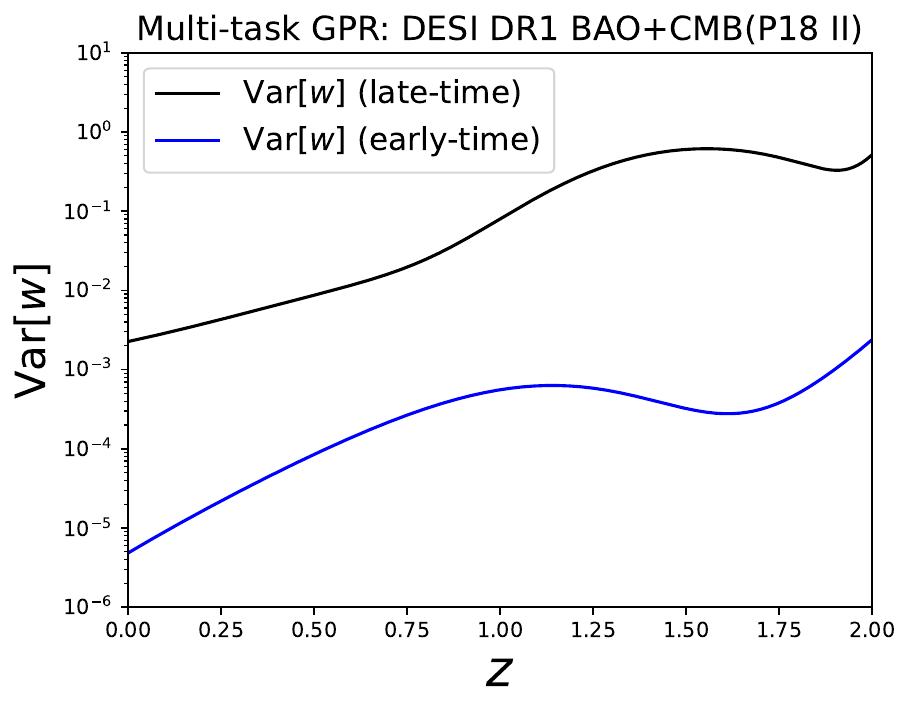}
\includegraphics[height=130pt,width=0.32\textwidth]{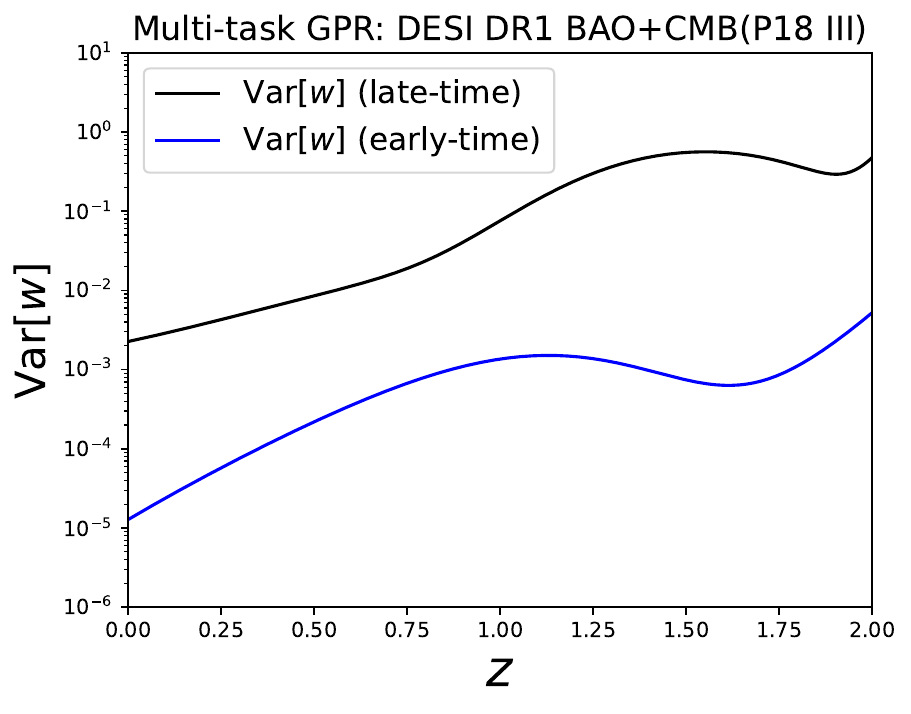}
\caption{
\label{fig:DESI_diff_cmb_priors_var_comparison}
Contribution of error in $w$ (through variance) from error in late-time data (black lines), compared to the error from early-time data (blue lines), for the three CMB distance priors.
}
\end{figure*}

The reason that the results are similar for three different CMB distance priors is that the main contribution to the error in $w$ is from the error in the late-time data, not from the error of early-time data -- the error in $w$ from late-time data dominates over that from early-time data. This can be seen if we plot the two contributions separately, by rewriting the total variance of $w$ as a sum:
\begin{align}
\text{Var}[w] &= \text{Var}[w] (\text{late})+\text{Var}[w] (\text{early}) , \\
\text{Var}[w] (\text{late}) &= \left(\!\frac{\partial w}{\partial \tilde{D}_H}\!\right)^{\!2}\! \text{Var}[\tilde{D}_H] + \left(\!\frac{\partial w}{\partial \tilde{D}'_H}\!\right)^{\!2}\! \text{Var}[\tilde{D}'_H] + 2 \frac{\partial w}{\partial \tilde{D}_H}\,\frac{\partial w}{\partial \tilde{D}'_H} \text{Cov}[\tilde{D}_H,\tilde{D}'_H] , \\
\text{Var}[w] (\text{early}) &= \left(\frac{\partial w}{\partial \gamma}\right)^2 \text{Var}[\gamma] .
\end{align}
Figure~\ref{fig:DESI_diff_cmb_priors_var_comparison} displays the late-time (black lines) and early-time (blue lines) variances for three different CMB distance priors. This shows that Var[$w$] (late) is $\sim3$ orders of magnitude larger than Var[$w$] (early). The standard deviation in the reconstructed $w$ is clearly dominated by errors in the late-time data. Hence the estimation of the standard deviation of $w$ hardly depends on the error in the early CMB data.

The discussions in this section have been done with certain combinations of data, but the conclusion and results are similar for other combinations of data considered in this study.

\section{Mean function dependence of GPR predictions}
\label{sec-gpr_different_means}

Here we explicitly show how the results depend on the different mean functions other than on the zero mean function \cite{Hwang:2022hla}. We consider three of the main cosmological models: $\Lambda$CDM, $w$CDM, and $w_0w_a$CDM.

When we reconstruct the errors, we use the full marginalization over all the parameters, i.e., {\em all parameters of the mean function and all hyper-parameters of the kernel covariance function}. In particular, this is important when the chosen mean function is close to the actual mean values of the data because only using best-fit values of these parameters from the optimization of the log marginal likelihood would have very small error bars in the estimation of any function using GP.

\begin{figure*}
\centering
\centering
\includegraphics[height=130pt,width=0.32\textwidth]{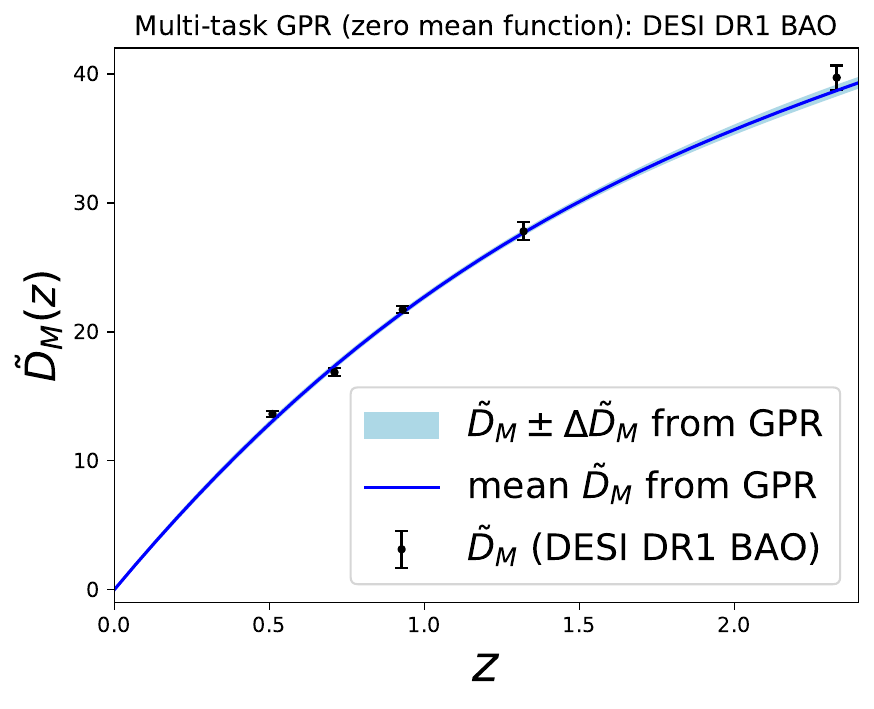}
\includegraphics[height=130pt,width=0.32\textwidth]{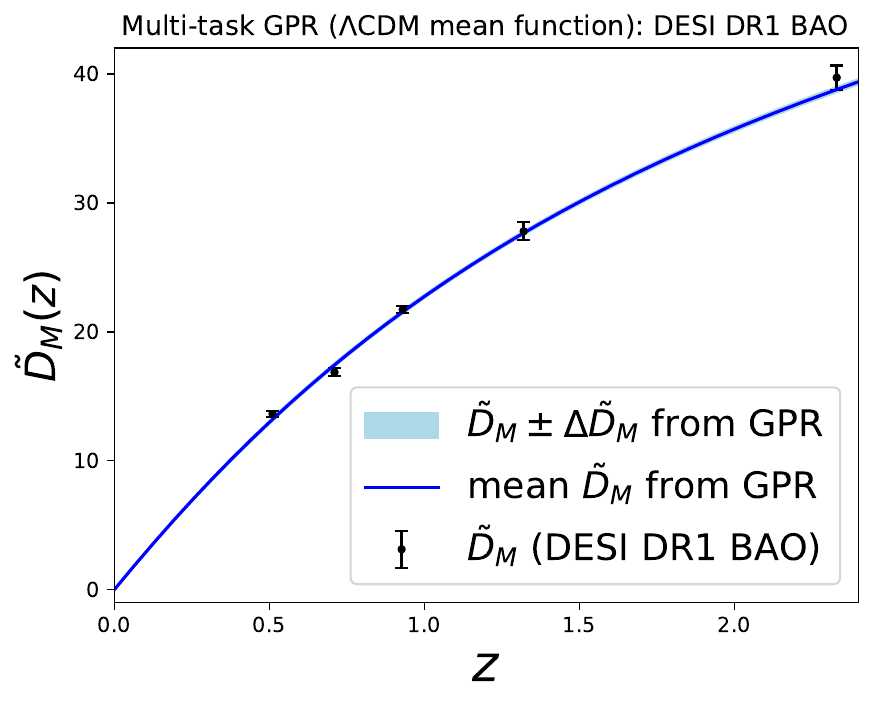}
\includegraphics[height=130pt,width=0.32\textwidth]{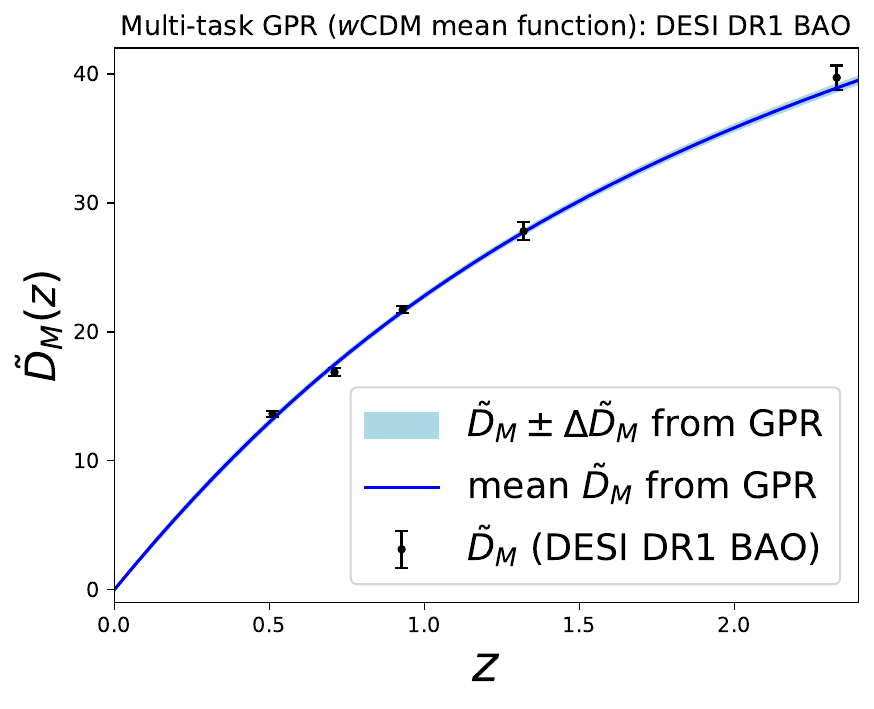} \\
\includegraphics[height=130pt,width=0.32\textwidth]{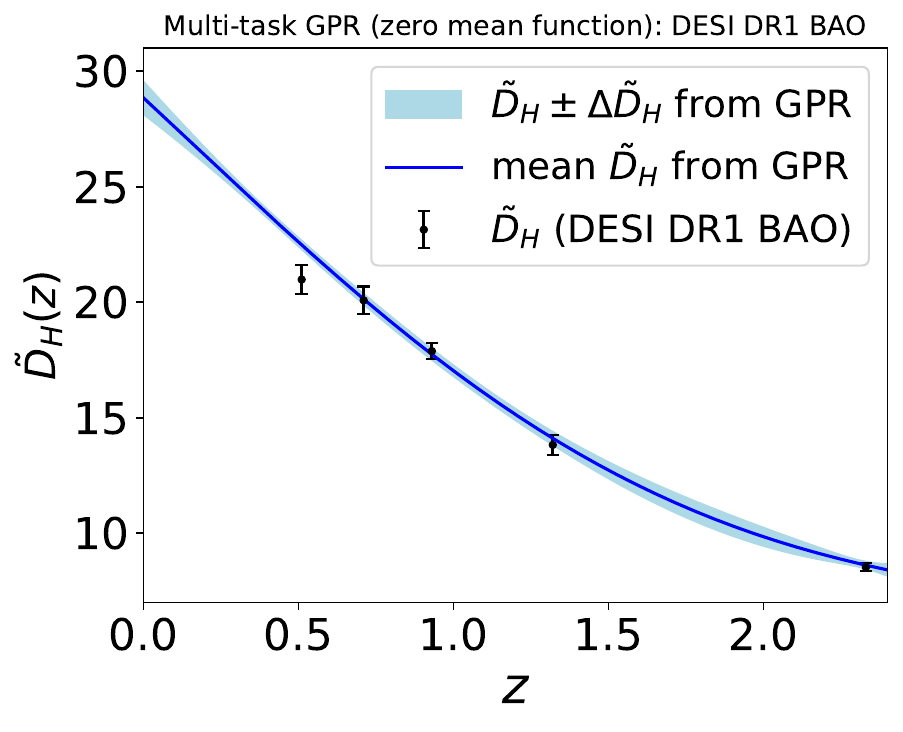}
\includegraphics[height=130pt,width=0.32\textwidth]{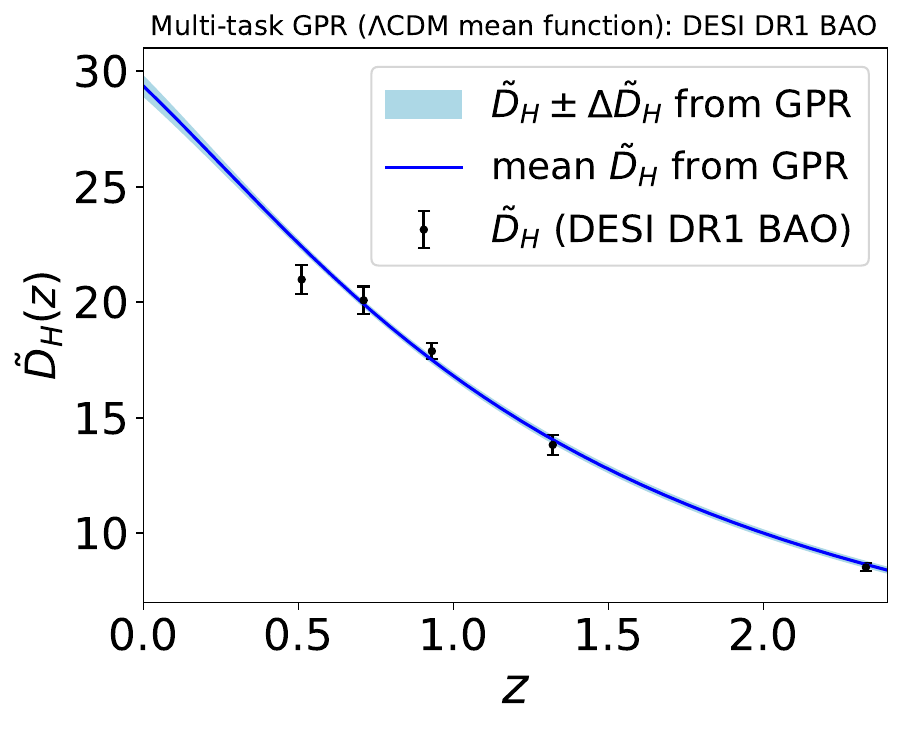}
\includegraphics[height=130pt,width=0.32\textwidth]{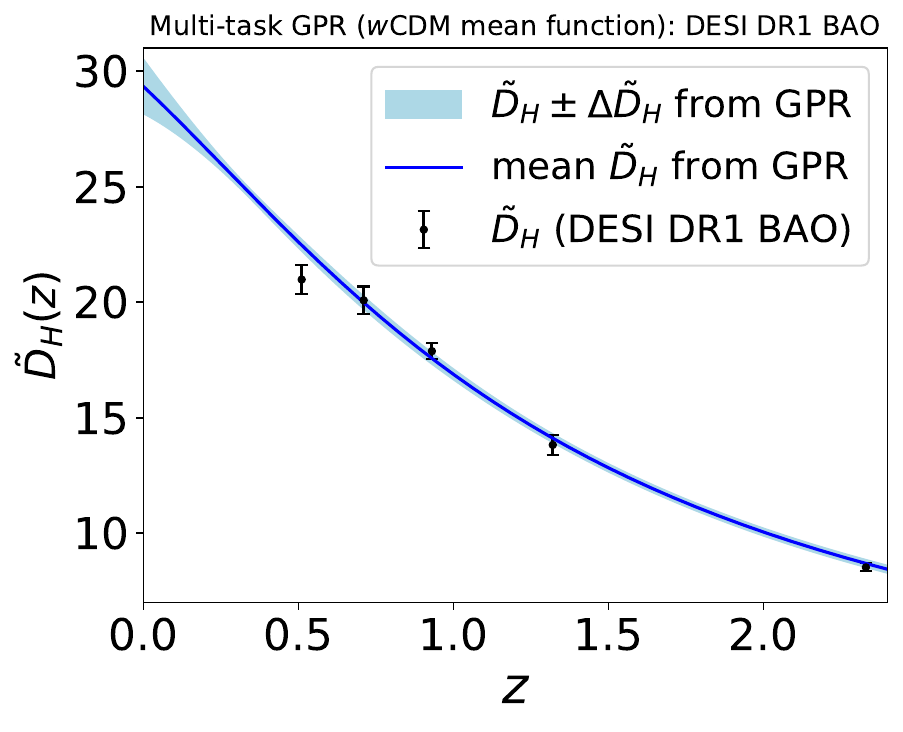} \\
\includegraphics[height=130pt,width=0.32\textwidth]{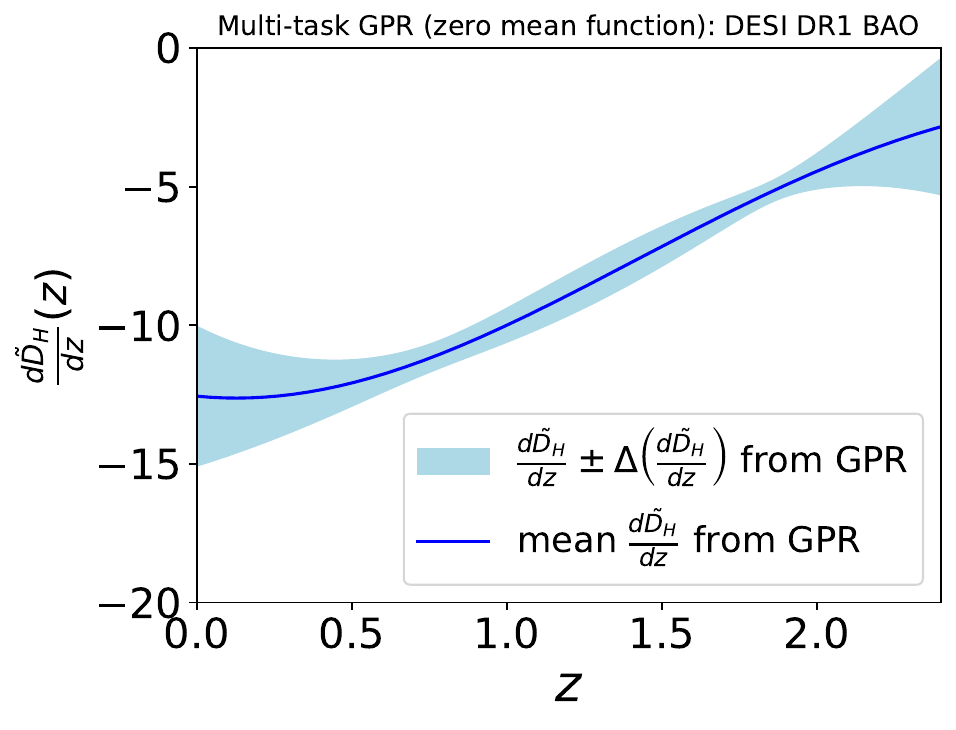}
\includegraphics[height=130pt,width=0.32\textwidth]{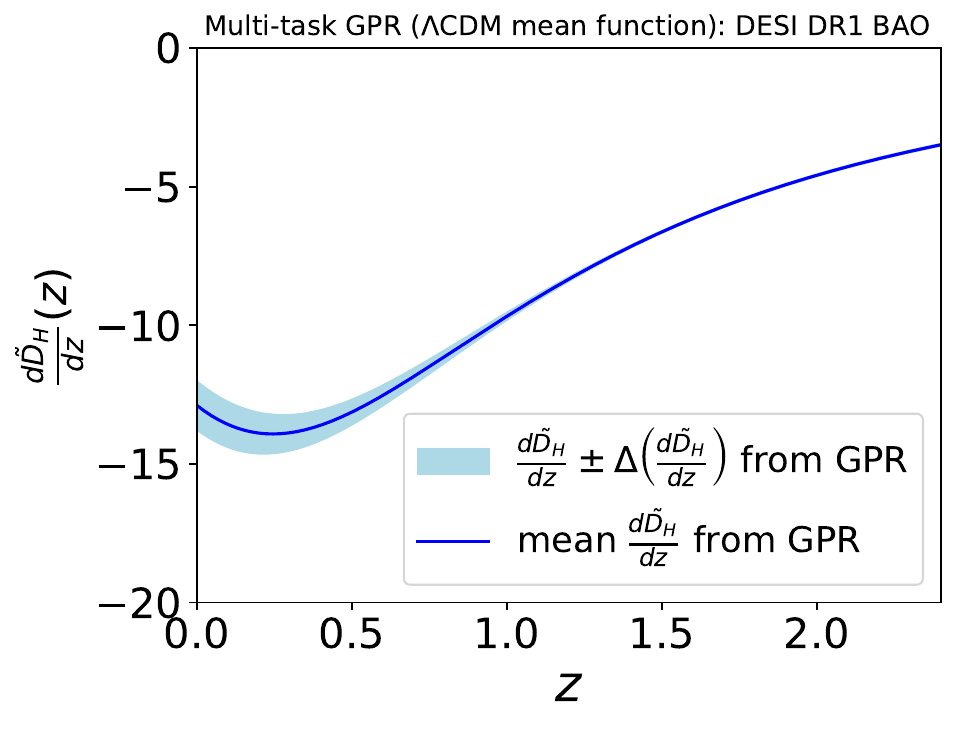}
\includegraphics[height=130pt,width=0.32\textwidth]{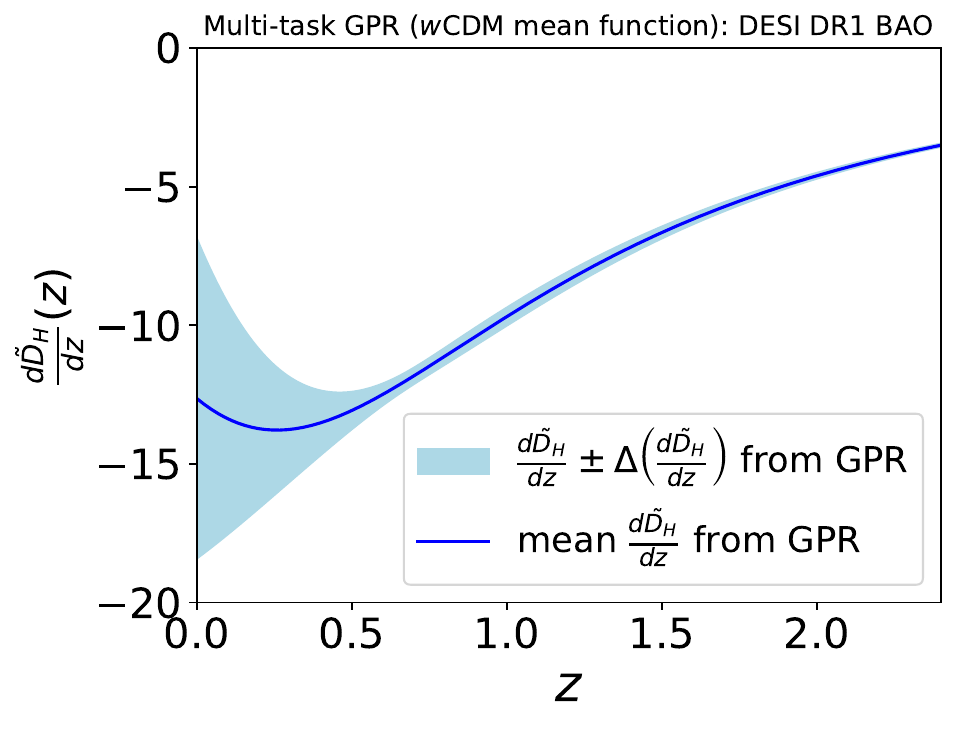}
\caption{
\label{fig:DESI_diff_mean}
Predictions of multi-task GPR for $\tilde{D}_M$ (top row), $\tilde{D}_H$ (middle row), and $\tilde{D}'_H$ (bottom row), for three mean functions: zero (left column), $\Lambda$CDM (middle column), and $w$CDM (right column), using DESI DR1 BAO data.
}
\end{figure*}

Figure~\ref{fig:DESI_diff_mean} displays the mean and 1$\sigma$ confidence intervals (standard deviations) of $\tilde{D}_M$ (top row), $\tilde{D}_H$ (middle row), and $\tilde{D}'_H$ (bottom row), for the three mean functions, i.e, zero  (left column), $\Lambda$CDM  (middle column), and $w$CDM  (right column). These are obtained using multi-task GPR  on DESI DR1 BAO data. We find no significant differences in these reconstructed functions for three different mean functions,  except for $\tilde{D}'_H$. The predictions of $\tilde{D}'_H$ slightly differ we show below that this has minimal impact on the reconstruction of the main quantity of interest, i.e., the dark energy equation of state.

\begin{figure*}
\centering
\centering
\includegraphics[height=110pt,width=0.24\textwidth]{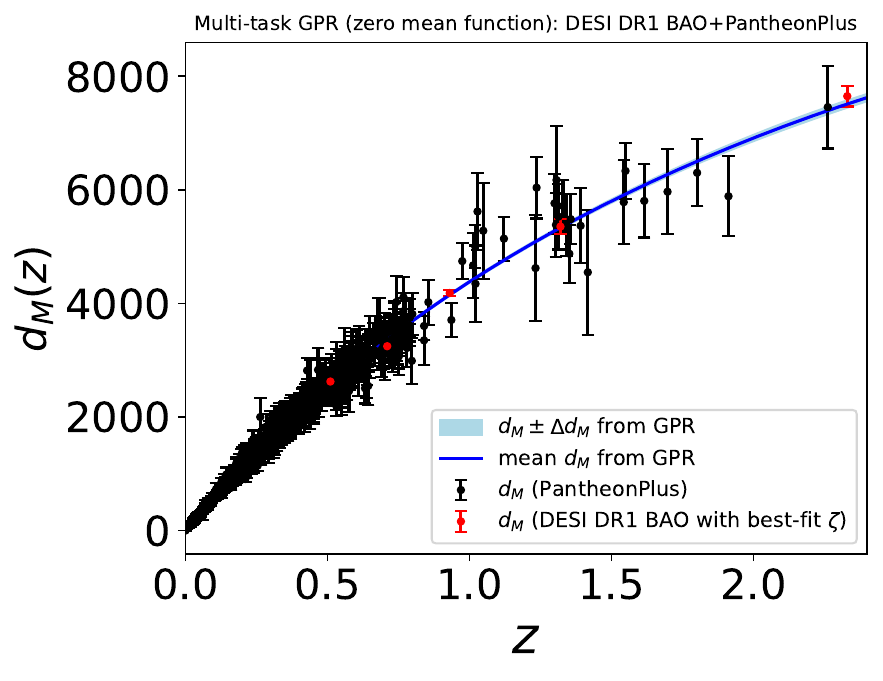}
\includegraphics[height=110pt,width=0.24\textwidth]{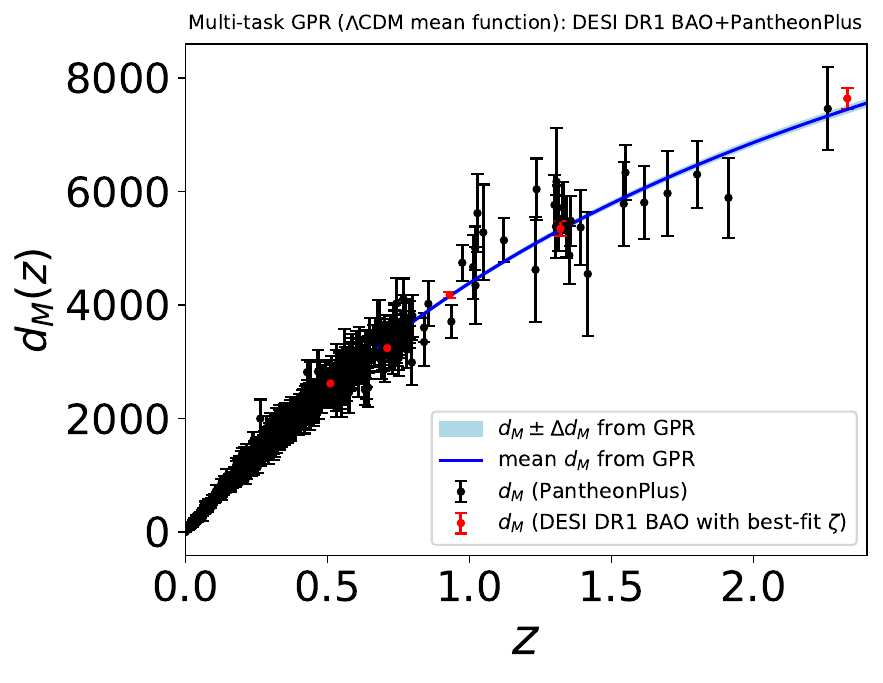}
\includegraphics[height=110pt,width=0.24\textwidth]{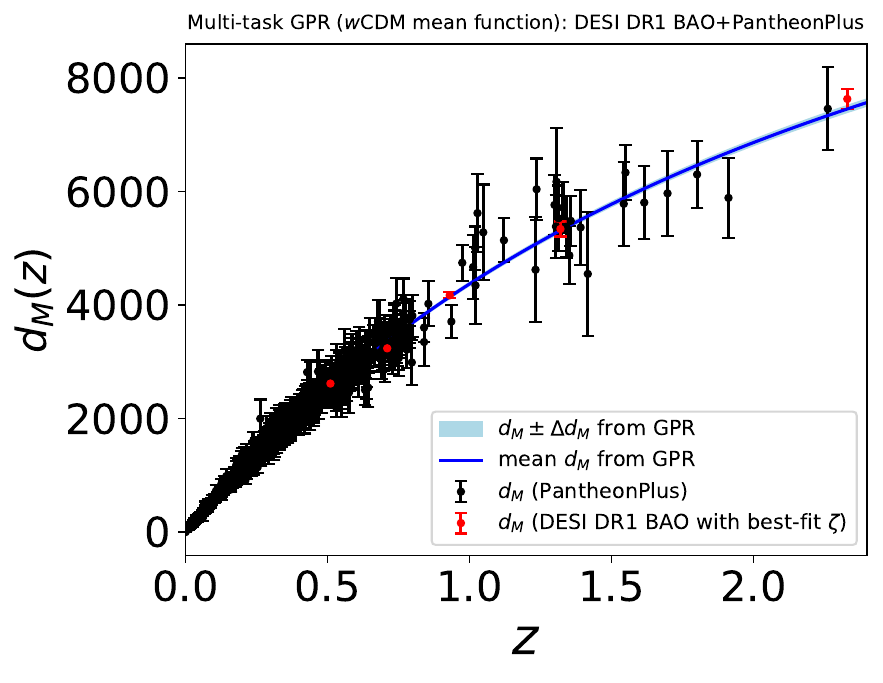}
\includegraphics[height=110pt,width=0.24\textwidth]{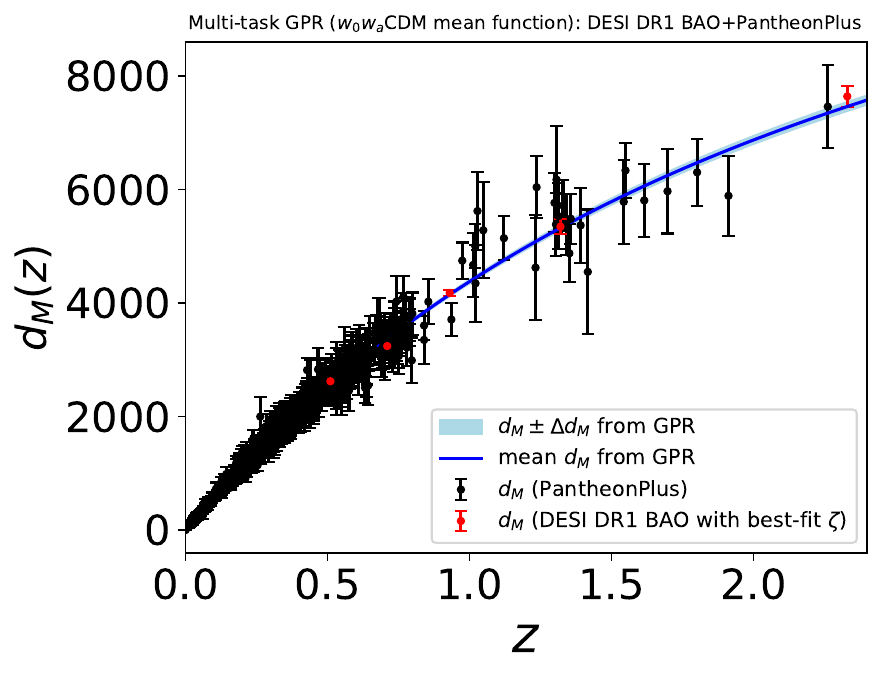} \\
\includegraphics[height=110pt,width=0.24\textwidth]{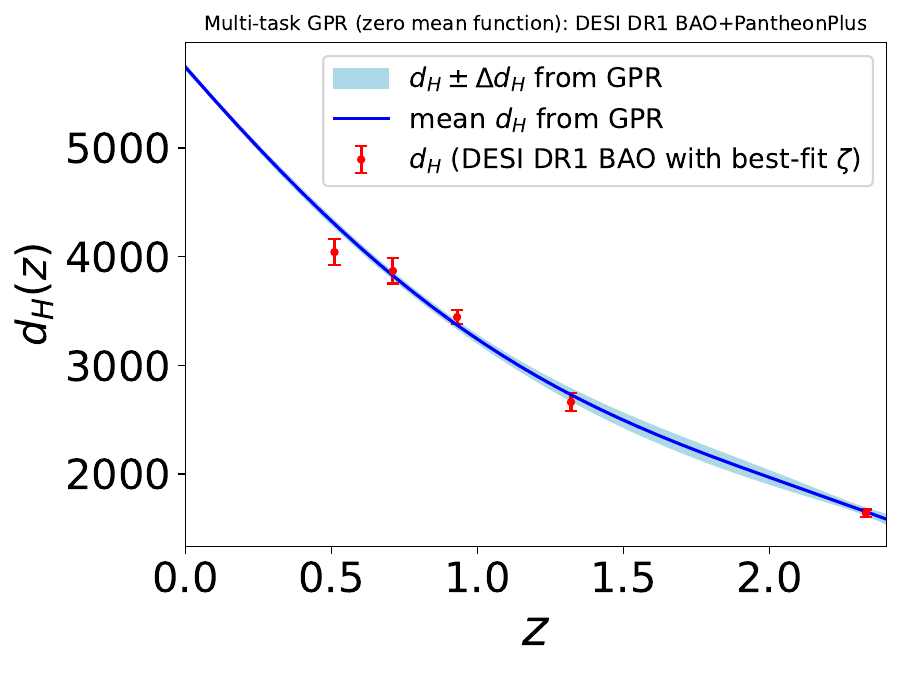}
\includegraphics[height=110pt,width=0.24\textwidth]{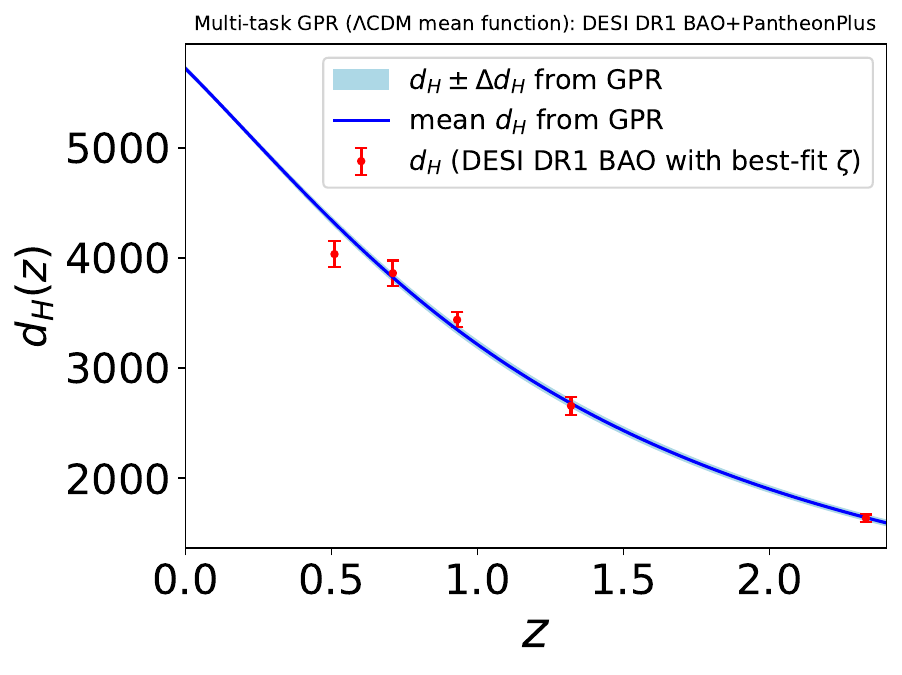}
\includegraphics[height=110pt,width=0.24\textwidth]{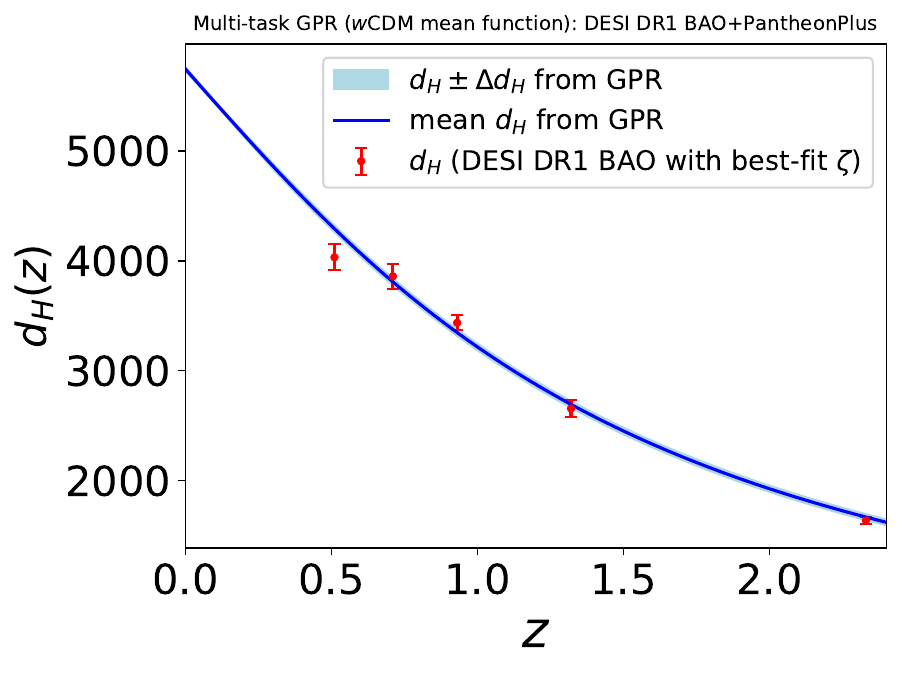}
\includegraphics[height=110pt,width=0.24\textwidth]{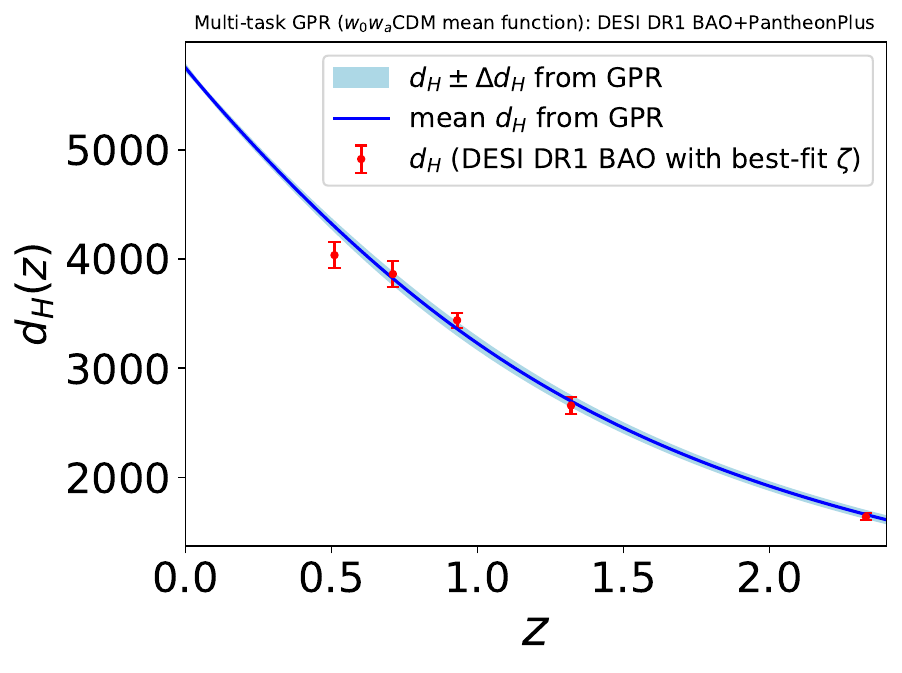} \\
\includegraphics[height=110pt,width=0.24\textwidth]{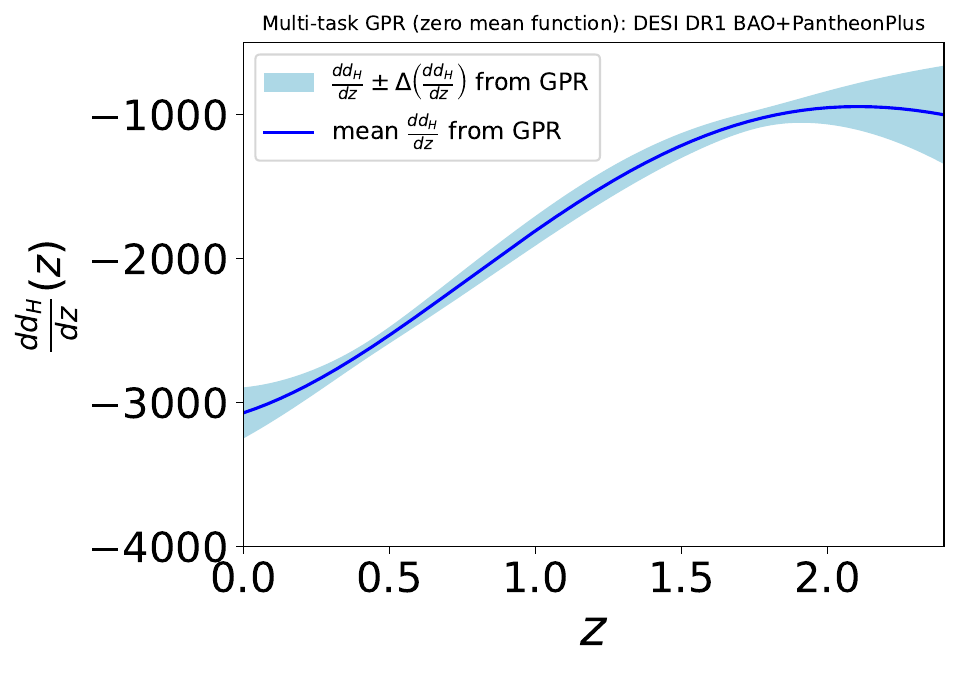}
\includegraphics[height=110pt,width=0.24\textwidth]{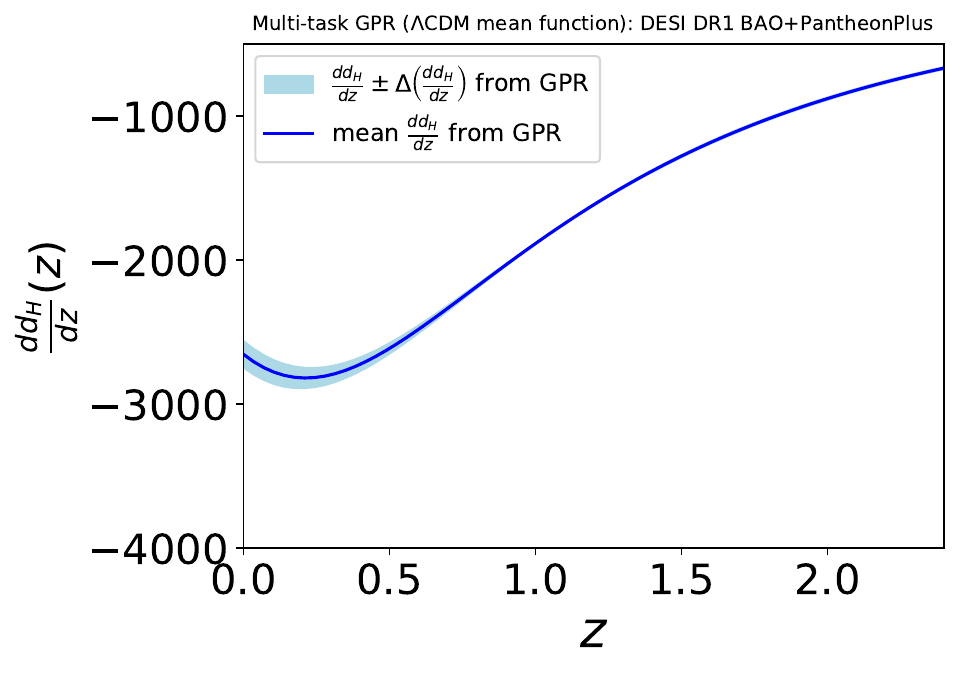}
\includegraphics[height=110pt,width=0.24\textwidth]{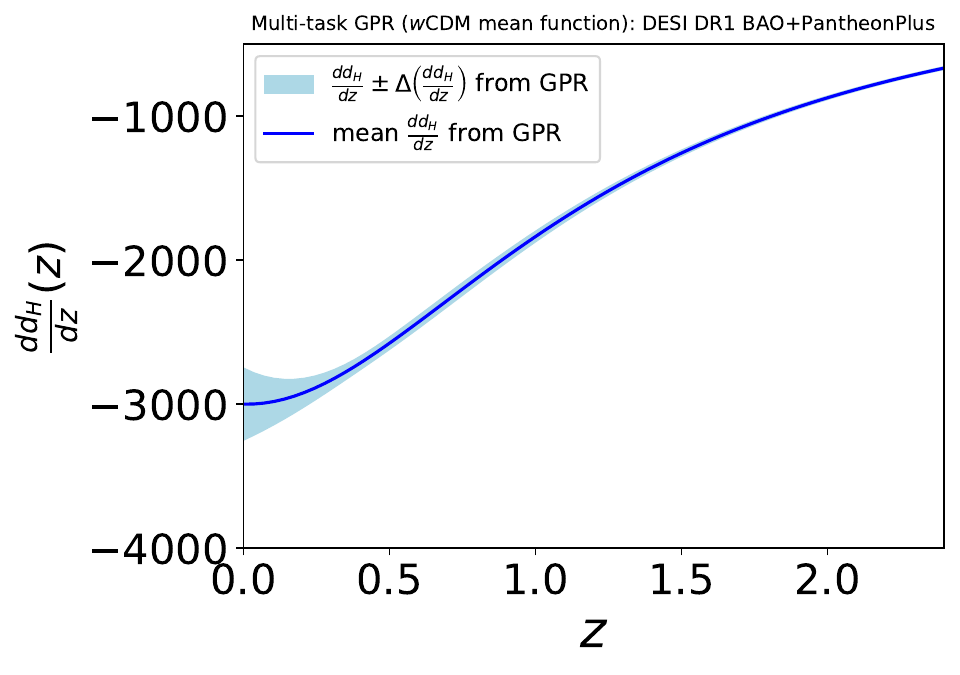}
\includegraphics[height=110pt,width=0.24\textwidth]{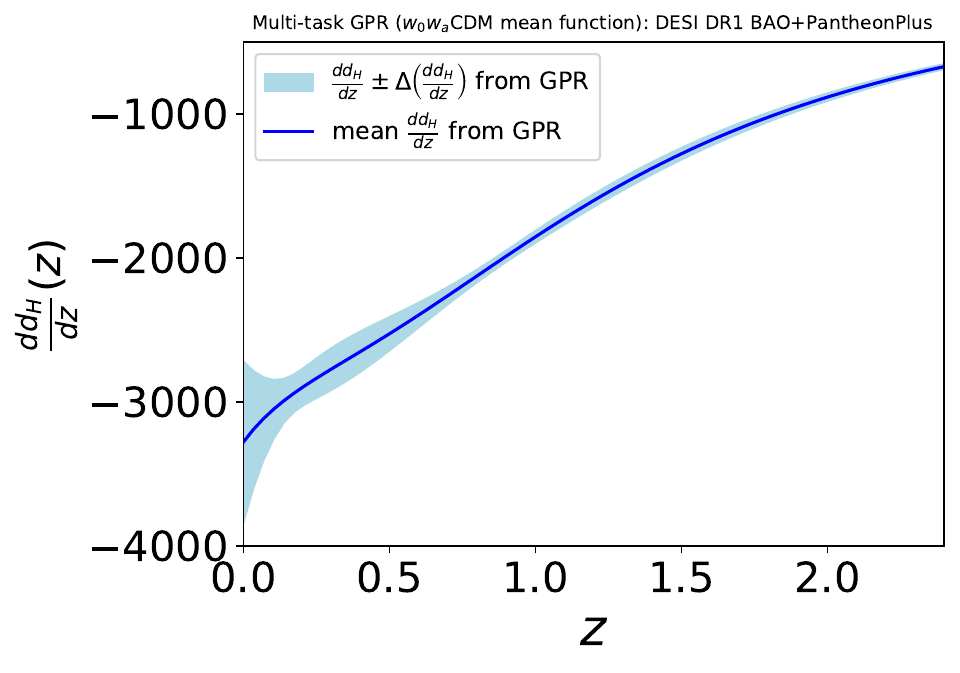}
\caption{
\label{fig:PP_DESI_diff_mean}
Predictions of multi-task GPR for $\tilde{D}_M$ (top row), $\tilde{D}_H$ (middle row), and $\tilde{D}'_H$ (bottom row) for four mean functions: zero (column 1), $\Lambda$CDM (column 2), $w$CDM (column 3), and $w_0w_a$CDM (column 4) using DESI DR1 BAO+PantheonPlus data.
}
\end{figure*}

In Fig.~\ref{fig:PP_DESI_diff_mean}, we repeat the plots in Fig.~\ref{fig:DESI_diff_mean}, but using the combinations of DESI DR1 BAO and PantheonPlus data, and with a fourth mean function,  $w_0w_a$CDM.\footnote{We did not include $w_0w_a$CDM in Fig.~\ref{fig:DESI_diff_mean}, since we do not find any proper constraints due to the poor constraining power of DESI DR1 BAO data alone on a general model like $w_0w_a$CDM.} We find a similar conclusion as for Fig.~\ref{fig:DESI_diff_mean}.

\begin{figure*}
\centering
\centering
\includegraphics[height=130pt,width=0.32\textwidth]{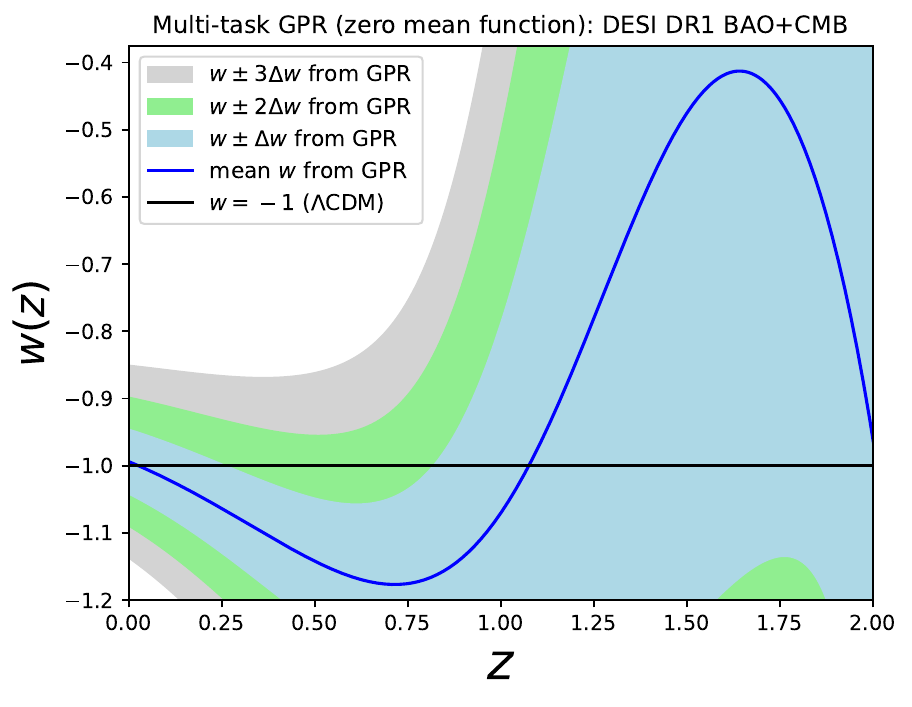}
\includegraphics[height=130pt,width=0.32\textwidth]{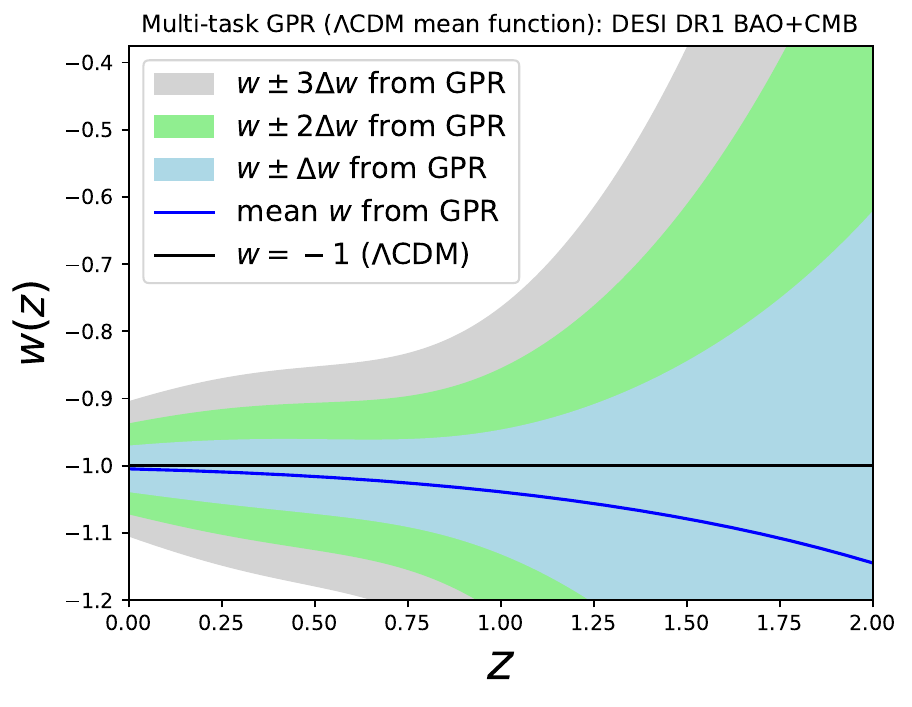}
\includegraphics[height=130pt,width=0.32\textwidth]{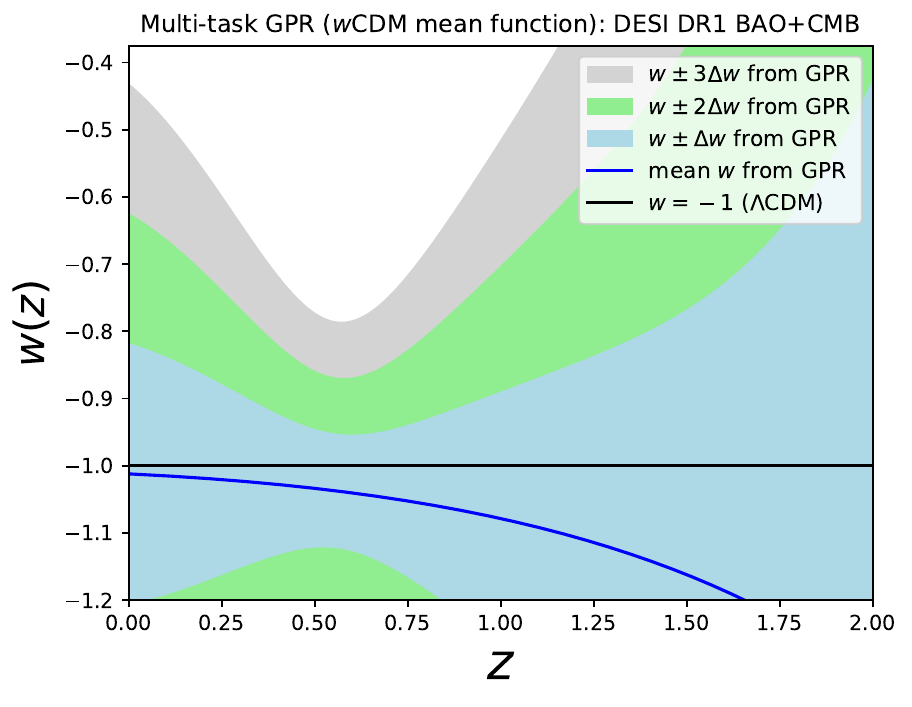}
\caption{
\label{fig:DESI_CMB_diff_mean_eos}
Reconstructed $w$ with the corresponding 1$\sigma$, 2$\sigma$, and 3$\sigma$ confidence regions obtained from DESI DR1 BAO+CMB data for three mean functions, as in Fig.~\ref{fig:DESI_diff_mean}.
}
\end{figure*}

From the reconstructed $\tilde{D}_H$ and $\tilde{D}'_H$, we compute the (effective) equation of state of dark energy $w$ (as we did before in the main text) with the addition of CMB data for each of the mean functions. For the DESI DR1 BAO and CMB combined data, we plot $w$ and the corresponding 1$\sigma$, 2$\sigma$, and 3$\sigma$ confidence regions in Fig.~\ref{fig:DESI_CMB_diff_mean_eos}. The color combinations are the same as in the main text. We find that the reconstructed mean values of $w$ are slightly different for the different mean functions, but the differences are significantly smaller than the estimated standard deviation corresponding to any of the mean functions. Thus in the sense of statistical confidence, the results are almost similar. This can be stated in another way: that we find no significant evidence of deviation from the $\Lambda$CDM model and this conclusion is the same for all three chosen mean functions.

\begin{figure*}
\centering
\centering
\includegraphics[height=110pt,width=0.24\textwidth]{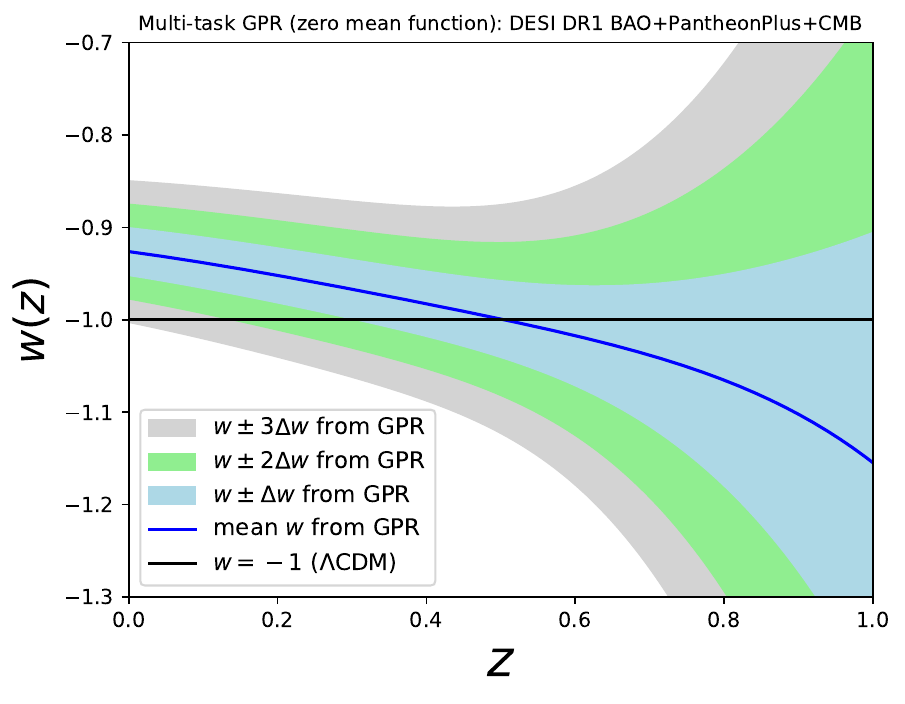}
\includegraphics[height=110pt,width=0.24\textwidth]{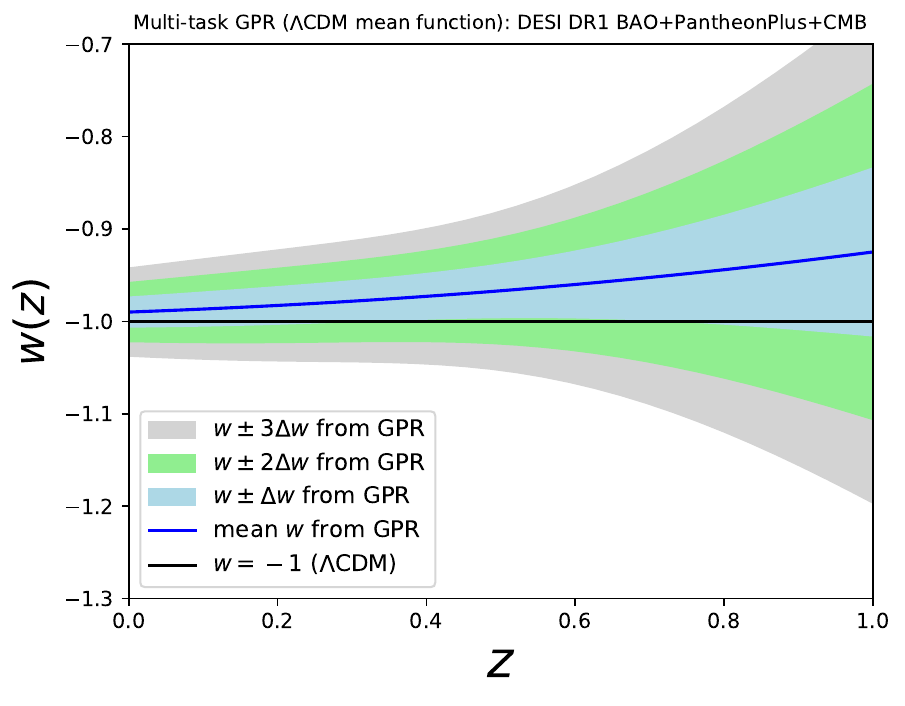}
\includegraphics[height=110pt,width=0.24\textwidth]{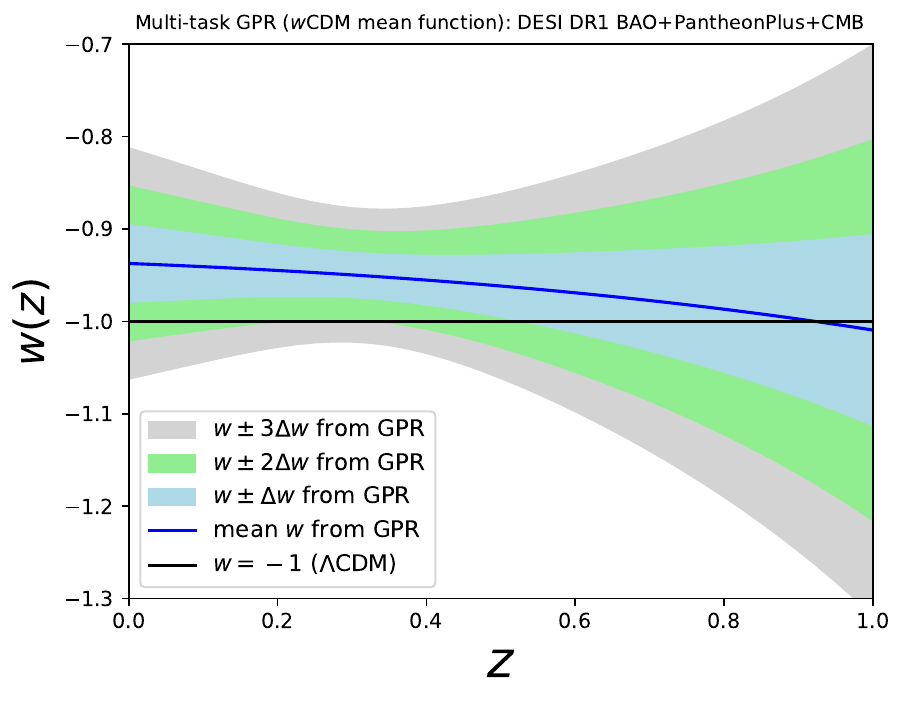}
\includegraphics[height=110pt,width=0.24\textwidth]{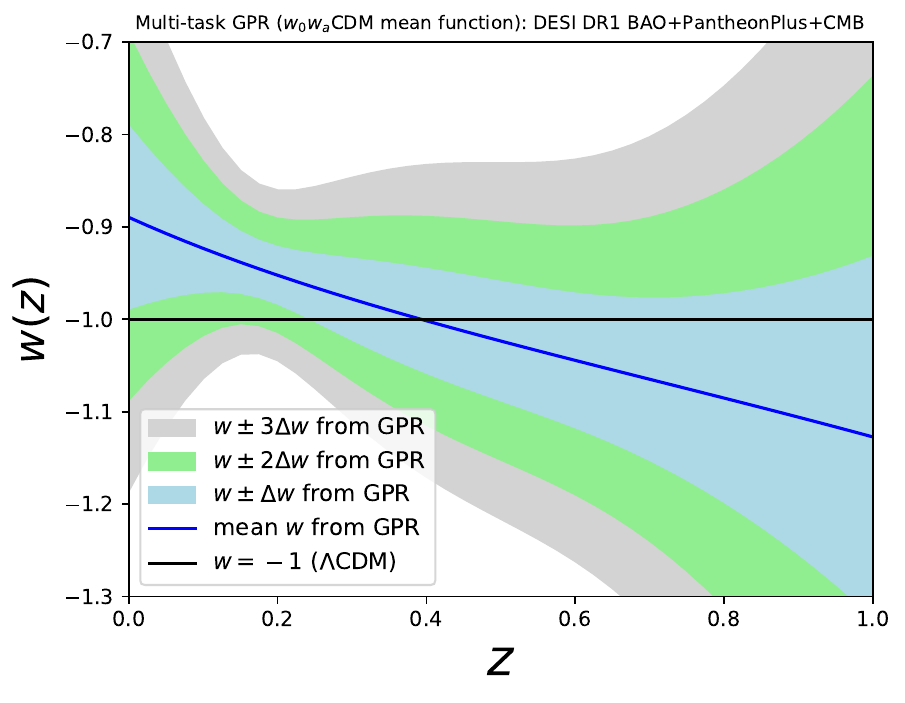}
\caption{
\label{fig:PP_DESI_CMB_diff_mean_eos}
Reconstructed $w$  from DESI DR1 BAO+PantheonPlus+CMB data, with 1, 2, 3$\sigma$ confidence regions, for four mean functions, as in Fig.~\ref{fig:PP_DESI_diff_mean}.
}
\end{figure*}

Figure~\ref{fig:PP_DESI_CMB_diff_mean_eos} repeats the plots in Fig.~\ref{fig:DESI_CMB_diff_mean_eos}, but now with the addition of the  PantheonPlus data. We also include the $w_0w_a$CDM mean function. Similar conclusions apply here: in almost the entire redshift region the $\Lambda$CDM model is either $\sim1\sigma$ away or well within it. This result is significantly independent of the chosen mean function. 

Our final comment is that there is no significant evidence of dynamical dark energy and this remains independent of any mean function.

\clearpage
\bibliographystyle{JHEP}
\bibliography{references}

\end{document}